%% file: muc-perf.tex
\documentclass[11pt,twoside,a4paper]{cernrep}
\usepackage{rep_common}
\usepackage{cancel}
\usepackage{gensymb}
\usepackage{graphicx}
\usepackage{siunitx} 
\usepackage{xspace}
\usepackage{cite}
\usepackage{float}
\usepackage{lineno}
\floatstyle{plaintop}
\restylefloat{table}
\usepackage{hyperref}
\usepackage{xstring,ragged2e}

\usepackage{graphicx,multirow,colortbl}

\input workshopsymbols.tex
\input papersymbols.tex

\newcommand\snowmass{\begin{center}\rule[-0.2in]{\hsize}{0.01in}\\\rule{\hsize}{0.01in}\\
\vskip 0.1in Submitted to the  Proceedings of the US Community Study\\ 
on the Future of Particle Physics (Snowmass 2021)\\ 
\rule{\hsize}{0.01in}\\\rule[+0.2in]{\hsize}{0.01in} \end{center}}

\newcommand{\sectionauthor}[1]{}

\newcommand{\note}[1]{}

\ifdefined\qty\else
  \ifdefined\NewCommandCopy
    \NewCommandCopy\qty\SI
  \else
    \NewDocumentCommand\qty{O{}mm}{\SI[#1]{#2}{#3}}
  \fi
\fi
\ifdefined\unit\else
  \ifdefined\NewCommandCopy
    \NewCommandCopy\unit\si
  \else
    \NewDocumentCommand\unit{O{}m}{\si[#1]{#2}}
  \fi
\fi

\begin{document}

\title{{\normalfont\bfseries\boldmath\huge
      \begin{center}
        Simulated Detector Performance at the Muon Collider
      \end{center}
      \vspace*{-35pt}
    }
    {\textnormal{\normalsize \snowmass
        \vspace*{-40pt}
      }}
    {\textnormal{\normalsize
        \abstract{
In this paper we report on the current status of studies on the expected performance for a detector designed to operate in a muon collider environment. Beam-induced backgrounds (BIB) represent the main challenge in the design of the detector and the event reconstruction algorithms. 
The current detector design aims to show that satisfactory performance can be achieved, while further optimizations are expected to significantly improve the overall performance.
We present the characterization of the expected beam-induced background, describe the detector design and software used for detailed event simulations taking into account BIB effects. 
The expected performance of charged-particle reconstruction, jets, electrons, photons and muons is discussed, including an initial study on heavy-flavor jet tagging. A simple method to measure the delivered luminosity is also described. Overall, the proposed design and reconstruction algorithms can successfully reconstruct the high transverse-momentum objects needed to carry out a broad physics program.
        }\\[30pt]}}
    {\textnormal{\normalsize\justifying
    This is one of the five reports submitted to Snowmass by the muon colliders community at large. The reports' preparation effort has been coordinated by the International Muon Collider Collaboration. Authors and Signatories have been collected with a
        \href{https://indico.cern.ch/event/1130036/}{subscription page}, and are defined as follows:
        \begin{itemize}
          \item An ``Author'' contributed to the results documented in the report in any form, including e.g.~by participating to the discussions of the community meetings and sending comments on the draft, or plans to contribute to the future work.
          \item
                A ``Signatory'' expresses support to the efforts described in the report and endorses the Collaboration plans.
        \end{itemize}
      }}
}

\input{authors}

\begin{titlepage}

\vspace*{-1.8cm}

\noindent
\begin{tabular*}{\linewidth}{lc@{\extracolsep{\fill}}r@{\extracolsep{0pt}}}
\vspace*{-1.2cm}\mbox{\!\!\!\includegraphics[width=.14\textwidth]{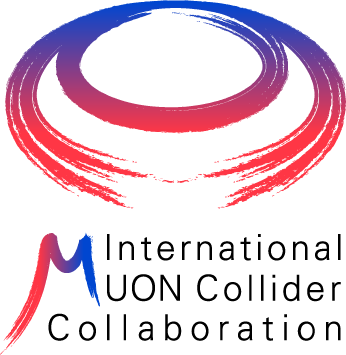}} & &  \\
 & & \today \\  
 & & \href{https://muoncollider.web.cern.ch}{https://muoncollider.web.cern.ch} \\ 
 & & \\
\hline
\end{tabular*}

\vspace*{0.3cm}
%
%
\maketitle
\vspace{\fill}

\end{titlepage}


\clearpage

\section*{Executive Summary}
\input{executive-summary}
\clearpage

\tableofcontents
\clearpage
\def\thefootnote{\fnsymbol{footnote}}
\setcounter{footnote}{0}

\section{Introduction}
\label{sec:introduction}
\sectionauthor{Simone Pagan Griso}
\input{introduction}

\section{Beam-Induced Background characterization}
\label{sec:bib}
\sectionauthor{Camilla Curatolo, Nazar Bartosik}
\input{bib}

\section{Overview of the Detector design}
\label{sec:detector}
\sectionauthor{Nazar Bartosik}
\input{detector}

\section{Detector Simulation Software}
\label{sec:software}
\sectionauthor{Nazar Bartosik}
\input{software}

\section{Charged particle reconstruction}
\label{sec:tracking}
\sectionauthor{Karol Krizka}
\input{tracking}

\section{Jets}
\label{sec:jets}
\sectionauthor{Lorenzo Sestini}
\input{jets}
\subsection{Heavy-flavor jet identification}
\label{sec:jets:heavyflavour}
\sectionauthor{Laura Buonincontri, Paola Mastrapasqua}
\input{jets_heavyflavour}

\section{Photons and Electrons}
\label{sec:phele}
\sectionauthor{Massimo Casarsa}
\subsection{Photon reconstruction and identification}
\input{photons}
\label{sec:photons}
\subsection{Electron reconstruction and identification}
\label{sec:electrons}
\input{electrons}

\section{Muons}
\label{sec:muons}
\sectionauthor{Ilaria Vai, Chiara Aim\`e, Cristina Riccardi, Paola Salvini, Nicol\`o Valle}
\input{muons}


\section{Luminosity measurement}
\label{sec:lumi}
\sectionauthor{Donatella Lucchesi, Laura Buonincontri}
\input{lumi}


\section{Conclusions}
\input{conclusions}


\bibliographystyle{JHEP}
\bibliography{muc-perf.bib}  

\end{document}

%% file: workshopsymbols.tex
\def\beq{\begin{equation}}
\def\eeq#1{\label{#1}\end{equation}}
\def\eeqn{\end{equation}}

\newenvironment{Eqnarray}%
   {\arraycolsep 0.14em\begin{eqnarray}}{\end{eqnarray}}
\def\beqa{\begin{Eqnarray}}
\def\eeqa#1{\label{#1}\end{Eqnarray}}
\def\eeqan{\end{Eqnarray}}

\def\lsim{\mathrel{\raise.3ex\hbox{$<$\kern-.75em\lower1ex\hbox{$\sim$}}}}
\def\gsim{\mathrel{\raise.3ex\hbox{$>$\kern-.75em\lower1ex\hbox{$\sim$}}}}

\def\del{\partial}
\def\Dslash{\not{\hbox{\kern-4pt $D$}}}
\def\dslash{\not{\hbox{\kern-2pt $\del$}}}
\def\pslash{\not{\hbox{\kern-2pt $p$}}}
\def\ETmiss{\not{\hbox{\kern-4pt $E$}}_T}

\def\Dlr{\mathrel{\raise1.5ex\hbox{$\leftrightarrow$\kern-1em\lower1.5ex\hbox{$D$}}}}

\def\Pl{{\mbox{\scriptsize Pl}}}

\def\ee{e^+e^-}

\def\MSB{{\bar{M \kern -2pt S}}}
\def\msb{{\bar{\scriptsize M \kern -1pt S}}}

\def\drb{{\bar{\scriptsize D \kern -1pt R}}}

\def\MeV{{\rm MeV}}
\def\GeV{{\rm GeV}}
\def\TeV{{\rm TeV}}



%% file: papersymbols.tex
\def\mumu{\ensuremath{\mu^{+}\mu^{-}}\xspace}
\def\ee{\ensuremath{e^{+}e^{-}}\xspace}

\def\pt{\ensuremath{p_\textrm{T}}\xspace}

%% file: authors.tex
\newcounter{instituteref}
\newcommand{\iinstitute}[2]{\refstepcounter{instituteref}\label{#1}$^{\ref{#1}}$\href{http://inspirehep.net/record/#1}{#2}}
\newcommand{\iauthor}[3]{\href{http://inspirehep.net/record/#1}{#2}$^{#3}$}
\author{Editors: \\
    \iauthor{1073143}{N.~Bartosik}{\ref{902889}},
    \iauthor{1252769}{K.~Krizka}{\ref{902953}},
    \iauthor{1048820}{S.~Pagan~Griso}{\ref{902953}}
    \\ \vspace*{4mm}Authors: \\
    \iauthor{1757334}{C.~Aim\`e}{\ref{943385},\ref{902885}},
    \iauthor{1060487}{A.~Apyan}{\ref{902682}},
    \iauthor{1067349}{M.A.~Mahmoud.}{\ref{912409}},
    \iauthor{1029828}{A.~Bertolin}{\ref{902884}},
    \iauthor{1015478}{A.~Braghieri}{\ref{902885}},
    \iauthor{1894439}{L.~Buonincontri}{\ref{902884},\ref{903113}},
    \iauthor{1707397}{S.~Calzaferri}{\ref{902885}},
    \iauthor{1057458}{M.~Casarsa}{\ref{902888}},
    \iauthor{}{L.~Castelli}{\ref{903113}},
    \iauthor{1014281}{M.G.~Catanesi}{\ref{902877}},
    \iauthor{1418744}{F.G.~Celiberto}{\ref{906718},\ref{912328}},
    \iauthor{1021757}{A.~Cerri}{\ref{1241166}},
    \iauthor{1037833}{G.~Chachamis}{\ref{905303}},
    \iauthor{1013275}{A.~Colaleo}{\ref{902660},\ref{902877}},
    \iauthor{1937290}{C.~Curatolo}{\ref{902882}},
    \iauthor{2049478}{G.~Da~Molin}{\ref{903113}},
    \iauthor{1012395}{S.~Dasu}{\ref{903349}},
    \iauthor{1012030}{D.~Denisov}{\ref{902689}},
    \iauthor{1012025}{H.~Denizli}{\ref{908452}},
    \iauthor{1031269}{B.~Di~Micco}{\ref{906528},\ref{907692}},
    \iauthor{1011508}{T.~Dorigo}{\ref{902884}},
    \iauthor{1404358}{F.~Errico}{\ref{902660},\ref{902877}},
    \iauthor{1010105}{A.~Ferrari}{\ref{1276460}},
    \iauthor{1719039}{D.~Fiorina}{\ref{902885}},
    \iauthor{1009009}{M.~Gallinaro}{\ref{905303}},
    \iauthor{1894454}{L.~Giambastiani}{\ref{903113},\ref{902884}},
    \iauthor{1262268}{A.~Gianelle}{\ref{902884}},
    \iauthor{1971617}{C.~Giraldin}{\ref{903113}},
    \iauthor{1006149}{M.~Herndon}{\ref{903349}},
    \iauthor{1067690}{T.R.~Holmes}{\ref{1623978}},
    \iauthor{1028687}{S.~Jindariani}{\ref{902796}},
    \iauthor{1345391}{G.K.~Krintiras}{\ref{902912}},
    \iauthor{1071846}{L.~Lee}{\ref{1623978}},
    \iauthor{1074984}{Q.~Li}{\ref{903603}},
    \iauthor{1000076}{R.~Lipton}{\ref{902796}},
    \iauthor{}{S.~Lomte}{\ref{903349}},
    \iauthor{999862}{K.R.~Long}{\ref{902868},\ref{903174}},
    \iauthor{999654}{D.~Lucchesi}{\ref{903113},\ref{902884}},
    \iauthor{1971307}{P.~Mastrapasqua}{\ref{910783}},
    \iauthor{1074063}{F.~Meloni}{\ref{902770}},
    \iauthor{2049482}{A.~Montella}{\ref{902888}},
    \iauthor{}{F.~Nardi}{\ref{903113},\ref{902884}},
    \iauthor{994095}{N.~Pastrone}{\ref{902889}},
    \iauthor{1772198}{A.~Pellecchia}{\ref{902660}},
    \iauthor{1050691}{K.~Potamianos}{\ref{903112}},
    \iauthor{992463}{E.~Radicioni}{\ref{902877}},
    \iauthor{1217056}{R.~Radogna}{\ref{902660},\ref{902877}},
    \iauthor{1020819}{C.~Riccardi}{\ref{943385},\ref{902885}},
    \iauthor{1028713}{L.~Ristori}{\ref{902796}},
    \iauthor{990505}{P.~Salvini}{\ref{902885}},
    \iauthor{}{I.~Sarra}{},
    \iauthor{989645}{D.~Schulte}{\ref{902725}},
    \iauthor{1019799}{A.~Senol}{\ref{908452}},
    \iauthor{1342183}{L.~Sestini}{\ref{902884}},
    \iauthor{1622677}{F.~M.~Simone}{\ref{902660},\ref{902877}},
    \iauthor{1074094}{R.~Simoniello}{\ref{902725}},
    \iauthor{}{A.~Stamerra}{\ref{902660},\ref{902877}},
    \iauthor{1071725}{X.~Sun}{\ref{1210798}},
    \iauthor{1078570}{M.~Swiatlowski}{\ref{903290}},
    \iauthor{1268914}{J.~Tang}{\ref{903702},\ref{903123}},
    \iauthor{1878399}{E.~A.~Thompson}{\ref{902770}},
    \iauthor{1265350}{I.~Vai}{\ref{902885}},
    \iauthor{1613622}{M.~Valente}{\ref{903290}},
    \iauthor{1643523}{N.~Valle}{\ref{943385},\ref{902885}},
    \iauthor{1071756}{R.~Venditti}{\ref{902660},\ref{902877}},
    \iauthor{1063935}{P.~Verwilligen}{\ref{902877}},
    \iauthor{1073818}{H.~Weber}{\ref{902858}},
    \iauthor{1971310}{A.~Zaza}{\ref{902660},\ref{902877}},
    \iauthor{1863481}{D.~Zuliani}{\ref{903113},\ref{902884}}
    \\ \vspace*{4mm} Signatories: \\
    \iauthor{1018987}{D.~Acosta}{\ref{903156}},
    \iauthor{1018902}{K.~Agashe}{\ref{902990}},
    \iauthor{1018633}{B.C.~Allanach}{\ref{907623}},
    \iauthor{1423596}{P.~Andreetto}{\ref{902884}},
    \iauthor{1018264}{F.~Anulli}{\ref{902887}},
    \iauthor{1049113}{A.~Apresyan}{\ref{902796}},
    \iauthor{1491320}{P.~Asadi}{\ref{1237813}},
    \iauthor{}{D.~Athanasakos}{\ref{910429}},
    \iauthor{1041900}{A.~Azatov}{\ref{904416},\ref{902888}},
    \iauthor{1028433}{J.J.~Back}{\ref{903734}},
    \iauthor{1424044}{L.~Bandiera}{\ref{905268}},
    \iauthor{1017330}{R.~J.~Barlow}{\ref{911708}},
    \iauthor{1037853}{E.~Barzi}{\ref{902796},\ref{903092}},
    \iauthor{2031609}{F.~Batsch}{\ref{902725}},
    \iauthor{1068289}{M.~Bauce}{\ref{902887},\ref{903168}},
    \iauthor{1016672}{J.~S.~Berg}{\ref{902689}},
    \iauthor{1016557}{J.~Berryhill}{\ref{902796}},
    \iauthor{1049763}{A.~Bersani}{\ref{902881}},
    \iauthor{1020223}{K.M.~Black}{\ref{903349}},
    \iauthor{1015872}{M.~Bonesini}{\ref{902882},\ref{907960}},
    \iauthor{1015812}{C.~Booth}{\ref{903196}},
    \iauthor{1794682}{S.~Bottaro}{\ref{903128},\ref{902886}},
    \iauthor{1031261}{D.~Bowring}{\ref{902796}},
    \iauthor{1015214}{A.~Bross}{\ref{902796}},
    \iauthor{1114205}{E.~Brost}{\ref{902689}},
    \iauthor{1077579}{D.~Buttazzo}{\ref{902886}},
    \iauthor{1670912}{B.~Caiffi}{\ref{902881}},
    \iauthor{1014742}{G.~Calderini}{\ref{926589},\ref{903119}},
    \iauthor{}{P.~Cameron}{\ref{902689}},
    \iauthor{1275234}{R.~Capdevilla}{\ref{908474},\ref{903282}},
    \iauthor{}{F.~Casaburo}{},
    \iauthor{1020772}{G.~Cavoto}{\ref{903168},\ref{902887}},
    \iauthor{1075318}{L.~Celona}{\ref{902879}},
    \iauthor{}{G.~Cesarini}{},
    \iauthor{1793525}{C.~Cesarotti}{\ref{902835}},
    \iauthor{1014143}{Z.~Chacko}{\ref{902990}},
    \iauthor{2023221}{A.~Chanc\'e}{\ref{912490}},
    \iauthor{2037614}{S.~Chen}{\ref{1471035}},
    \iauthor{1069708}{Y.-T.~Chien}{\ref{1275736}},
    \iauthor{1272180}{M.~Chiesa}{\ref{943385},\ref{902885}},
    \iauthor{1013241}{G.~Collazuol}{\ref{902884},\ref{903113}},
    \iauthor{}{D.~J.~Colling}{\ref{902868}},
    \iauthor{1862239}{M.~Costa}{\ref{903128},\ref{902886}},
    \iauthor{1046385}{N.~Craig}{\ref{903307}},
    \iauthor{}{L.~M.~Cremaldi}{},
    \iauthor{1060042}{A.~Crivellin}{\ref{903370},\ref{905405}},
    \iauthor{1024481}{D.~Curtin}{\ref{903282}},
    \iauthor{1067364}{R.~T.~D'Agnolo}{\ref{1087875}},
    \iauthor{2052018}{M.~Dam}{\ref{902882}},
    \iauthor{1076225}{H.~Damerau}{\ref{902725}},
    \iauthor{1021004}{J.~de~Blas}{\ref{903836}},
    \iauthor{1651018}{E.~De~Matteis}{\ref{907142}},
    \iauthor{1012237}{A.~Deandrea}{\ref{1743848}},
    \iauthor{1012143}{J.~Delahaye}{\ref{902725}},
    \iauthor{1019723}{A.~Delgado}{\ref{903085}},
    \iauthor{1039487}{C.~Densham}{\ref{903174}},
    \iauthor{1011983}{R.~Dermisek}{\ref{902874}},
    \iauthor{1395010}{K.~F.~Di~Petrillo}{\ref{902796}},
    \iauthor{1246709}{J.~Dickinson}{\ref{902796}},
    \iauthor{1054778}{J.~Duarte}{\ref{903305}},
    \iauthor{1064320}{P.~Everaerts}{\ref{903349}},
    \iauthor{1010523}{L.~Everett}{\ref{903349}},
    \iauthor{1069878}{S.~Farinon}{\ref{902881}},
    \iauthor{1648215}{J.~F.~Somoza}{\ref{902725}},
    \iauthor{1010065}{G.~Ferretti}{\ref{902825}},
    \iauthor{1009979}{F.~Filthaut}{\ref{903075}},
    \iauthor{1894571}{M.~Forslund}{\ref{910429}},
    \iauthor{1052115}{R.~Franceschini}{\ref{906528},\ref{907692}},
    \iauthor{1046463}{P.~Franchini}{\ref{902948},\ref{903170}},
    \iauthor{1019509}{M.~Frigerio}{\ref{1508424}},
    \iauthor{1009120}{E.~Gabrielli}{\ref{903287},\ref{902888}},
    \iauthor{1068164}{I.~Garcia~Garcia}{\ref{903889}},
    \iauthor{1946817}{F.~Garosi}{\ref{904416}},
    \iauthor{}{A.S.~Giannakopoulou}{\ref{903237}},
    \iauthor{1075917}{D.~Giove}{\ref{907142}},
    \iauthor{1029806}{L.~Gladilin}{},
    \iauthor{1706734}{A.~Glioti}{\ref{1471035}},
    \iauthor{1008112}{S.~Goldfarb}{\ref{902999}},
    \iauthor{1037882}{H.M.~Gray}{\ref{903299},\ref{902953}},
    \iauthor{1059457}{L.~Gray}{\ref{902796}},
    \iauthor{1198373}{A.~Greljo}{\ref{902668}},
    \iauthor{1007486}{C.~Grojean}{\ref{902770},\ref{902858}},
    \iauthor{1274618}{J.~Gu}{\ref{903628}},
    \iauthor{1055424}{J.~Haley}{\ref{903094}},
    \iauthor{1259916}{C.~Han}{\ref{903702}},
    \iauthor{1006825}{T.~Han}{\ref{903130}},
    \iauthor{2044726}{J.~Hauptman}{\ref{902893}},
    \iauthor{1383268}{B.~Henning}{\ref{1471035}},
    \iauthor{1912097}{K.~Hermanek}{\ref{902874}},
    \iauthor{1515880}{S.~Homiller}{\ref{902835}},
    \iauthor{1475406}{S.~Jana}{\ref{902841}},
    \iauthor{2049476}{H.~Jia}{\ref{903349}},
    \iauthor{}{C.~Jolly}{\ref{903174}},
    \iauthor{1051663}{Y.~Kahn}{\ref{902867}},
    \iauthor{1003695}{D.~M.~Kaplan}{\ref{902865}},
    \iauthor{1653554}{I.~Karpov}{\ref{902725}},
    \iauthor{1067609}{D.~Kelliher}{\ref{903174}},
    \iauthor{1002991}{W.~Kilian}{\ref{903203}},
    \iauthor{1020007}{K.~Kong}{\ref{902912}},
    \iauthor{1019544}{P.~Koppenburg}{\ref{903832}},
    \iauthor{1077491}{G.~Krnjaic}{\ref{902796}},
    \iauthor{1001375}{P.~Kyberd}{\ref{903940}},
    \iauthor{999784}{R.~LOSITO}{\ref{902725}},
    \iauthor{}{S.~Levorato}{},
    \iauthor{1064657}{W.~Li}{\ref{903156}},
    \iauthor{1996476}{R.~L.~Voti}{\ref{902887}},
    \iauthor{1256188}{Z.~Liu}{\ref{903010}},
    \iauthor{1074693}{M.~Liu}{\ref{903142}},
    \iauthor{1700371}{Q.~Lu}{\ref{902835}},
    \iauthor{1355155}{L.~Ma}{\ref{904187}},
    \iauthor{1514492}{Y.~Ma}{\ref{903130}},
    \iauthor{999053}{F.~Maltoni}{\ref{910783},\ref{902674}},
    \iauthor{998923}{B.~Mansouli\'e}{\ref{912490}},
    \iauthor{1668860}{L.~Mantani}{\ref{907623}},
    \iauthor{1670119}{S.~Mariotto}{\ref{903009},\ref{907142}},
    \iauthor{1078065}{D.~Marzocca}{\ref{902888}},
    \iauthor{998430}{K.~Matchev}{\ref{902804}},
    \iauthor{1054925}{A.~Mazzacane}{\ref{902796}},
    \iauthor{1076700}{A.~Mazzolari}{\ref{905268}},
    \iauthor{1751811}{N.~McGinnis}{\ref{903290}},
    \iauthor{1025277}{P.~Meade}{\ref{910429}},
    \iauthor{997877}{B.~Mele}{\ref{902887}},
    \iauthor{1022138}{P.~Merkel}{\ref{902796}},
    \iauthor{1461119}{C.~Merlassino}{\ref{903112}},
    \iauthor{1066275}{R.~K.~Mishra}{\ref{902835}},
    \iauthor{1070072}{A.~Mohammadi}{\ref{903349}},
    \iauthor{}{R.~Mohapatra}{},
    \iauthor{997065}{N.~Mokhov}{\ref{902796}},
    \iauthor{996989}{P.~Montagna}{\ref{943385},\ref{902885}},
    \iauthor{1064125}{R.~Musenich}{\ref{902881}},
    \iauthor{995835}{M.S.~Neubauer}{\ref{902867}},
    \iauthor{995826}{D.~Neuffer}{\ref{902796}},
    \iauthor{995794}{H.~Newman}{\ref{902711}},
    \iauthor{995460}{Y.~Nomura}{\ref{903299}},
    \iauthor{1070110}{I.~Ojalvo}{\ref{16750}},
    \iauthor{1063052}{G.~Ortona}{\ref{902889}},
    \iauthor{1077958}{P.~Panci}{\ref{903129},\ref{902886}},
    \iauthor{994435}{M.~Palmer}{\ref{902689}},
    \iauthor{1067971}{R.~Paparella}{\ref{907142}},
    \iauthor{1023838}{P.~Paradisi}{\ref{1513358},\ref{902884}},
    \iauthor{1067962}{A.~Perloff}{\ref{902748}},
    \iauthor{993440}{F.~Piccinini}{\ref{902885}},
    \iauthor{1021028}{M.~Pierini}{\ref{902725}},
    \iauthor{1651162}{M.~Prioli}{\ref{907142}},
    \iauthor{1024769}{M.~Procura}{\ref{903326}},
    \iauthor{992222}{R.~Rattazzi}{\ref{1471035}},
    \iauthor{1214912}{D.~Redigolo}{\ref{902880}},
    \iauthor{992031}{L.~Reina}{\ref{902803}},
    \iauthor{1021811}{J.~Reuter}{\ref{902770}},
    \iauthor{991702}{R.A.~Rimmer}{\ref{904961}},
    \iauthor{1056642}{F.~Riva}{\ref{902813}},
    \iauthor{1040385}{T.~Robens}{\ref{902678}},
    \iauthor{1054170}{C.~Rogers}{\ref{903174}},
    \iauthor{1947424}{M.~Romagnoni}{\ref{905268}},
    \iauthor{991185}{L.~Rossi}{\ref{903009},\ref{907142}},
    \iauthor{1054727}{R.~Ruiz}{\ref{902756}},
    \iauthor{1912150}{R.~Ryne}{\ref{1189711}},
    \iauthor{1072232}{F.~Sala}{\ref{908583}},
    \iauthor{1885424}{J.~Salko}{\ref{902668}},
    \iauthor{1077871}{E.~Salvioni}{\ref{1513358},\ref{902884}},
    \iauthor{990367}{J.~Santiago}{\ref{909079},\ref{903836}},
    \iauthor{989950}{J.~Schieck}{\ref{903324},\ref{904536}},
    \iauthor{1022121}{R.~Schwiehorst}{\ref{903006}},
    \iauthor{1055618}{D.~Sertore}{\ref{907142}},
    \iauthor{1071696}{V.~Sharma}{\ref{903349}},
    \iauthor{988978}{V.~Shiltsev}{\ref{902796}},
    \iauthor{1019902}{J.~Shu}{\ref{903895}},
    \iauthor{988660}{L.~Silvestris}{\ref{902877}},
    \iauthor{1889775}{K.~Skoufaris}{\ref{902725}},
    \iauthor{1046340}{P.~Snopok}{\ref{902865}},
    \iauthor{988143}{F.J.P.~Soler}{\ref{902823}},
    \iauthor{1066476}{M.~Sorbi}{\ref{903009},\ref{907142}},
    \iauthor{1319078}{G.~Stark}{\ref{1218068}},
    \iauthor{1057643}{M.~Statera}{\ref{907142}},
    \iauthor{1058749}{D.~Stratakis}{\ref{902796}},
    \iauthor{1077402}{N.~Strobbe}{\ref{903010}},
    \iauthor{1071880}{J.~Stupak}{\ref{1273509}},
    \iauthor{987285}{S.~Su}{\ref{902647}},
    \iauthor{987128}{R.~Sundrum}{\ref{912511}},
    \iauthor{1454316}{A.~Sytov}{\ref{905268}},
    \iauthor{1020371}{A.~Taffard}{\ref{903302}},
    \iauthor{1257546}{A.~Tesi}{\ref{902880}},
    \iauthor{1054400}{J.~Thaler}{\ref{1237813}},
    \iauthor{1064514}{R.~Torre}{\ref{902881}},
    \iauthor{985810}{L.~Tortora}{\ref{907692}},
    \iauthor{1031757}{Y.~Torun}{\ref{902865}},
    \iauthor{1778841}{S.~Trifinopoulos}{\ref{902888}},
    \iauthor{2025179}{R.~U.~Valente}{\ref{907142}},
    \iauthor{1077738}{N.~Vignaroli}{\ref{902883}},
    \iauthor{1863232}{L.~Vittorio}{\ref{903128},\ref{902886}},
    \iauthor{984555}{P.~Vitulo}{\ref{943385},\ref{902885}},
    \iauthor{1077733}{E.~Vryonidou}{\ref{902984}},
    \iauthor{1054127}{C.~Vuosalo}{\ref{903349}},
    \iauthor{984146}{L.-T.~Wang}{\ref{902729}},
    \iauthor{1260509}{C.G.~Whyte}{\ref{904214}},
    \iauthor{1511975}{Y.~Wu}{\ref{903094}},
    \iauthor{1037622}{A.~Wulzer}{\ref{903113}},
    \iauthor{1618109}{K.~Xie}{\ref{903130}},
    \iauthor{982905}{A.~Yamamoto}{\ref{902916}},
    \iauthor{1019845}{K.~Yonehara}{\ref{902796}},
    \iauthor{1024759}{H.-B.~Yu}{\ref{903304}},
    \iauthor{1064691}{M.~Zanetti}{\ref{903113}},
    \iauthor{}{J.~Zhang}{},
    \iauthor{1066114}{Y.~J.~Zheng}{\ref{902912}},
    \iauthor{981974}{A.~Zlobin}{\ref{902796}},
    \iauthor{1037623}{J.~Zurita}{\ref{907907}}
    \vspace*{1cm}} \institute{\small
\iinstitute{902889}{{INFN Sezione di Torino, Italy}};
\iinstitute{902953}{{Physics Division, Lawrence Berkeley National Laboratory, United States}};
\iinstitute{943385}{Universit{\`a} di Pavia, Italy};
\iinstitute{902682}{Department of Physics, Brandeis University, United States};
\iinstitute{912409}{{Center for High Energy Physics (CHEP-FU), Fayoum University, Egypt}};
\iinstitute{902884}{INFN Sezione di Padova,Italy};
\iinstitute{902885}{{INFN Sezione di Pavia, Italy}};
\iinstitute{902888}{INFN Sezione di Trieste, Italy};
\iinstitute{903113}{Dipartimento di Fisica e Astronomia, Universit'a di Padova, Italy};
\iinstitute{902877}{INFN Sezione di Bari, Italy};
\iinstitute{906718}{European Centre for Theoretical Studies in Nuclear Physics and Related Areas (ECT*), Italy};
\iinstitute{1241166}{MPS School, University of Sussex, United Kingdom};
\iinstitute{905303}{Laborat{\' o}rio de Instrumenta\c{c}{\~ a}o e F{\' \i}sica Experimental de Part{\' \i}culas (LIP), Portugal};
\iinstitute{902660}{{Department of Physics, Universit{\`a} degli Studi di Bari, Italy}};
\iinstitute{902882}{{INFN Sezione di Milano, Italy}};
\iinstitute{903349}{University of Wisconsin, United States};
\iinstitute{902689}{Brookhaven National Laboratory, United States};
\iinstitute{908452}{Department of Physics, Bolu Abant Izzet Baysal University, Turkey};
\iinstitute{906528}{Dipartimento di Matematica e Fisica, Universit\`a Roma Tre, Italy};
\iinstitute{1276460}{Helmholtz-Zentrum Dresden-Rossendorf, Germany};
\iinstitute{1623978}{University of Tennessee, United States};
\iinstitute{902796}{Fermi National Accelerator Laboratory, United States};
\iinstitute{902912}{Department of Physics and Astronomy, United States};
\iinstitute{903603}{Peking University, China};
\iinstitute{902868}{Imperial College London, United Kingdom};
\iinstitute{910783}{Centre for Cosmology, Particle Physics and Phenomenology (CP3), Universit\'e Catholique de Louvain, Belgium};
\iinstitute{902770}{Deutsches Elektronen-Synchrotron DESY, Germany};
\iinstitute{903112}{Department of Physics, Oxford University, United Kingdom};
\iinstitute{902725}{CERN, Switzerland};
\iinstitute{1210798}{State Key Laboratory of Nuclear Physics and Technology, Peking University, China};
\iinstitute{903290}{TRIUMF, Canada};
\iinstitute{903702}{Sun Yat-sen University, China};
\iinstitute{902858}{{Institut f{\"u}r Physik, Humboldt-Universit{\"a}t zu Berlin, Germany}};
\iinstitute{912328}{INFN-TIFPA Trento Institute of Fundamental Physics and Applications, Italy};
\iinstitute{907692}{INFN Sezione di Roma Tre, Italy};
\iinstitute{903174}{STFC, United Kingdom};
\iinstitute{903123}{Institute of High-Energy Physics, China};
\iinstitute{903156}{Physics \& Astronomy Department, Rice University, United States};
\iinstitute{902990}{Maryland Center for Fundamental Physics, University of Maryland, United States};
\iinstitute{907623}{DAMTP, University of Cambridge, United Kingdom};
\iinstitute{902887}{INFN Sezione di Roma, Italy};
\iinstitute{1237813}{Center for Theoretical Physics, Massachusetts Institute of Technology, United States};
\iinstitute{910429}{YITP, Stony Brook, United States};
\iinstitute{904416}{SISSA International School for Advanced Studies, Italy};
\iinstitute{903734}{Department of Physics, University of Warwick, United Kingdom};
\iinstitute{905268}{INFN Sezione di Ferrara, Italy};
\iinstitute{911708}{The University of Huddersfield, United Kingdom};
\iinstitute{902881}{INFN Sezione di Genova, Italy};
\iinstitute{903196}{{Department of Physics and Astronomy, University of Sheffield, United Kingdom}};
\iinstitute{903128}{{Scuola Normale Superiore, Italy}};
\iinstitute{902886}{INFN Sezione di Pisa, Italy};
\iinstitute{926589}{CNRS/IN2P3, France};
\iinstitute{908474}{Perimeter Institute, Canada};
\iinstitute{903168}{Dipartimento di Fisica, Sapienza Univ. Roma, Italy};
\iinstitute{902879}{{INFN Sezione di Catania, Italy}};
\iinstitute{902835}{Department of Physics, Harvard University, United States};
\iinstitute{912490}{IRFU, CEA, Université Paris-Saclay, France};
\iinstitute{1471035}{{Theoretical Particle Physics Laboratory (LPTP), Institute of Physics, EPFL, Switzerland}};
\iinstitute{1275736}{Physics and Astronomy Department, Georgia State University, United States};
\iinstitute{903307}{University of California, Santa Barbara, United States};
\iinstitute{903370}{University of Zurich, Switzerland};
\iinstitute{903282}{Department of Physics, University of Toronto, Canada};
\iinstitute{1087875}{Universit\`e Paris Saclay, CNRS, CEA, Institut de Physique Th\`eorique, France};
\iinstitute{903836}{CAFPE and Departamento de F\'isica Te\'orica y del Cosmos, Universidad de Granada, Spain};
\iinstitute{907142}{{Laboratori Acceleratori e Superconduttività Applicata (LASA), INFN, Italy}};
\iinstitute{1743848}{IP2I, Universit\'e Lyon 1, CNRS/IN2P3, France};
\iinstitute{903085}{University of Notre Dame, United States};
\iinstitute{902874}{Physics Department, Indiana University, United States};
\iinstitute{903305}{University of California San Diego, United States};
\iinstitute{902825}{Chalmers University of Technology, Sweden};
\iinstitute{903075}{Radboud University and Nikhef, The Netherlands};
\iinstitute{902948}{{University of Lancaster, Department of Physics, United Kingdom}};
\iinstitute{1508424}{Laboratoire Charles Coulomb, CNRS and University of Montpellier, France};
\iinstitute{903287}{Physics Department, University of Trieste, Italy};
\iinstitute{903889}{Kavli Institute for Theoretical Physics, University of California, United States};
\iinstitute{903237}{SUNY at Stony Brook, United States};
\iinstitute{902999}{School of Physics, University of Melbourne, Australia};
\iinstitute{903299}{UC Berkeley, United States};
\iinstitute{902668}{Albert Einstein Center for Fundamental Physics, Institute for Theoretical Physics, University of Bern, Switzerland};
\iinstitute{903628}{Department of Physics, Fudan University, China};
\iinstitute{903094}{Oklahoma State University, United States};
\iinstitute{903130}{University of Pittsburgh, United States};
\iinstitute{902893}{Iowa State University, United States};
\iinstitute{902841}{Max-Planck-Institut f{\"u}r Kernphysik, Germany};
\iinstitute{902867}{Department of Physics, University of Illinois at Urbana-Champaign, United States};
\iinstitute{902865}{Illinois Institute of Technology, United States};
\iinstitute{903203}{Department of Physics, University of Siegen, Germany};
\iinstitute{903832}{Nikhef National Institute for Subatomic Physics, The Netherlands};
\iinstitute{903940}{{College of Engineering, Design and Physical Sciences, Brunel University, United Kingdom}};
\iinstitute{903010}{School of Physics and Astronomy, University of Minnesota, United States};
\iinstitute{903142}{Purdue University, United States};
\iinstitute{904187}{Shandong University, China};
\iinstitute{903009}{{Dipartimento di Fisica Aldo Pontremoli, Universit\'a degli Studi di Milano, Italy}};
\iinstitute{902804}{Physics Department, University of Florida, United States};
\iinstitute{902711}{California Institute of Technology, United States};
\iinstitute{16750}{Princeton University, United States};
\iinstitute{903129}{Pisa University, Italy};
\iinstitute{1513358}{{Dipartamento di Fisica e Astronomia ``G.~Galilei", Università di Padova, Italy}};
\iinstitute{902748}{{Department of Physics, University of Colorado, United States}};
\iinstitute{903326}{University of Vienna, Faculty of Physics, Austria};
\iinstitute{902880}{INFN Sezione di Firenze, Italy};
\iinstitute{902803}{Florida State University, United States};
\iinstitute{904961}{JLab, United States};
\iinstitute{902813}{D\'epartment de Physique Th\'eorique, Universit\'e de Gen\`eve, Switzerland};
\iinstitute{902678}{Rudjer Boskovic Institute, Croatia};
\iinstitute{902756}{Institute of Nuclear Physics -- Polish Academy of Sciences {\rm (IFJ PAN)}, Poland};
\iinstitute{1189711}{Lawrence Berkeley National Laboratory, United States};
\iinstitute{411233}{International Institute of Physics, Universidade Federal do Rio Grande do Norte, Brazil};
\iinstitute{908583}{Laboratoire de Physique Th\'eorique et Hautes \'Energies, Sorbonne Universit\'e, CNRS, France};
\iinstitute{909079}{{CAFPE}, Spain};
\iinstitute{903324}{Institut f\"ur Hochenergiephysik der \"Osterreichischen Akademie der Wissenschaften, Austria};
\iinstitute{903006}{MIchigan State University, United States};
\iinstitute{903895}{CAS Key Laboratory of Theoretical Physics, Insitute of Theoretical Physics, Chinese Academy of Sciences, P.R.China};
\iinstitute{902823}{School of Physics and Astronomy, University of Glasgow, United Kingdom};
\iinstitute{1218068}{SCIPP, UC Santa Cruz, United States};
\iinstitute{1273509}{University of Oklahoma, United States};
\iinstitute{902647}{University of Arizona, United States};
\iinstitute{912511}{Maryland Center for Fundamental Physics, University of Maryland, United States};
\iinstitute{903302}{University of California Irvine, United States};
\iinstitute{902883}{Universit{\'a} di Napoli ``Federico II" and INFN Napoli, Italy};
\iinstitute{902984}{University of Manchester, United Kingdom};
\iinstitute{902729}{Department of Physics, University of Chicago, United States};
\iinstitute{904214}{Physics, SUPA, United Kingdom};
\iinstitute{902916}{{High Energy Accelerator Research Organization KEK, Japan}};
\iinstitute{903304}{{Department of Physics and Astronomy, University of California, United States}};
\iinstitute{907907}{{Instituto de F{\'i}sica Corpuscular, CSIC-Universitat de Val{\'e}ncia, Spain}};
\iinstitute{1111512}{Institut f{\"u}r Allgemeine Elektrotechnik, Universit{\"a}t Rostock, Germany};
\iinstitute{903092}{Ohio State University, United States};
\iinstitute{907960}{Dipartimento di Fisica, Universit\`a Milano Bicocca, Italy};
\iinstitute{903119}{LPNHE, Sorbonne Universit\'e, France};
\iinstitute{905405}{Paul Scherrer Institute, Switzerland};
\iinstitute{903170}{{Royal Holloway University of London, Department of Physics, United Kingdom}};
\iinstitute{902674}{Dipartimento di Fisica e Astronomia, Universit\`a di Bologna, Italy};
\iinstitute{904536}{Atominstitut, Technische Universit\"at Wien, Austria}
}

%% file: executive-summary.tex
\note{ALL: Please READ AND COMMENT. 90\% of our readers will only read this part.}

Despite the success of the Standard Model (SM) of particle physics, several observations point to the existence of physics beyond what is currently known. Theoretical explanations of such phenomena most probably require new states or forces that manifest at energies higher than what is currently accessible at world leading facilities. The lack of a firm indication of the exact energy scale of those phenomena further motivates an energy frontier research program capable of reaching to the highest possible energy.
In the landscape of possible future energy-frontier accelerators, a circular $\mu^+\mu^-$ collider, also referred to simply as muon collider, is a particularly attractive option for the future of energy frontier exploration. Such a machine has the potential to deliver a vast physics program~\cite{sm21-imcc-physics-summary} in a relatively compact accelerator complex.

The aim of this paper is to describe the expected performance of a multi-purpose Muon Collider Detector (MCD) designed to reconstruct the products of a muon-muon collisions with high accuracy. Several operating center-of-mass energies are possible for the considered machine; in this paper a center-of-mass energy of $1.5$~TeV is explored. 

One of the biggest challenges for a MCD is to successfully disentangle the products of the $\mu^+\mu^-$ collisions from an intense beam-induced background (BIB) coming primarily from secondary and tertiary interactions of the muon decay products. For this reason an essential part of the Machine Detector Interface (MDI) at a Muon Collider is a pair of tungsten nozzles cladded with borated polyethylene, which reduce the rate and energy of BIB particles that reach the detector by several orders of magnitude. The nozzles limit the maximum angular acceptance of the MCD to about $10^\circ$ from the beamline axis.

The MCD design presented in this paper follows the classical cylindrical layout typical for multipurpose detectors of symmetric collisions. It consists of a silicon-based Vertex Detector, with innermost barrel layer at $22$~mm from the beamline, followed by silicon-based Inner and Outer Trackers. High granularity electromagnetic and hadronic calorimeters and an outer muon spectrometer complement the main elements of the detector.

A detailed GEANT4 based simulation package is used to simulate the response of the detector. A fast digitization and detailed reconstruction algorithms were developed to study the expected performance of the MCD in the event reconstruction. The BIB is overlaid to the $\mu^+\mu^-$ process of interest.

Three approaches are studied for track reconstruction at the MCD: a Conformal Tracking (CT) algorithm developed for the clean environment of the electron-positron colliders and supplemented with either a Region of Interest approach or a filter based on double pixel layers in the innermost layers of the detector, and a Combinatorial Kalman Filter (CKF) algorithm developed for the busy environment of hadron colliders. The results show a robust track reconstruction for charged particles with the momentum above $1~$GeV throughout the detector acceptance, with the potential for extending towards softer particles.

Jets are reconstructed using a particle-flow approach and a $k_\textrm{T}$-based clustering. An energy calibration is derived and applied throughout the studies. The performance is evaluated on samples of light, $b-$ and $c-$jets, resulting in a reconstruction efficiency ranging from 82\% at $p_\textrm{T}\approx 20$~GeV to 95\% at higher $p_\textrm{T}$ with sub-percent fake rate in the region $\vert\eta\vert < 1.5$. The jet energy resolution ranges from about 50\% to about $15\%$, depending on $p_\textrm{T}$. The resulting di-jet invariant mass resolution allows good separation between hadronically-decaying Higgs and $Z$ bosons. 

A simple heavy-flavor tagging based on finding a displaced-vertex yields a $b-$jet identification efficiency ranging from 50\% to 80\%, depending on $p_\textrm{T}$, with a mis-tag rate of about $1-5\%$ (20\%) for light ($c$-)jets. It is expected that more advanced algorithms would be able to perform significantly better. 

Prototypes of electron and photon reconstruction algorithms able to cope with BIBs have been developed, yielding a successful reconstruction of high-$p_\textrm{T}$ electrons and photons with relatively small loss of efficiency and energy resolution.
Muon reconstruction is performed by combining a stand-alone track reconstructed in the muon system and a track reconstructed in the tracker. BIBs mostly affect the forward endcap region. Filtering hits based on their properties as well as based on standalone muon tracks allows for negligible impact of BIB in the performance of muon reconstruction.
Reconstruction of other high-\pT~objects, as $\tau$ leptons, is still in progress, but it is expected to pose challenges similar to the ones that have been already solved for track and jet reconstruction as well as heavy-flavor jets identification. 

The precise determination of the delivered luminosity is of crucial importance for the planned physics program. A method based on $\mu-$Bhabha scattering measurement shows that a statistical precision of about $0.2\%$ can be achieved for the integrated luminosity delivered per year of data-taking.

In conclusion, this manuscript uses detailed simulation of a proposed Muon Collider detector that satisfies the basic requirements of reconstructing high-$p_\textrm{T}$ objects needed to deliver an ambitious physics program. The effects of BIB is taken into account, showing that its impact can be successfully mitigated.

%% file: introduction.tex
The Standard Model (SM) of particle physics has been an extremely successful theory to describe measurements at the energy frontier. In the last decades, a vast physics programs at the energy frontier has produced discoveries that led to the current incarnation of the Standard Model (SM) theory and stress-tested it with precision measurements. Nevertheless, observations, as well as theoretical motivations, unambiguously point to physics phenomena Beyond the Standard Model (BSM). Theoretical explanations of such phenomena almost always require new states or forces that manifest at energies higher than what currently accessible at world leading facilities. Despite the lack of a clear energy target that guarantees with certainty a discovery, multiple indirect indications point to energy scales of several TeV as the most likely energy regime where BSM phenomena would manifest. It is therefore of paramount importance to develop a physics program for the energy frontier that allows us to explore such an energy regime in the future, possibly with the ability of adapting the target energy depending on future insights from accelerator and non-accelerator based experiments. 

In the landscape of possible future energy-frontier accelerators~\cite{Shiltsev:2019rfl}, a circular $\mu^+\mu^-$ collider, also referred to simply as muon collider, is a particularly attractive option for the future of energy frontier exploration~\cite{Long:2020wfp}. Such a machine has the potential to deliver a vast physics program~\cite{sm21-imcc-physics-summary,sm21-imcc-physics-3TeV,Buttazzo:2018qqp,Franceschini:2021dxn,Capdevilla:2021fmj,AlAli:2021let,Buttazzo:2020uzc,Han:2020pif,Capdevilla:2020qel,Ruhdorfer:2019utl,Chiesa:2020awd,Costantini:2020stv,Capdevilla:2021rwo,Bartosik:2020xwr,Yin:2020afe,Kalinowski:2020rmb,Liu:2021jyc,Han:2021udl,Bottaro:2021srh,Li:2021lnz,Asadi:2021gah,Sahin:2021xzt,Chen:2021rid,Haghighat:2021djz,Bottaro:2021snn,Sen:2021fha,Han:2021lnp,Bandyopadhyay:2021pld,Dermisek:2021mhi,Qian:2021ihf,Chiesa:2021qpr,Liu:2021akf,Buttazzo:2021lzi,DiLuzio:2018jwd,Han:2020uak,Chen:2021pqi,Capdevilla:2021kcf,Cesarotti:2022ttv} in a relatively compact accelerator complex~\cite{sm21-imcc-accelerator}. Such an option has been studied in the past~\cite{Neuffer:1994bt,Foster:1995ru,Palmer:1996gs,Gallardo:1996aa,Ankenbrandt:1999cta,Snopok:2007vyn,Alexahin:2012zzc,Alexahin:2013ojp}, and it sparked renewed interest thanks to advances in accelerator technology as well as given the current needs of high-energy exploration of the field.

The aim of this paper is to describe the expected performance of a multi-purpose muon collider detector (MCD). The study focuses on the reconstruction algorithms and performance of high-\pt objects that will be used needed for a successful physics program. The results are built upon and significantly extend the ones reported in Ref.~\cite{muc-perf-2020}. The current design aims at showing that satisfactory performance can be achieved in a muon collider environment. Further optimization of the detector design and reconstruction algorithms are expected to significantly improve the performance of a MCD and will be subject of future work.

Several operating center-of-mass energies are possible for a future muon collider~\cite{sm21-imcc-accelerator}. In this paper, unless otherwise noted, a center-of-mass energy of $1.5$~TeV is assumed. 
Table~\ref{tab:acc-params} reports additional accelerator parameters that have been assumed for the simulations used for the results presented in this manuscript. 
Most of the results presented in this manuscript do not simulate the spread of collisions in the transverse and longitudinal directions, unless otherwise stated; it is expected that this has a minor impact on the performance, and it is discussed explicitly when relevant.

\begin{table}[h]
    \centering
    \begin{tabular}{l|c|c}
        Parameter & Symbol & Value \\
        \hline\hline
        Center-of-mass energy & $\sqrt{s}$ &  $1.5$~TeV\\
        Muons per bunch      & $N_\mu$ & $2\cdot 10^{12}$\\
        Normalised transverse emittance & $\epsilon_{TN}$ & 25 $\pi \,\mu$m rad\\
        Normalised longitudinal emittance &$\epsilon_{LN}$ & 7.5 MeV m \\
        IP relative energy spread & $\delta_E$   & 0.1 \%\\ 
        IP beta function &$\beta^{*}_{x,y}$ & 1 cm \\
        IP transverse beam size & $\sigma_{x,y}$ & 6 $\mu$m \\
        IP longitudinal beam size & $\sigma_{z}$ & 10 mm   \\
    \end{tabular}
    \caption{Representative set of muon collider parameters used in the detailed simulation presented in this paper.}
    \label{tab:acc-params}
    \label{tab:collider_parameters}
\end{table}

One of the biggest challenges for a MCD is to successfully disentangle the products of the $\mu^+\mu^-$ collisions from an intense beam-induced background (BIB) coming primarily from secondary and tertiary interactions of the muon decay products~\cite{Mokhov:1996tq,Mokhov:2011zzd,Collamati:2021sbv}. Section~\ref{sec:bib} summarizes the main characteristics expected for such a background and also includes a discussion on the expected background composition for higher center-of-mass energy muon colliders. 

BIB has a big influence on the detector design, which is summarised in Section~\ref{sec:detector}. The technical challenges for the MCD and the most promising technologies for fulfilling the physics requirements of the detector are described in a separate manuscript~\cite{sm21-imcc-det-technologies}, which also covers the readout and trigger schema.

Results in this manuscript are obtained using detailed simulations of the MCD based on a \textit{Geant4}~\cite{GEANT4:2002zbu} framework. The software setup is described in Section~\ref{sec:software}. All simulations, unless otherwise stated, include overlaying on top of the desired $\mu^+\mu^-$ collision the model of BIB described in Section~\ref{sec:bib}.

Sections~\ref{sec:tracking}, \ref{sec:jets}, \ref{sec:phele} and \ref{sec:muons} describe the reconstruction algorithms and expected performance for the basic objects needed to carry a comprehensive physics program. Reconstruction of other objects, as $\tau$ leptons or missing momentum, is still in progress, but it is expected to pose challenges similar to the ones that have been already solved. A method for measuring the delivered luminosity with high accuracy is described in Section~\ref{sec:lumi}.

%% file: bib.tex
The unstable nature of muons (lifetime $\tau_{\mu} =$~\qty{2.2}{ \micro\second} at rest) makes the beam-induced background (BIB) a much more challenging issue at a Muon Collider than it is at facilities that use stable-particle beams.
In fact, $2 \cdot 10^{12}$ muons per bunch in a \qty{750}{\GeV} muon beam lead to about $4 \cdot 10^5$ muon decays per meter.
Interactions of the decay products with the accelerator lattice produce even larger amounts of particles that eventually reach the detector, making the reconstruction of clean $\mu^+ \mu^-$ collision events nearly impossible without dedicated BIB mitigation. A discussion of the expected radiation levels due to BIB can be found in Ref.~\cite{sm21-imcc-det-technologies}.
For this reason an essential part of the Machine Detector Interface (MDI) at a Muon Collider is a pair of tungsten (W) nozzles cladded with borated polyethylene (BCH), which reduce the rate and energy of BIB particles reaching the detector by several orders of magnitude.

\begin{figure}
    \centering
    \includegraphics[width=1\textwidth]{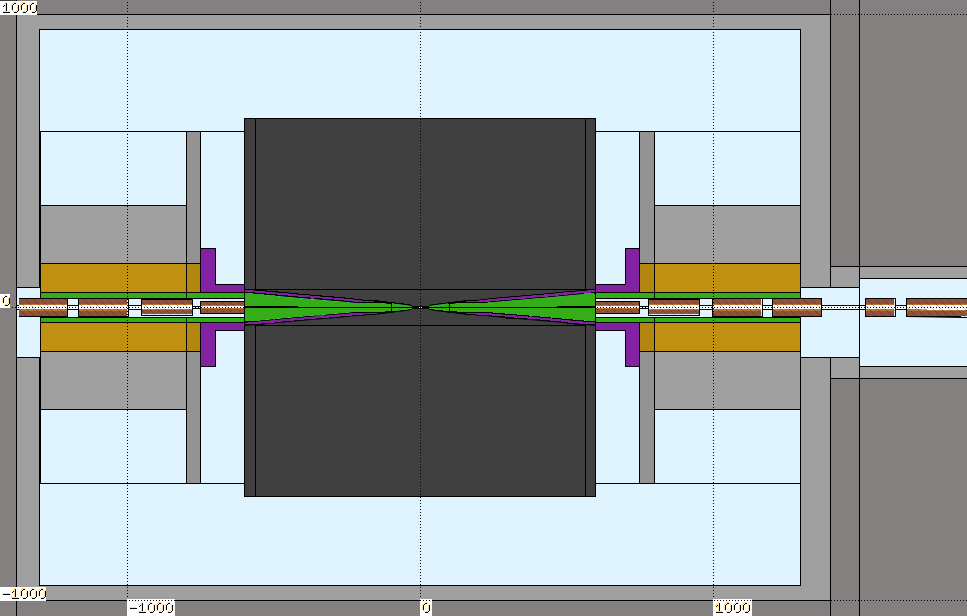}
    \includegraphics[width=1\textwidth]{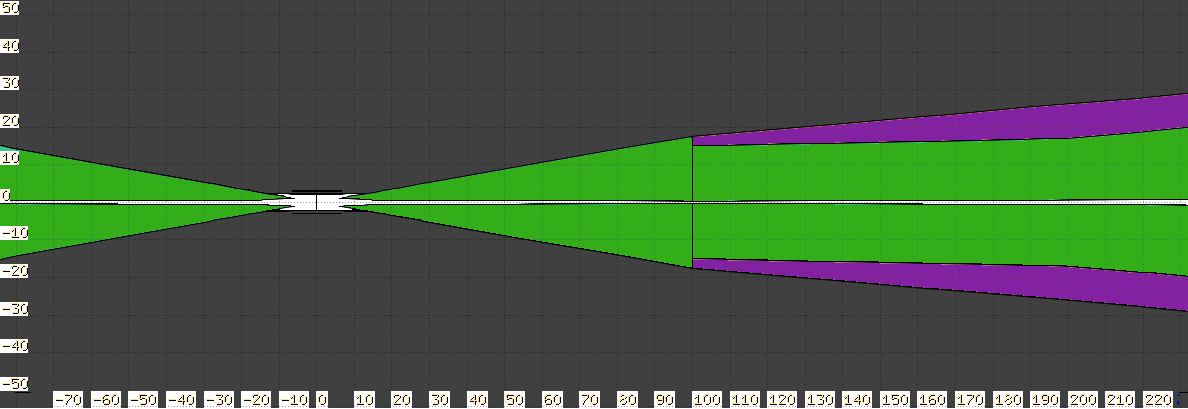}
    \caption{Cross-sectional view of the accelerator lattice including the MDI as designed by MAP collaboration for the $\sqrt{s} = $1.5~TeV Muon Collider and visualised with FLUKA simulation software. Distinct colours represent different materials of the MDI: tungsten (green), borated polyethylene (violet), iron (mustard), concrete (gray). The black box in the center encloses the detector volume, which is excluded from the standalone BIB simulation process. Dimensions are reported in centimeters.}
    \label{fig:bib-mdi}
\end{figure}

A dedicated design of the MDI has been developed by MAP collaboration~\cite{Mokhov:2011zzd} using the MARS15 software~\cite{mars15} to simulate muon decays within \qty{\pm 200}{\metre} from the interaction point (IP) in a collider ring with the parameters summarised in Table~\ref{tab:collider_parameters}.
The exact shape and positioning of the nozzles was optimised specifically for the $\sqrt{s} =$~\qty{1.5}{\TeV} case, including the \qty{10}{\degree} opening angle and \qty{12}{\centi\metre} distance between the tips of the nozzles, as shown in Fig.~\ref{fig:bib-mdi}.
The same approach has been adopted by the Muon Collider collaboration based on the FLUKA simulation software to characterize BIB at higher center-of-mass energies and to study other MDI designs.

In order to include the effects of the BIB on detector performance the particles induced by the muon decays are collected at the outer surface of the MDI and before entering the detector volume, which is represented by a black box on Fig.~\ref{fig:bib-mdi}.
This allows to later simulate their interaction with the detector together with particles from the \mumu collision.
Thus, BIB sample in this article refers to the collection of particles originating from the muon decays before any interaction with the detector material.

\subsection{BIB properties at 1.5~TeV}
\label{sec:bib-prop}

\begin{figure}
    \centering
    \includegraphics[width=0.4\textwidth]{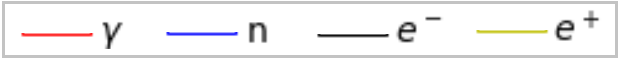}
    \includegraphics[width=1\textwidth]{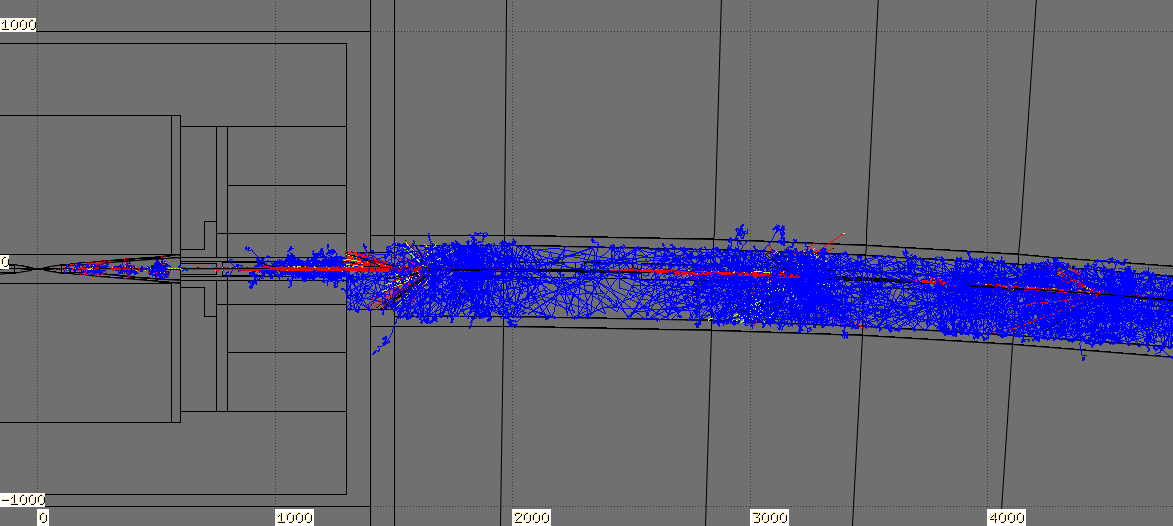}
    \includegraphics[width=1\textwidth]{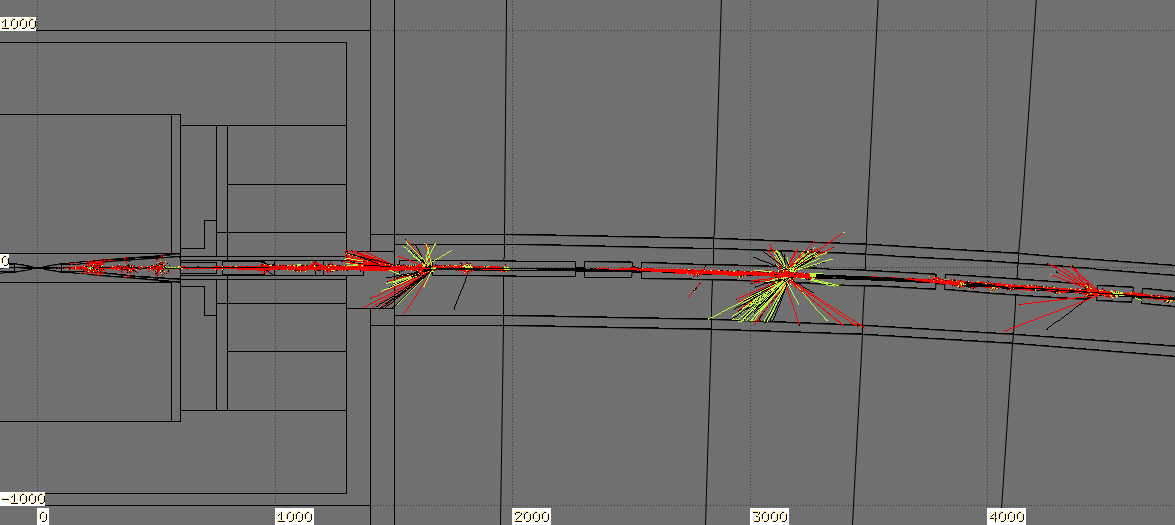}
    \includegraphics[width=1\textwidth]{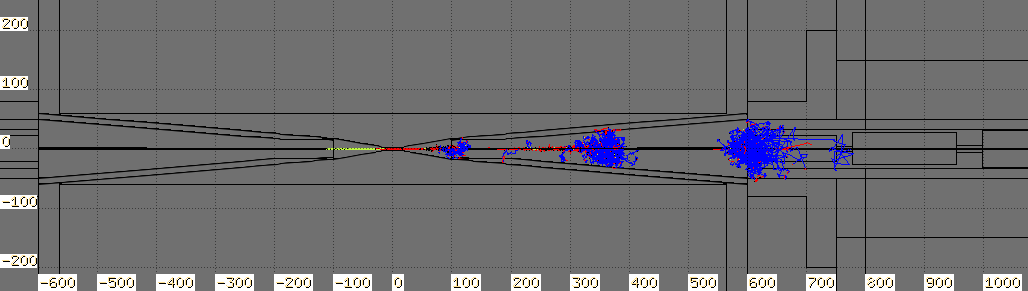}
    \caption{The top picture shows the tracks of secondary particles for a few $\mu^{-}$ decays arriving from the right, while in the middle picture neutrons are excluded. The bottom plot illustrates the tracks in the case of a single $\mu^{-}$ decay in the proximity of the IP. Different particle types are separated by colour: photons (red), neutrons (blue), $e^-$ (black), $e^+$ (yellow).}
    \label{fig:bib-mdi_tracks}
\end{figure}

The exact BIB sample used for detector-performance studies at $\sqrt{s} =$~\qty{1.5}{\TeV} presented in this paper has been produced in MARS15 software, because the FLUKA-based workflow was not fully validated at that time.
Yet FLUKA interface is used for visualisation of the accelerator geometry and BIB trajectories in the MDI region thanks to its advanced graphical capabilities.
An example of simulated particle decays and their interaction with the accelerator lattice is shown in Fig.~\ref{fig:bib-mdi_tracks}, demonstrating multiple points along the beamline where BIB particles originate from.

\begin{figure}
  \centering
    \includegraphics[width=0.32\textwidth]{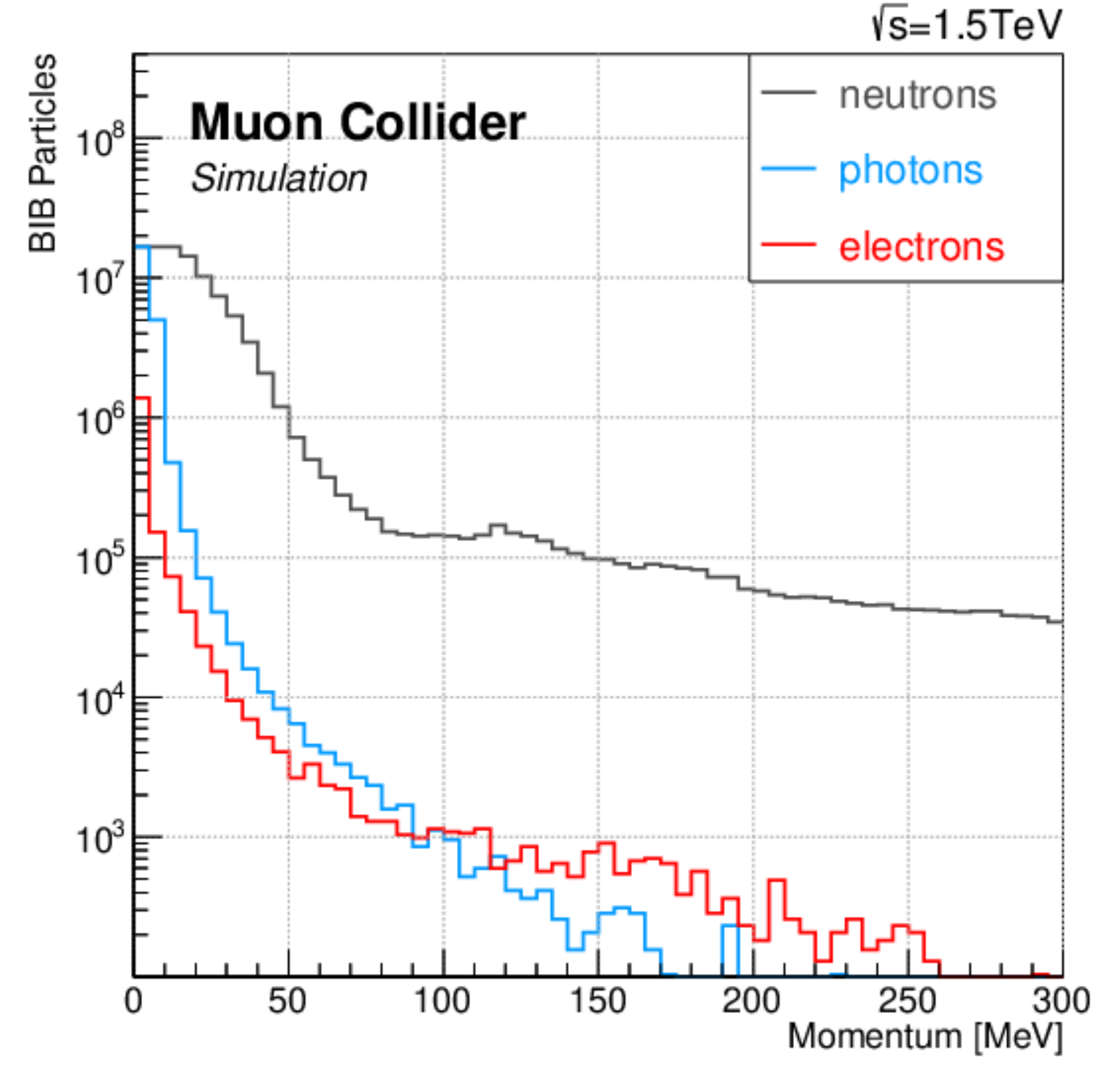} \hfill
    \includegraphics[width=0.32\textwidth]{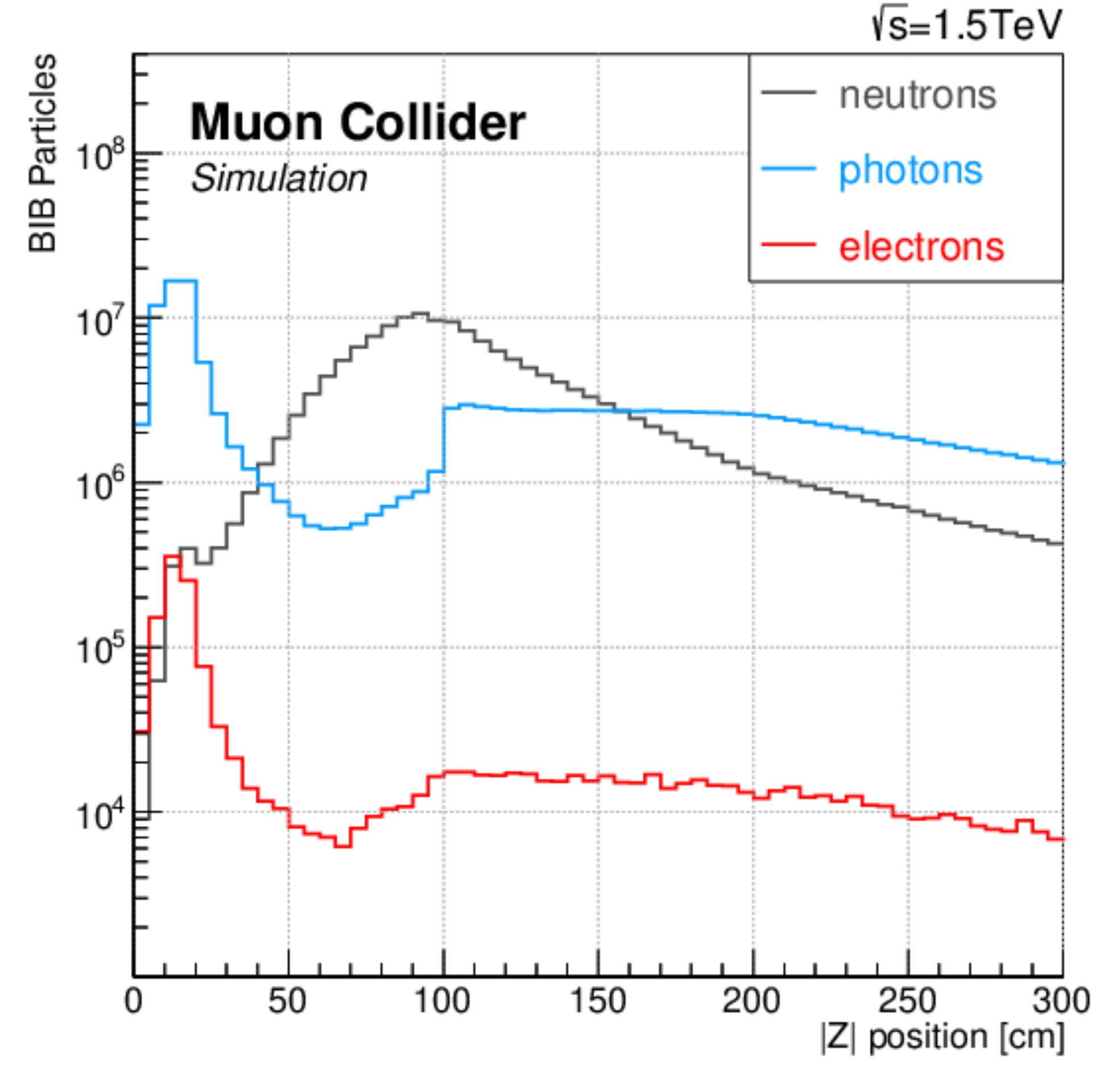} \hfill
    \includegraphics[width=0.32\textwidth]{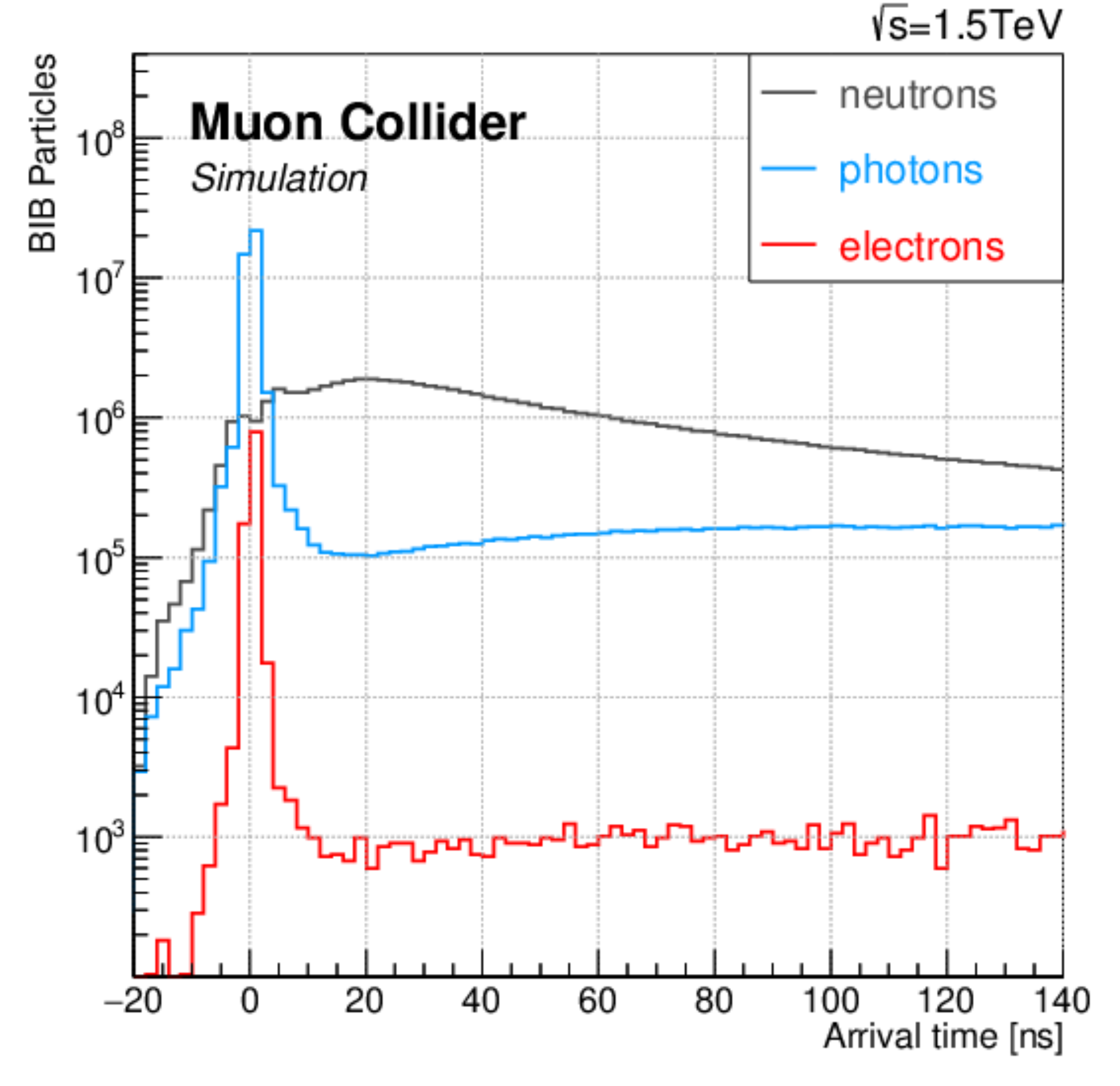}
    \caption{Kinematic properties of BIB particles entering the detector region: momentum (left), position along the beam line (middle) and arrival time with respect to the bunch crossing (right).}
    \label{fig:bib-props}
\end{figure}

BIB particles at a Muon Collider have a number of characteristic features that set them apart from typical backgrounds at \ee or $pp$ colliders, namely their low momentum, displaced origin and asynchronous time of arrival.
Distributions of these three properties are presented in Fig.~\ref{fig:bib-props} for the three dominant particle types exiting from the MDI surface: neutrons, photons and electrons/positrons.

The most distinctive aspect of BIB particles at the Muon Collider is their extremely large number and low momentum: about $4 \cdot 10^{8}$ particles exiting the MDI in a single bunch crossing depositing energy to the detector in a diffused manner.
The separation between the tips of the tungsten nozzles leads to most of the BIB particles exiting at a significant distance from the interaction point.
Finally, there is a substantial spread in the arrival time of the BIB particles with respect to the bunch crossing, ranging from a few nanoseconds for electrons and photons to microseconds for neutrons due to their smaller velocity.

Each of these aspects has different implications for the BIB signatures in different parts of the detector, which depend on the position, spatial granularity and timing capabilities of the corresponding sensitive elements.
Thus, a careful choice of detector technologies and reconstruction techniques allows to mitigate the negative effects of the BIB, as demonstrated in the later sections of this paper.

\input{bib_fluka}

%% file: bib_fluka.tex
\subsection{Simulation in FLUKA}

We report in this section the most relevant BIB features computed at $\sqrt{s}=1.5$~TeV by the Monte Carlo multi-particle transport code FLUKA \cite{Ferrari:2005zk, BOHLEN2014211}. The complex FLUKA geometry is assembled by means of the LineBuilder \cite{Mereghetti:1481554} using the optics file provided by the MAP collaboration. The accelerator elements have been defined in Fluka Elements Database following the information contained in this file and in MAP publications \cite{Alexahin:2011zz, 2018JInst..13P9004D}. The results obtained by FLUKA are benchmarked against those provided by the MAP collaboration and the detailed comparison is described in Ref. \cite{Collamati:2021sbv}.

The results presented below are computed for one beam, given the symmetric nature of the $\mu^+$$\mu^-$ collider. In particular, the primary $\mu^-$ beam is simulated according to parameters reported in Table \ref{tab:acc-params} travelling counterclockwise starting \qty{200}{\m} away from the IP.

\begin{figure}[htb]
	\centering
	\includegraphics[width=0.4\textwidth]{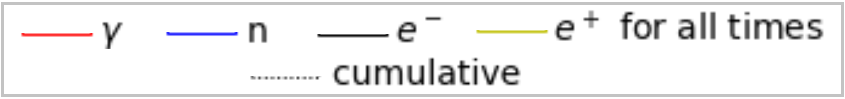}
	\includegraphics[width=0.75\textwidth]{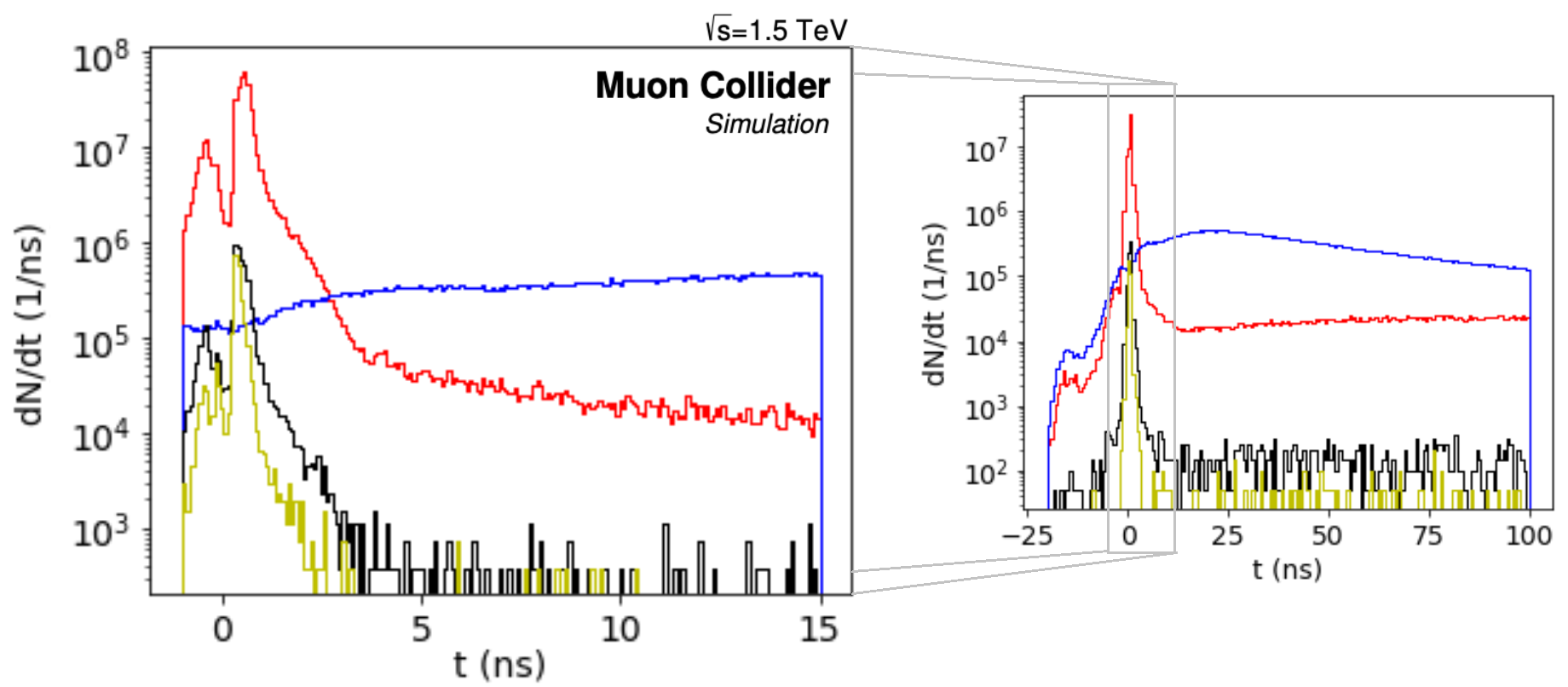}	\includegraphics[width=0.55\textwidth]{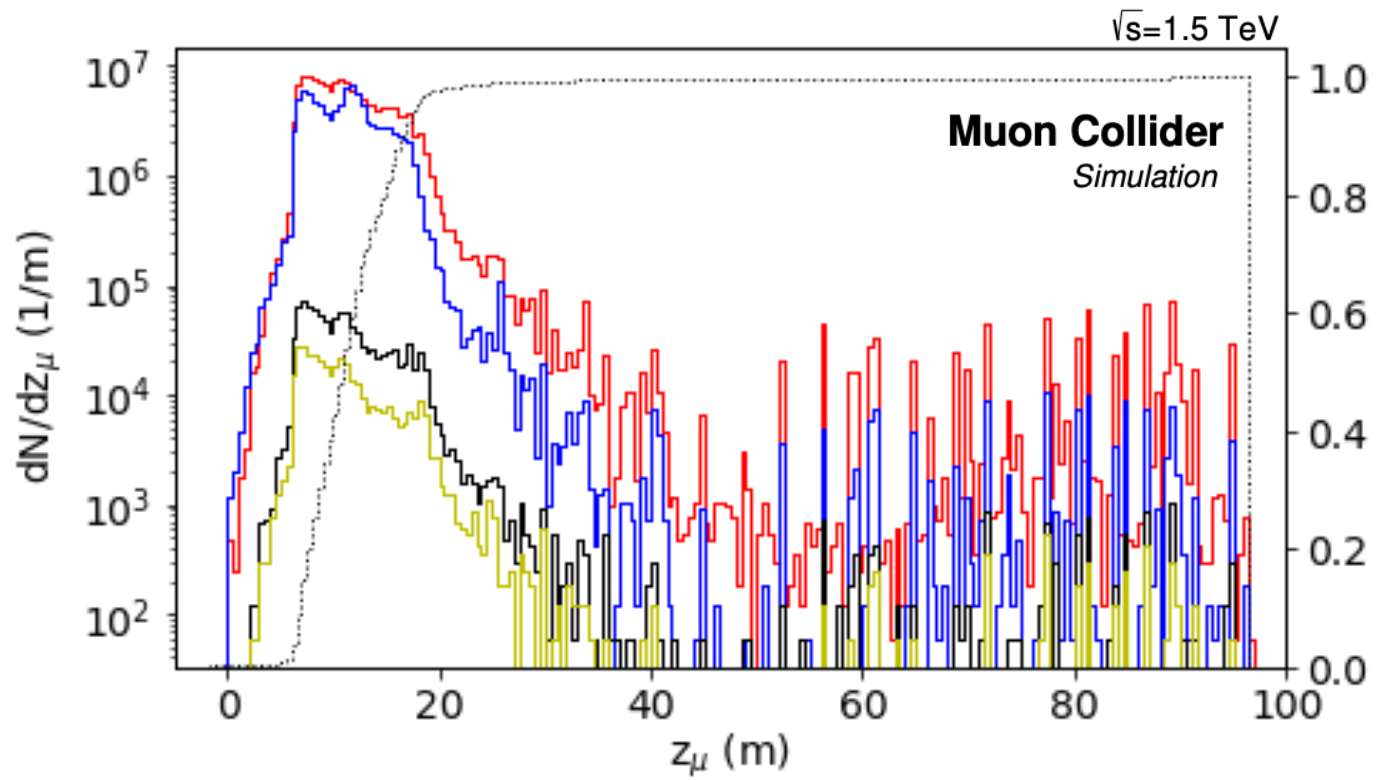}
	\caption{\label{fig:bibtime} Time distribution of BIB particles exiting the machine (top). Longitudinal distribution of primary $\mu^-$ decay generating BIB particles exiting the machine and cumulative function normalized to 1 of the total distribution in black dotted line (bottom). Results by FLUKA considering primary $\mu^-$ decays within \qty{100}{\m} from the IP. }
\end{figure}

The major contributors to BIB are photons, neutrons and electrons/positrons. The time at which BIB exits the machine in the IR is spread over a wide range but the major part is concentrated around the beam crossing time ($t = 0$), as shown by the top panel of Fig.~\ref{fig:bibtime}. 

The bottom panel of Fig.~\ref{fig:bibtime} reports the longitudinal distribution of primary $\mu^-$ decays generating the most relevant BIB families: the cumulative function shows it is enough to consider decays within $\sim\qty{25}{\m}$ from the IP. On the contrary, simulations show that to correctly account for the secondary $\mu^\pm$, it is necessary to consider primary decays up $\sim\qty{100}{\m}$ from the IP.

\begin{figure}[htb]
\centering
\includegraphics[width=0.45\textwidth]{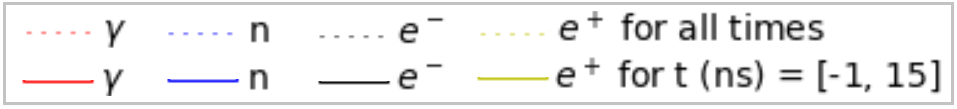}
\includegraphics[width=1\textwidth]{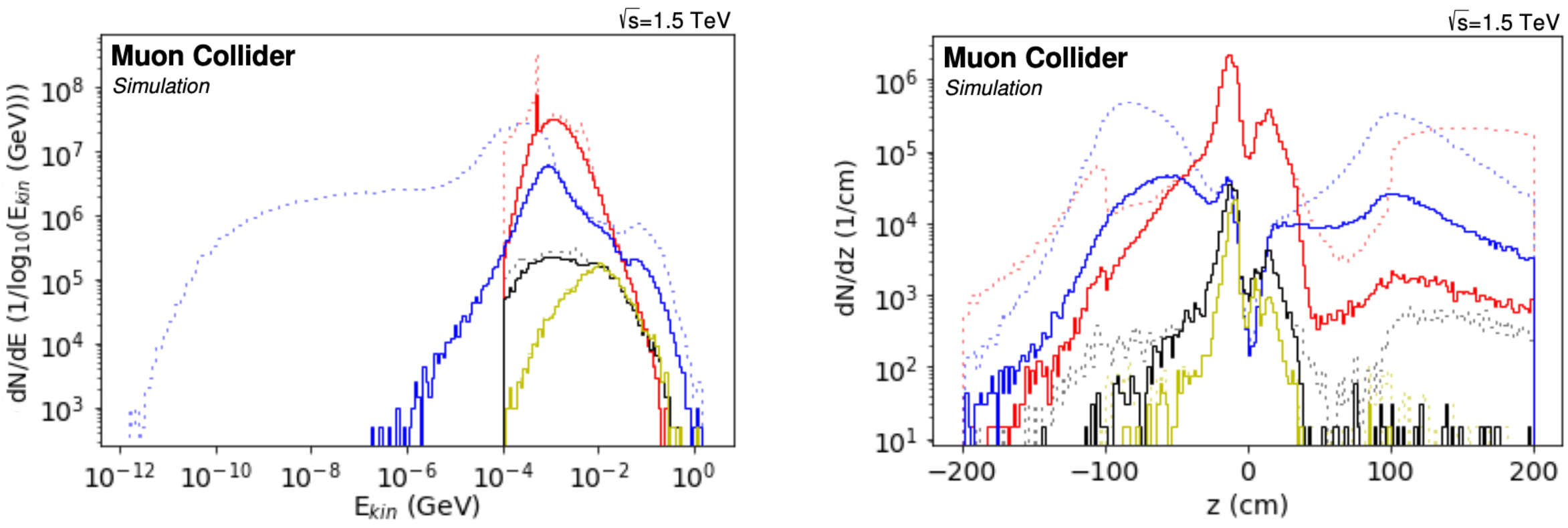}
\caption{\label{fig:bibenergy} Lethargy plot (left) and longitudinal exit coordinate distribution (right) of BIB particles, divided by particle type. Results by FLUKA considering primary $\mu^-$ decays within 100~m from the IP. No time cut is applied to distributions represented in dotted lines while in solid line only particles exiting the machine between -1 and \qty{15}{\ns} are considered.}
\end{figure}

The kinetic energy distribution of most relevant BIB particle types is reported in Fig.~\ref{fig:bibenergy}. Energy cutoffs have been applied in the simulation at 100 keV for $\gamma$, $e^\pm$, $\mu^\pm$, charged hadrons and at 10$^{-14}$ GeV for neutrons. The nozzles act in a very significant way in cutting out the high energy BIB component: as we can notice the BIB particles entering the detector hall have kinetic energy below few GeVs. Only charged hadrons and secondary muons can reach much higher energies but their number is quite low, in the order of 10$^{4}$ and 10$^{3}$, with respect to 10$^{7}$ photons, neutrons and 10$^{5}$ electrons, positrons. 

Most of the BIB exits the machine in the region around the IP and by considering a time cut within -1 and 15 ns, which is the most relevant for the detector measures, a big portion of photons and neutrons is removed, as displayed by the left panel of Fig.~\ref{fig:bibenergy}.

\newpage
\clearpage
\subsection{Implications of higher beam energies}

FLUKA simulations at $\sqrt{s}=3$~TeV and $\sqrt{s}=10$~TeV are currently under development. Since the MDI has not yet been optimized for those energies, the one designed for $\sqrt{s}=1.5$~TeV has been adopted. In both cases the preliminary results show a BIB with intensity of the same level as in the $\sqrt{s}=1.5$~TeV configuration characterized by spatial and temporal structures very similar to those presented in the previous section. A careful optimization of machine lattice and MDI is expected to further suppress BIB in the detector region.

%% file: detector.tex
\begin{figure}
    \centering
    \includegraphics[width=0.99\textwidth]{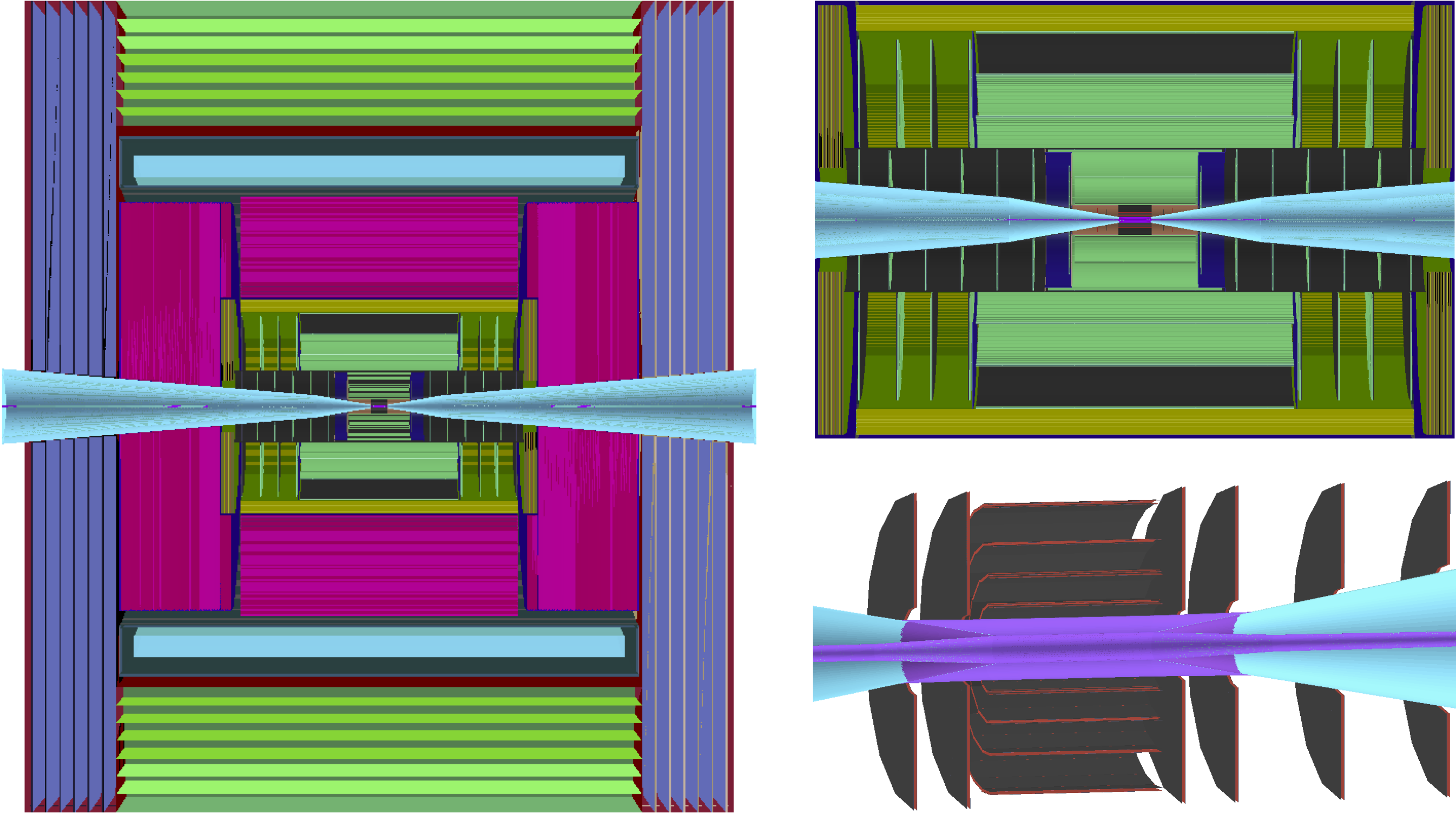}
    \caption{Rendering of the MCD geometry used for the presented simulation studies, including the cone-shaped shielding nozzles (cyan) and the beryllium beampipe (violet). Shown are the R-Z cross sections of the full detector geometry (left) and two zoomed-in portions: up to ECAL (top right) and up to Vertex Detector (bottom right). Muon Detector (violet and green) surrounds the solenoid (cyan), which encloses the HCAL (magenta), ECAL (yellow) and the Tracking Detector (green and black).}
    \label{fig:detector-geometry}
\end{figure}

The Muon Collider Detector (MCD) follows the classical cylindrical layout typical for multipurpose detectors of symmetric collisions and the specific geometry used for simulation studies in this work has the reference code \texttt{MuColl\_v1}.
The rendering of the detector geometry is presented in Fig.~\ref{fig:detector-geometry}, with the dimensions of each subsystem summarised in Table~\ref{tab:detector-dimensions}.
A cylindrical coordinate system is used with its centre placed at the nominal interaction point. The $Z$ axis is defined as the moving direction of the $\mu^+$ beam. The $X$ axis is defined to point towards the inner part of the ring and the $Y$ axis therefore pointing upwards. Cylindrical coordinates are often used with $R$, $\theta$ and $\phi$ denoting the radial distance from the interaction point, the polar and azimuthal angles respectively. Pseudo-rapidity $\eta = -\log\big[\tan(\theta/2)\big]$ is also used in some cases for convenience.

Starting from the Be beampipe with a radius of \qty{22}{\milli\metre}, the Vertex Detector is the closest to the IP with its innermost layer having a radius of only \qty{30}{\milli\metre}. It is followed by the Inner and Outer Trackers. 
The three sub-systems complete the all-silicon Tracking Detector, which operates in the strong magnetic field of \qty{3.57}{\tesla} provided by the superconducting solenoid, to reconstruct trajectories and transverse momenta ($p_\text{T}$) of charged particles.
High-granularity sampling ECAL and HCAL calorimeters are arranged outside of the Tracking Detector and provide measurements of electromagnetic and hadronic showers respectively, with fine longitudinal segmentation: 40 layers in ECAL and 60 layers in HCAL.
Outside of the superconducting solenoid is the Muon Detector based on RPC technology with its iron yoke closing the magnetic field. It is worth noting that accurate timing information from all sub-systems is an essential component of the detector, as it allows rejecting a large fraction of BIB.

\begin{table}
    \centering
    \begin{tabular}{r|l||c|c|l}
        \textbf{Subsystem} & \textbf{Region} & \textbf{R dimensions [cm]} & \textbf{$|$Z$|$ dimensions [cm]} & \textbf{Material} \\
        \hline\hline
        Vertex Detector & Barrel & $3.0 - 10.4$     & $65.0$            & Si \\
                        & Endcap & $2.5 - 11.2$     & $8.0 - 28.2$      & Si \\
        \hline
        Inner Tracker   & Barrel & $12.7 - 55.4$    & $48.2 - 69.2$     & Si \\
                        & Endcap & $40.5 - 55.5$    & $52.4 - 219.0$    & Si \\
        \hline
        Outer Tracker   & Barrel & $81.9 - 148.6$   & $124.9$           & Si \\
                        & Endcap & $61.8 - 143.0$   & $131.0 - 219.0$   & Si \\
        \hline\hline
        ECAL            & Barrel & $150.0 - 170.2$  & $221.0$           & W + Si \\
                        & Endcap & $31.0 - 170.0$   & $230.7 - 250.9$   & W + Si \\
        \hline
        HCAL            & Barrel & $174.0 - 333.0$  & $221.0$           & Fe + PS \\
                        & Endcap & $307.0 - 324.6$  & $235.4 - 412.9$   & Fe + PS \\
        \hline\hline
        Solenoid        & Barrel & $348.3 - 429.0$  & $412.9$           & Al \\
        \hline\hline
        Muon Detector   & Barrel & $446.1 - 645.0$  & $417.9$           & Fe + RPC \\
                        & Endcap & $57.5 - 645.0$   & $417.9 - 563.8$   & Fe + RPC \\
    \end{tabular}
    \caption{Boundary dimensions of individual subsystems of the Muon Collider Detector concept as defined in the geometry \texttt{MuColl\_v1}.}
    \label{tab:detector-dimensions}
\end{table}

The design of the \texttt{MuColl\_v1} detector is largely based on the \texttt{CLICdet} geometry developed by CLIC collaboration, with the details about technologies used for each component documented in Ref.~\cite{clicdet}.
The only subsystem with significant changes to its design is the Tracking Detector, which is described in the following.



\subsection{Tracking Detector}
\begin{figure}
    \centering
    \includegraphics[width=0.49\textwidth]{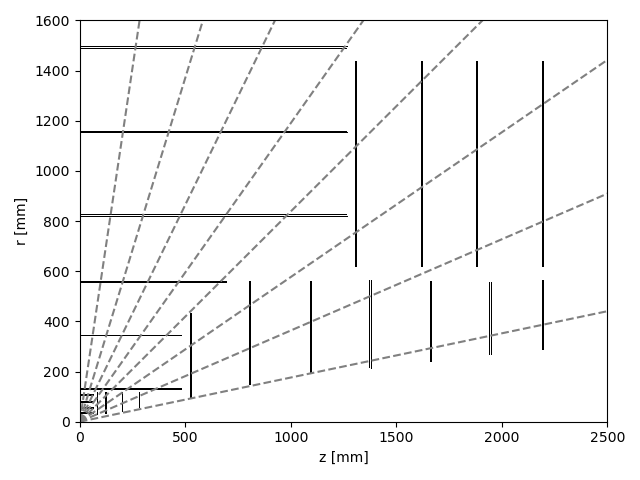}
    \includegraphics[width=0.49\textwidth]{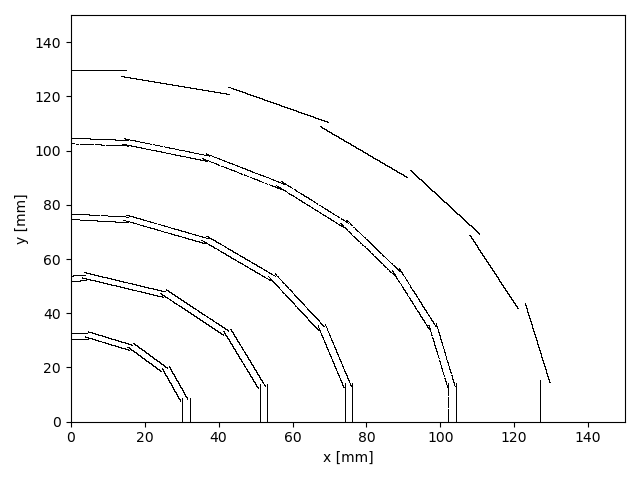}
    \caption{View of the tracking detector projected on $Z-R$ (left) and transverse plane (right). The transverse plane view is zoomed into the Vertex Detector to demonstrate the double-layer arrangement. \note{update with sub-detectors color coded}}
    \label{fig:det-tracker}
\end{figure}

The tracking detector is assumed to follow the typical design of an all-silicon tracker. The geometry is shown in Fig.~\ref{fig:det-tracker}. It consists of three sub-detector; the Vertex Detector, Inner and Outer Trackers. All three are split into a central barrel section and a forward end-cap section. The Vertex Detector is made from 4 double-layers with a \qty{2}{\mm} gap to improve secondary vertex resolution. The forward coverage is limited by the presence of the tungsten nozzle shield.

\begin{figure}
    \centering
    \includegraphics[width=0.65\textwidth]{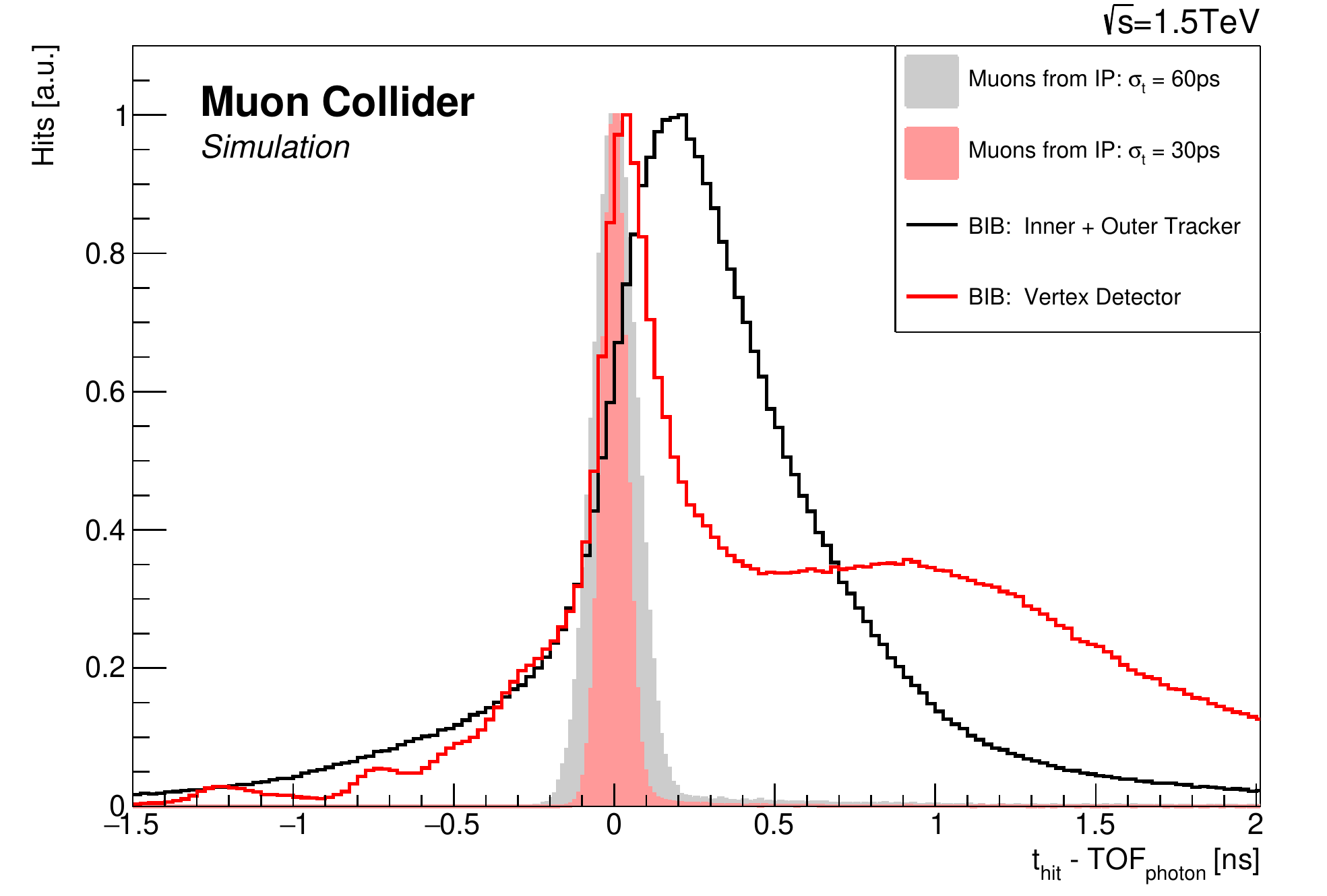}
    \caption{Comparison of hit-time distributions in the Tracking Detector between BIB particles (solid lines) and signal muons (filled areas) corrected by the time of flight of a photon from the IP, taking into account time resolution of each sub-detector.}
    \label{fig:det-hittiming}
\end{figure}

The Tracking Detector returns 4-dimensional hit coordinates; the three positional dimensions and precision timing information.  The timing information plays a critical role in reducing the effective hit density from BIB. Fig.~\ref{fig:det-hittiming} shows the timing distribution (corrected for time-of-flight) of hits from a \mumu collision compared to those induced by BIB. The spatial and timing resolutions assumed in the detector simulations are summarized in Table~\ref{tab:det-resolutions}.

\begin{table}[htb]
    \centering
    \begin{tabular}{l|c|c|c}
                        &  \textbf{Vertex Detector} & \textbf{Inner Tracker} & \textbf{Outer Tracker} \\
    \hline\hline
    Cell type           &  pixels & macropixels & microstrips \\
    Cell Size           &  \qty{25}{\um} $\times$ \qty{25}{\um} & \qty{50}{\um} $\times$ \qty{1}{\mm} & \qty{50}{\um} $\times$ \qty{10}{\mm} \\
    Sensor Thickness    &  \qty{50}{\um} & \qty{100}{\um} & \qty{100}{\um} \\
    Time Resolution     &  \qty{30}{\ps} & \qty{60}{\ps} & \qty{60}{\ps} \\
    Spatial Resolution  &  \qty{5}{\um} $\times$ \qty{5}{\um} & \qty{7}{\um} $\times$ \qty{90}{\um} & \qty{7}{\um} $\times$ \qty{90}{\um} \\
    \end{tabular}
    \caption{Assumed spatial and time resolution in different sub-systems of the Tracking Detector. There is no difference between the barrel and end-cap regions.}
    \label{tab:det-resolutions}
\end{table}

%% file: software.tex
Full simulation of a single \mumu collision event involves several stages:

\begin{enumerate}
  \item generation of all stable particles entering the detector;
  \item simulation of their interaction with the passive and sensitive material of the detector;
  \item simulation of the detector's response to these interactions;
  \item application of data-processing and object-reconstruction algorithms that would happen in a real experiment.
\end{enumerate}

The first stage of generating stable input particles is handled by standalone software, such as Monte Carlo event generators for the \mumu interaction and FLUKA or MARS15 for the BIB particles.
The rest of the simulation process is performed inside the iLCSoft framework~\cite{ilcsoft} previously used by the CLIC experiment~\cite{clic.cdr} and now forked for developments of Muon Collider studies~\cite{mucolsoft}.
Particle interactions with the detector material are simulated in GEANT4~\cite{GEANT4:2002zbu}, while detector response and event reconstruction are handled inside the modular Marlin framework~\cite{marlin}.
The detector geometry is defined using the DD4hep detector description toolkit~\cite{dd4hep}, which provides a consistent interface with both the GEANT4 and Marlin environments.
More detail about the software structure and computational optimization methods used for simulating such a large number of BIB particles is documented in Ref.~\cite{bib-optimisation}.

\subsection{Detector digitization}
\label{sec:software-digitization}

Response of each sensitive detector to the corresponding energy deposits returned by GEANT4 is simulated by dedicated digitization modules implemented as individual Marlin processors.

The Tracking Detector uses Gaussian smearing to account for the spatial and time resolutions of local hit coordinates on the sensor surface and the time coordinate.
The assumed resolution values are listed in Table~\ref{tab:det-resolutions}.
Acceptance time intervals, individually configured for each sub-detector, are used for replicating the finite readout time windows in the electronics of a real detector and to reject hits from asynchronous BIB particles.
The result of this simplified approach is one-to-one correspondence between the GEANT4 hits and digitized hits, which ignores the effect of charge distribution across larger area due to the Lorentz drift and shallow crossing angles with respect to the sensor surface.
These effects are taken into account in the more realistic tracker digitization code that is currently under development and will allow stronger BIB suppression based on cluster-shape analysis.

ECAL and HCAL detectors are digitized using realistic segmentation of sensitive layers into cells by summing all energy deposits in a single cell over the configured integration time of \qty{\pm 250}{\pico\second}.
Time of the earliest energy deposit is consequently assigned to the whole digitized hit.
The same digitization approach is used also for the Muon Detector.

%% file: tracking.tex
\begin{figure}
    \centering
    \includegraphics[width=0.7\textwidth]{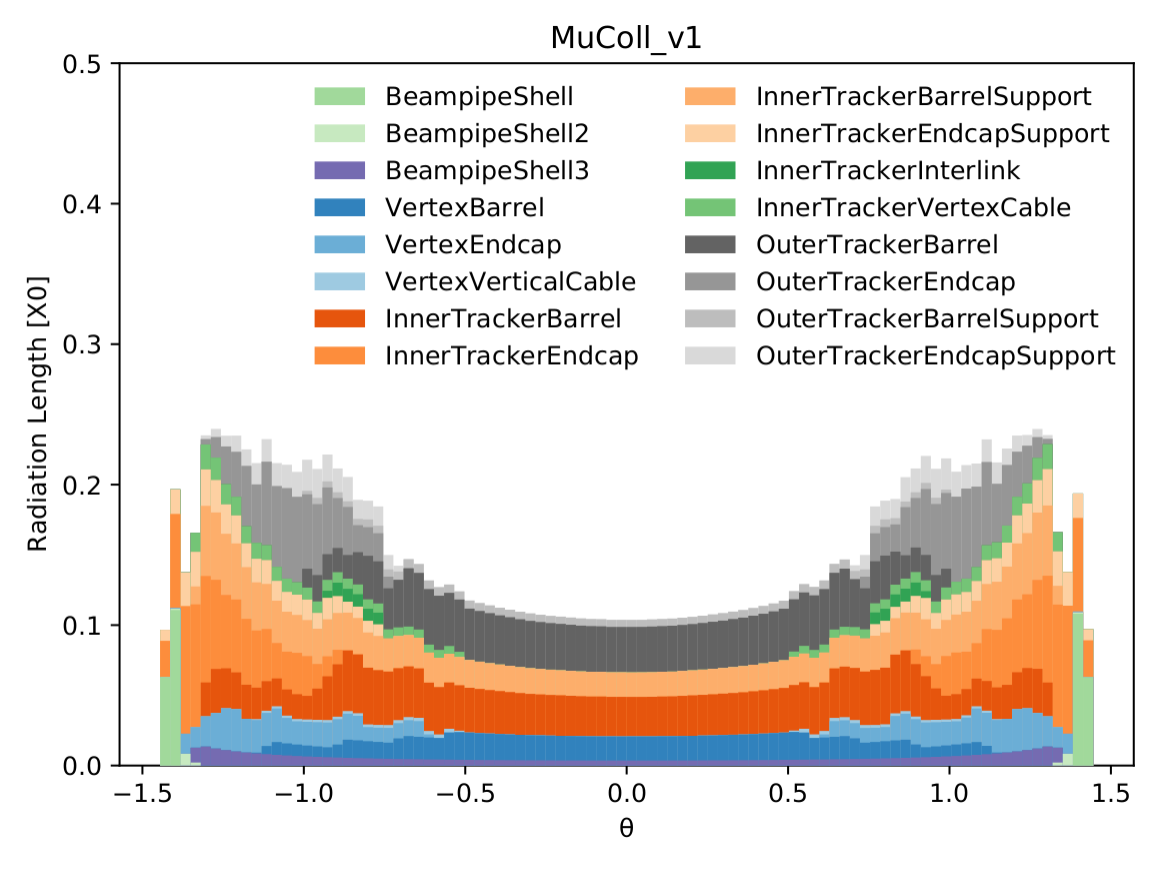}
    \caption{Radiation length of the tracking detectors, as seen along a line defined by the nominal interaction point and the polar angle $\theta$.}
    \label{fig:tracking:x0-tracker}
\end{figure}

In the magnetic field of MCD a charged particle will follow a helix trajectory aligned with the $Z$ axis. The radius of curvature in the transverse plane is proportional to the magnetic field (\qty{3.57}{\tesla}) and the transverse momentum of the particle.  Deviations from a perfect helix can occur due to multiple-scattering, ionizing  energy losses and bremsstrahlung. The first two are a direct function of the detector material. The amount of radiation lengths that a particle traverses through the tracking detector when starting from the nominal collision point is shown in Fig.~\ref{fig:tracking:x0-tracker}.

The charged particle trajectory is reconstructed from hits positions in the silicon tracking detector. The reconstructed object is called a track. A track consists of a set of hits (one per layer) and five fitted parameters describing the helix. A track reconstruction algorithm can roughly be broken up into two steps: pattern recognition to identify the hits belonging to a single track and fitting the hit coordinates by a track model to deduce the relevant track parameters.

\begin{figure}
    \centering
    \includegraphics[width=0.7\textwidth]{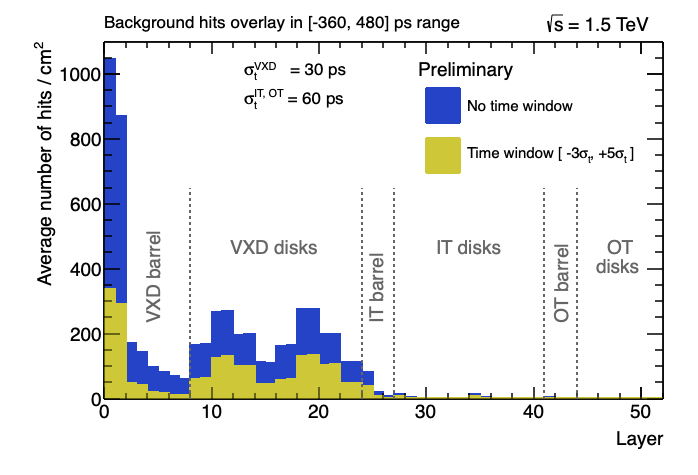}
    \caption{Hit density in the different layers of the tracking detectors in a single event with full BIB overlay. The density before (blue) and after (yellow) applying the timing cut is shown.}
    \label{fig:tracking:timing}
\end{figure}

Track reconstruction at the Muon Collider is complicated by the presence of a huge number of hits in the silicon sensor originating from the Beam Induced Background (BIB-hits). The density of BIB hits is ten times larger than the expected contribution from pile-up events at a High Luminosity LHC detector. Table~\ref{tab:tracking:density} compares the hit density between the MCD and ATLAS Inner TracKer (ITk) upgrade for HL-LHC operation~\cite{CERN-LHCC-2017-021, CERN-LHCC-2017-005}. The increase in possible hit combinations creates a challenge for the hit pattern recognition step. It is crucial to reduce the amount of hits at the input to a track reconstruction algorithm through alternate means, such as precision timing information. The BIB hits are out-of-time with hard collision hits after the time-of-flight correction has been applied. By applying a $-3\sigma/+5\sigma$ time window, the hit density can be reduced by a factor of two as seen in Fig.~\ref{fig:tracking:timing}.

\begin{table}[htb]
    \centering
    \begin{tabular}{l||c|c}
        \textbf{ATLAS ITk Layer} & \textbf{ITk Hit Density [\unit{\mm^2}]} & \textbf{MCD Equiv. Hit Density [\unit{\mm^2}]} \\
        \hline
        \hline
         Pixel Layer 0 & 0.643 & 3.68 \\
         Pixel Layer 1 & 0.022 & 0.51 \\
         Strips Layer 1 & 0.003 & 0.03
    \end{tabular}
    \caption{Comparison of the hit density in the tracking detector between the ATLAS ITk upgrade for HL-LHC and the MCD with full BIB overlay. The hit densities for layers at equivalent radii are shown. The MCD numbers are after timing cuts.}
    \label{tab:tracking:density}
\end{table}

The spatial distribution of BIB-hits is also unique. They are different from hits created by pile-up. Pile-up hits come from real charged particle tracks originating from multiple vertices in the collision region. On the other hand, BIB-hits come from a diffuse shower of soft particles originating from the nozzles. The compatibility of a track with a trajectory originating from the luminous region provides an important handle for differentiating "real" tracks of charged particles produced in the primary collision and "fake" tracks generated from random combinations of BIB-hits.

The remainder of this section describes three approaches that were studied for track reconstruction at the MCD. The first two use the Conformal Tracking (CT) algorithm developed for the clean environment of the electron-positron colliders~\cite{Brondolin:2019awm}. However in the presence of BIB, the CT algorithm takes weeks to reconstruct a single event and is impractical for large-scale production of simulated data. To ease the computational effort, the input hits are first reduced by either defining a Region of Interest (Section~\ref{sec:trk-roi}) or by exploiting the double-layered Vertex Detector to select only hit pairs pointing to the collision region (section~\ref{sec:trk-dl}). The third approach (Section~\ref{sec:trk-ckf}) uses the Combinatorial Kalman Filter (CKF)~\cite{Billoir:1989mh, Billoir:1990we, Mankel:1997dy} algorithm developed for the busy environment of hadron colliders. It can perform track reconstruction in a reasonable time without requiring any additional filtering of input hits.

It should be noted that the CT and CKF algorithms have very different software implementations that are responsible for much of the difference in their performance. The CKF algorithm is implemented using the A Common Tracking Software (ACTS)\cite{Ai:2021ghi} library that is heavily optimized for efficient computing. The same is not true for the CT algorithm implemented directly in iLCSoft with less emphasis on computational efficiency. It is possible that part of the computational improvements come from code quality alone. For example, the ACTS Kalman Filter implementation is a factor 200 faster than the default iLCSoft implementation given the same inputs.  This demonstrates the advantage of an experiment-independent track reconstruction library developed by a dedicated team with strong computing expertise.

The expected performance is assessed using a sample of single muons originating at the interaction point. BIB-hits are overlaid, unless explicitly noted. Two set of samples are used; one set is generated with the muon having a fixed momentum $(p)$ of $1$, $10$ and $100$~GeV and a uniform distribution in $\theta$. The second set is generated at discrete values of $\theta=13^\circ,30^\circ,89^\circ$ and uniform transverse momentum distribution in the $1-100$~GeV range. The chosen $\theta$ values correspond to particles expected to leave hits entirely in the endcap system, some in the barrel and some in the endcap, and entirely in the barrel region, respectively.

\subsection{Conformal Tracking in Regions of Interest}
\label{sec:trk-roi}
The Conformal Tracking algorithm is based on Ref~\cite{Brondolin:2019awm} and implementation in iLCSoft. The algorithm is designed and optimized to find charged particles in very clean environments, as the ones in \ee colliders. The conformal mapping technique~\cite{Hansroul:1988wa} is combined with cellular automaton approach~\cite{Glazov:1993ur} to increase the acceptance of non-prompt particles.  

Due to the very large hit multiplicity from BIB-hits, running such an algorithm for the full event is prohibitive in terms of CPU and memory resources. Instead a region-of-interest approach is used, where hits to be considered are pre-selected based on existing objects reconstructed in either the calorimeters or the muon system. To assess tracking performance only hits within a cone of $\Delta R < 0.5$ around the signal muon were selected as input to the CT algorithm, where $\Delta R = \sqrt{\Delta \phi^2 + \Delta \eta^2}$.

\begin{figure}
    \centering
    \includegraphics[width=0.49\textwidth]{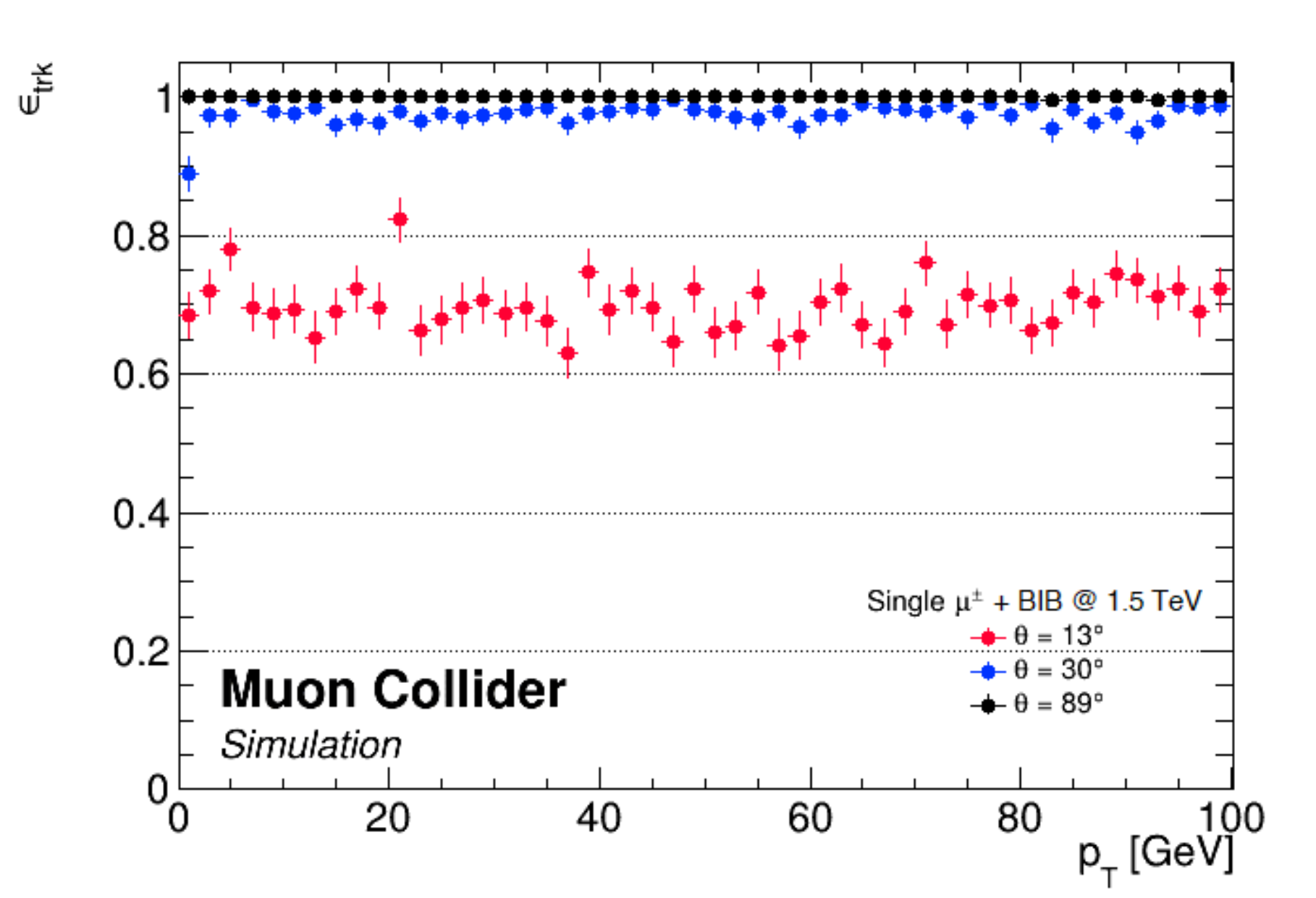}
    \includegraphics[width=0.49\textwidth]{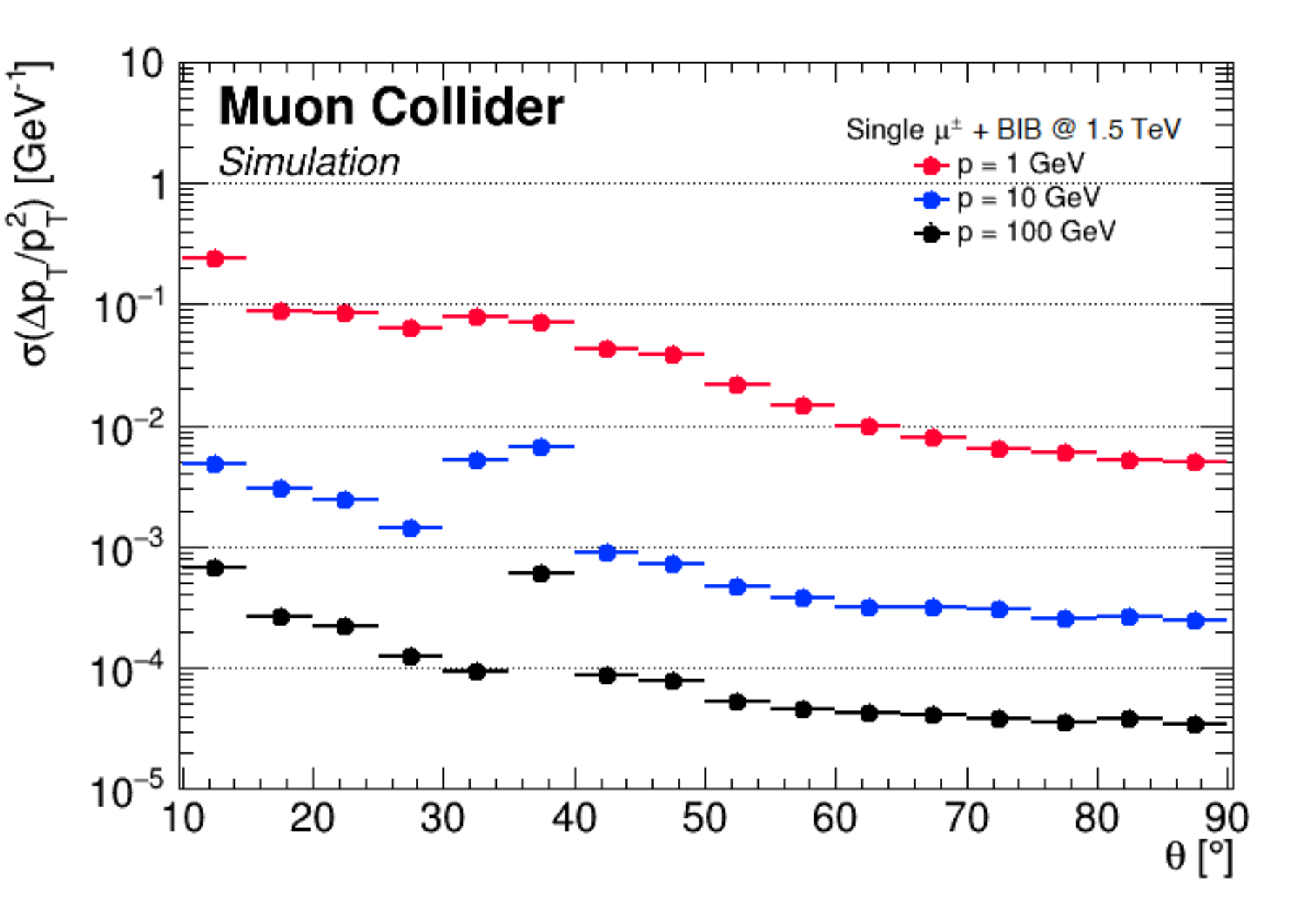}
    \caption{Tracking performance for single-muon events overlaid with BIB. Left: track reconstruction efficiency is shown as a function of $\pt$. Right: momentum resolution $\Delta\pt$ is shown divided by the $\pt^2$ as a function of $\theta$.\note{Replace with PDF version and final style}}
    \label{fig:tracking:perf-roi}
\end{figure}

Fig.~\ref{fig:tracking:perf-roi}~(left) shows the reconstruction efficiency as a function of $\pt$; a muon is considered reconstructed if at least half of the hits associated to the track have been originated by the muon. Optimal reconstruction efficiency is achieved throughout the \pt spectrum, with a somewhat smaller efficiency for very forward particles due to their proximity with the nozzle and much larger expected occupancy in the region. The latter is expected to be recovered by a more dedicated tuning of the algorithm or by using one of the algorithms described in the next sections. 

Fig.~\ref{fig:tracking:perf-roi}~(right) shows transverse momentum resolution as a function of polar angle $\theta$. The resolution is computed by comparing reconstructed and generated $\pt$ and shown divided by $\pt^2$. A localized degradation of the resolution can be seen around $\theta\approx 35^\circ$, corresponding to the barrel-endcap transition; in addition, the feature is enhanced by the non-physical lack of spread of the muon originating point. It is expected that a more realistic simulation of the luminous region as well as future optimizations of the tracker layout will mitigate such localized degradation to negligible levels.

\begin{figure}
    \centering
    \includegraphics[width=0.49\textwidth]{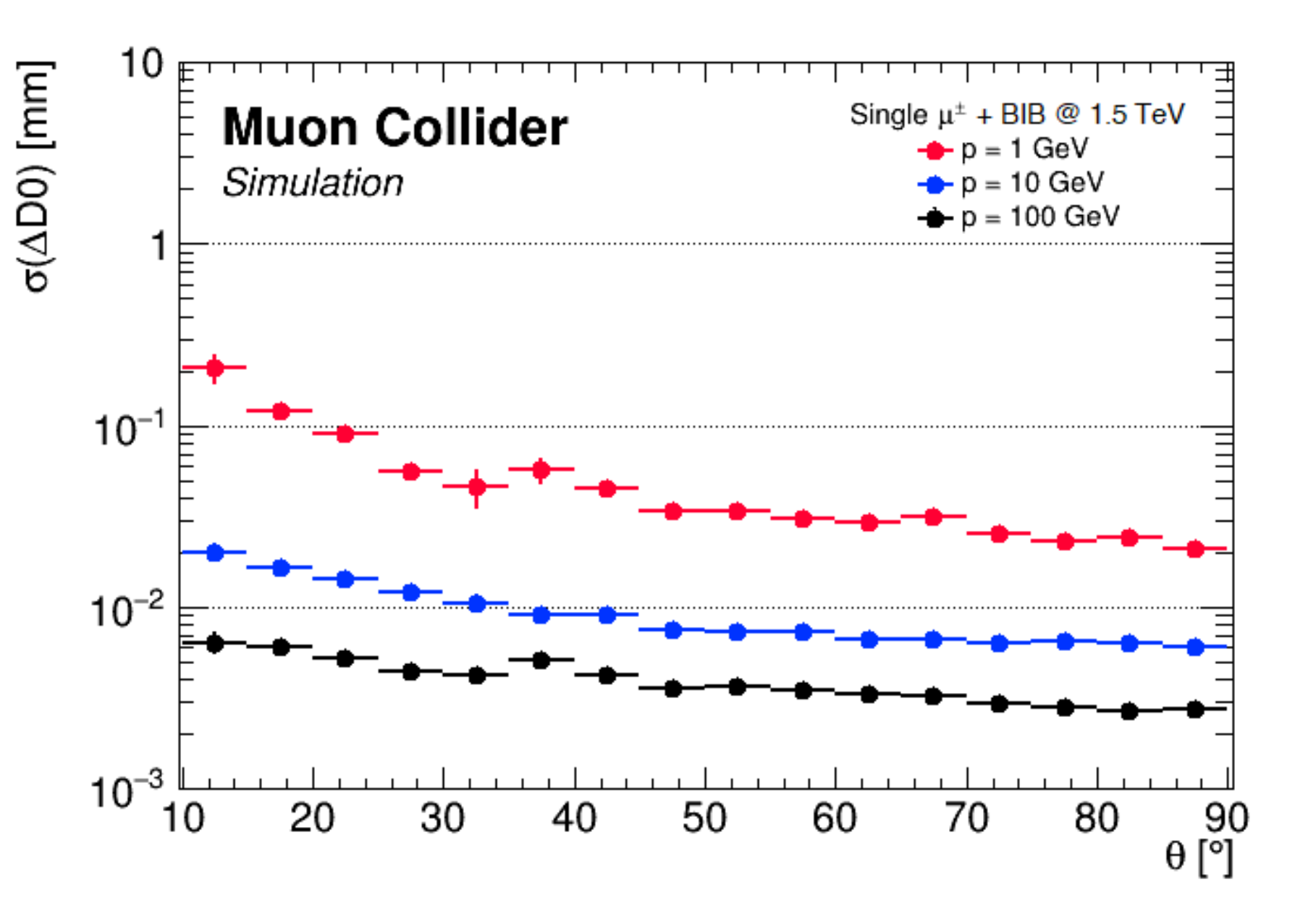}
    \includegraphics[width=0.49\textwidth]{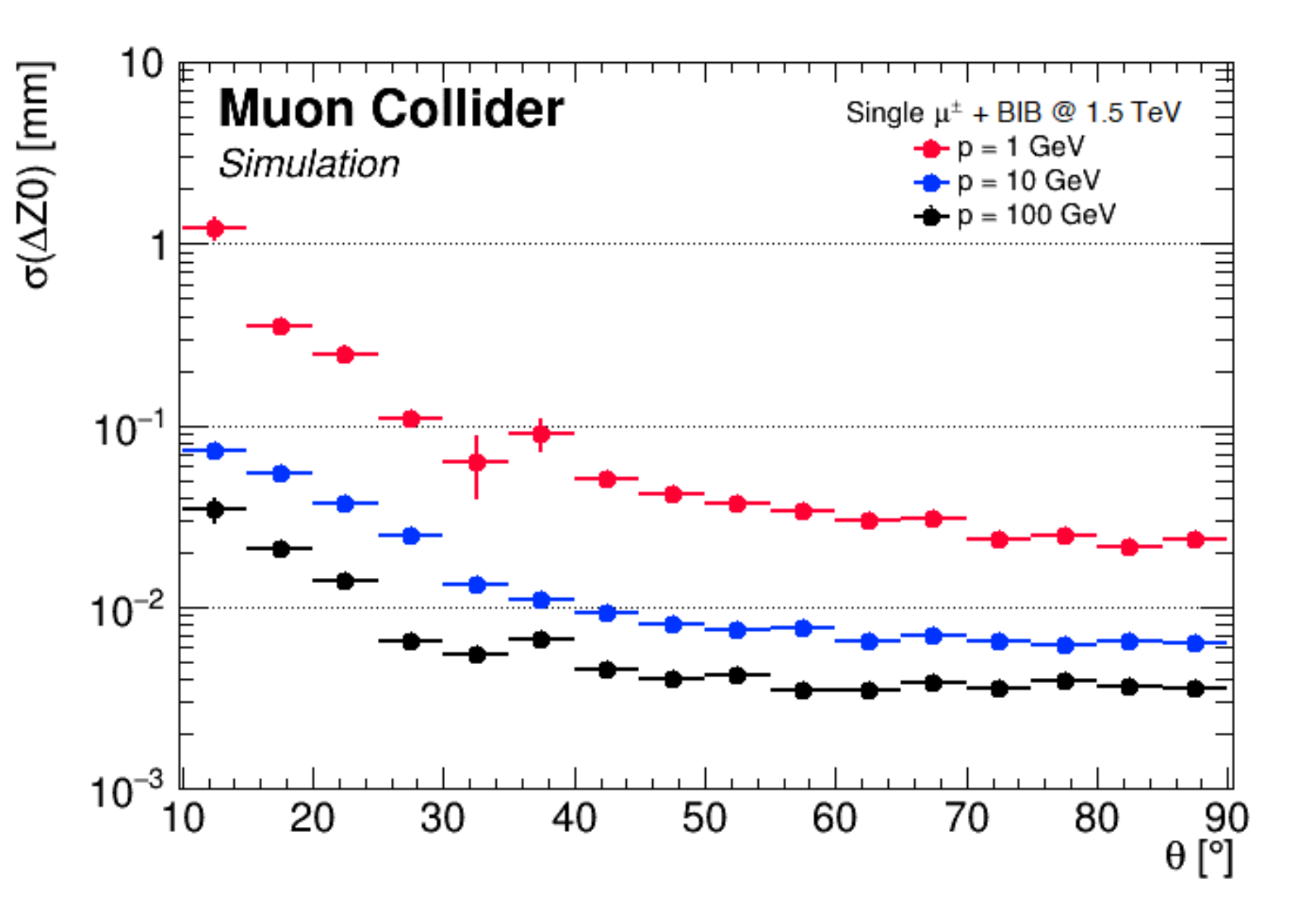}
    \caption{Tracking performance for single-muon events overlaid with BIB. Left: impact parameter resolution as a function of $\theta$. Right: longitudinal impact parameter resolution as a function of $\theta$}
    \label{fig:tracking:perf-roi_2}
\end{figure}

Fig.~\ref{fig:tracking:perf-roi_2}~shows the resolution on the impact parameter $D_0$ (left) and the resolution on the longitudinal impact parameter $Z_0$ (right) as a function of the polar angle $\theta$. Similarly to the case of the resolution on $\pt$, the resolution on $D_0$ and $Z_0$ slightly degrades in the barrel-endcap transition region, around $\theta \approx 35^\circ$. 

\note{TODO: likely will add expected resolution of $d_0$ and maybe $z_0$.}

\subsection{Conformal Tracking with Double Layers}
\label{sec:trk-dl}
\begin{figure}
    \centering
    \includegraphics[width=0.6\textwidth]{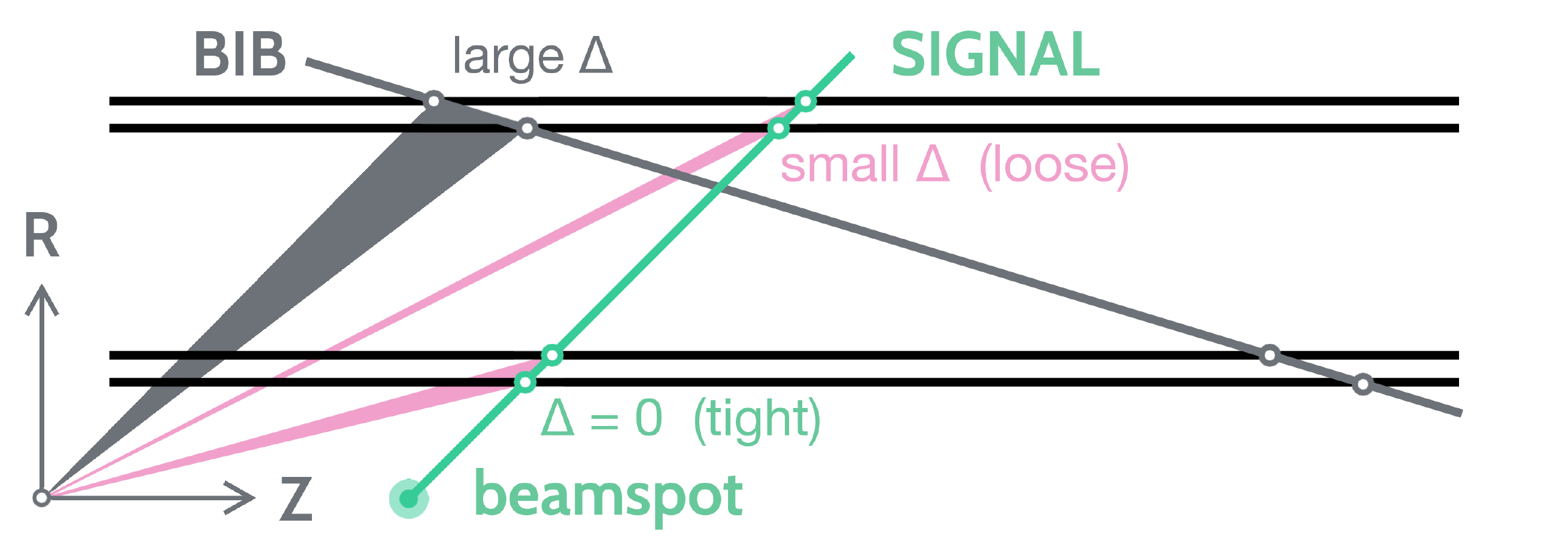}
    \caption{Illustration of the doublet-layer filtering used for the rejection of BIB-induced hits in the Vertex Detector. Horizontal black lines represent double layers of pixel sensors that are crossed by signal (green) and BIB (grey) particle tracks. Hit doublets created by a signal particle are perfectly aligned with the beamspot position, but have a sizeable angular difference when measured from the center of the detector. Hit doublets created by BIB particles are characterized by larger angular difference due to their shallow crossing angle and more displaced origin.}
    \label{fig:tracking:dl-howto}
\end{figure}

The Vertex Detector is constructed using double-layers (DL). A double-layer consists of two silicon detector layers separated by a small distance (\qty{2}{mm} for MCD). This concept will also be used by the CMS Phase-II tracking detector\cite{CERN-LHCC-2017-009} to reduce the hit combinatorics for a fast track reconstruction in their trigger system. It works by selecting only those hits that can form a pair with a hit from the neighboring layer that in aligned with the interaction point (IP). If there is no 2nd hit in the double-layer to form a consistent doublet the hit is discarded. This approach is particularly effective for rejecting BIB hits, because BIB electrons are very likely to either stop in the first layer due to their very low momentum, or to cross the double-layer at shallow angles, creating doublets that are not aligned with the IP.
The DL filtering implemented in our simulation software is based on the angular distance between the two hits of a doublet when measured from the interaction point, as demonstrated in Fig.~\ref{fig:tracking:dl-howto}. For simplicity the two variables used for filtering are the polar ($\Delta\theta$) and azimuthal ($\Delta\phi$) angle differences.

In practice there are several limitations to the precision of alignment that can be imposed by the DL filtering while maintaining high efficiency for signal tracks.
The first is driven by the finite spatial resolution of the pixel sensors, which limits the minimum resolvable displacement between the two hits of a doublet.

Its position needs to be known beforehand and any uncertainty will result in an inefficiency. The latter point is also important for displaced particles (ie: secondary vertices of $b$-mesons) that do not originate from the IP.

\begin{figure}
    \centering
    \includegraphics[width=0.47\textwidth]{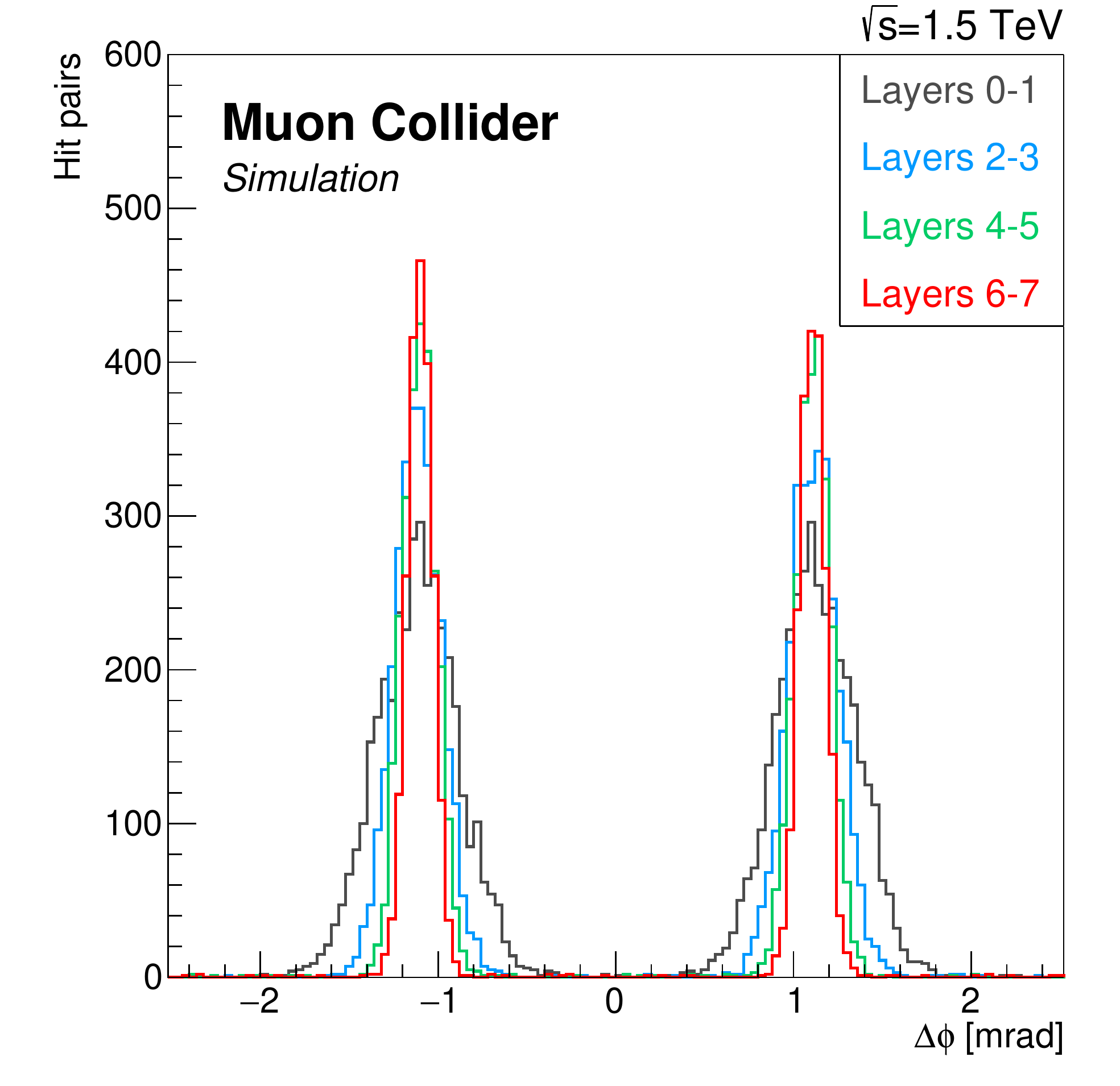} \hfill
    \includegraphics[width=0.47\textwidth]{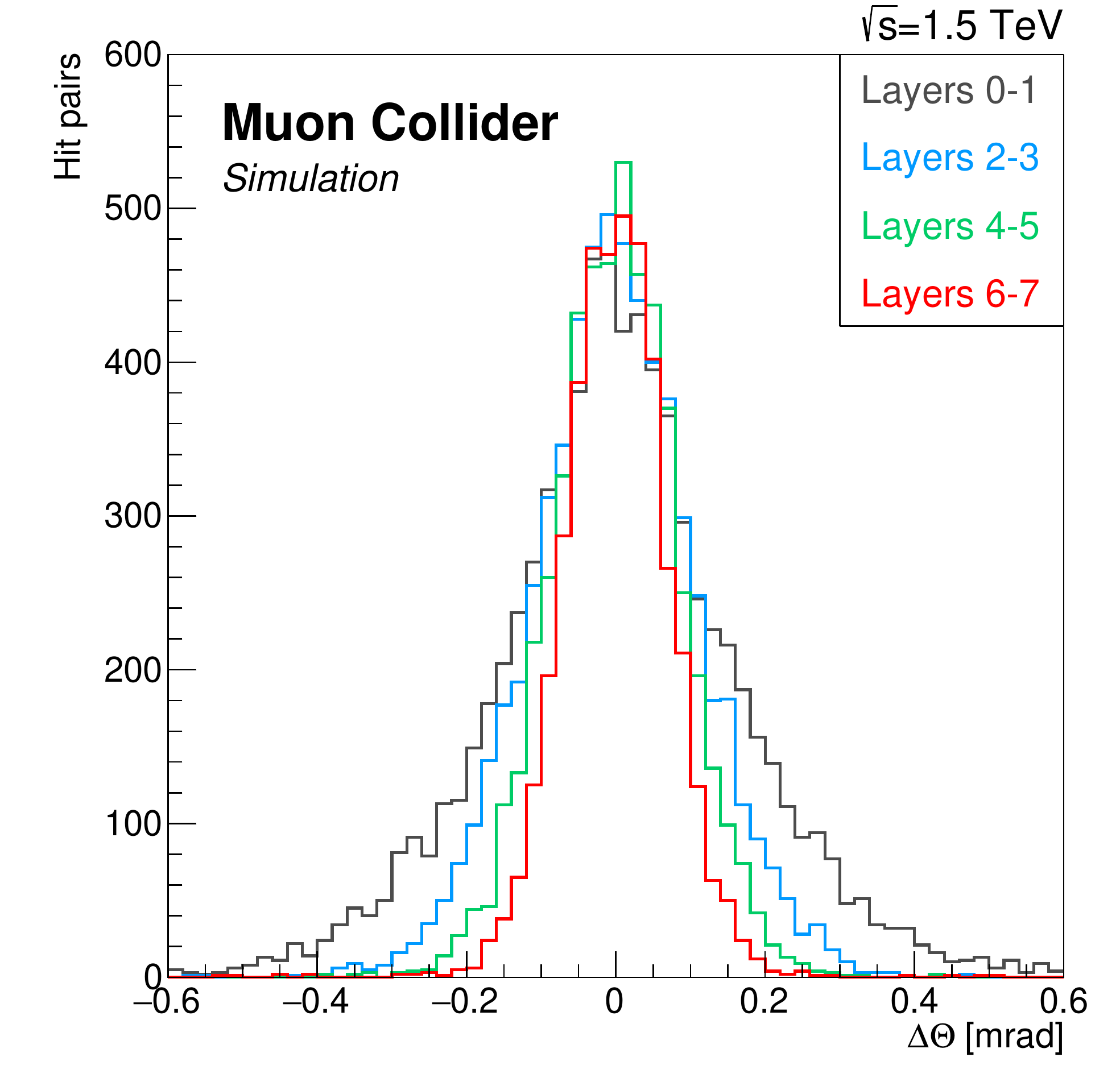}
    \caption{Expected $\Delta\phi$ (left) and $\Delta\theta$ (right) distributions for hit doublets in the barrel region of the Vertex Detector. Simulated are single-muon tracks with $\pt = \qty{1}{\GeV}$ radiated from the nominal interaction point in the centre of the detector. Angular spread is increasing in layers at smaller radii.}
    \label{fig:tracking:perf-dl-noBS}
\end{figure}

\begin{figure}
    \centering
    \includegraphics[width=0.47\textwidth]{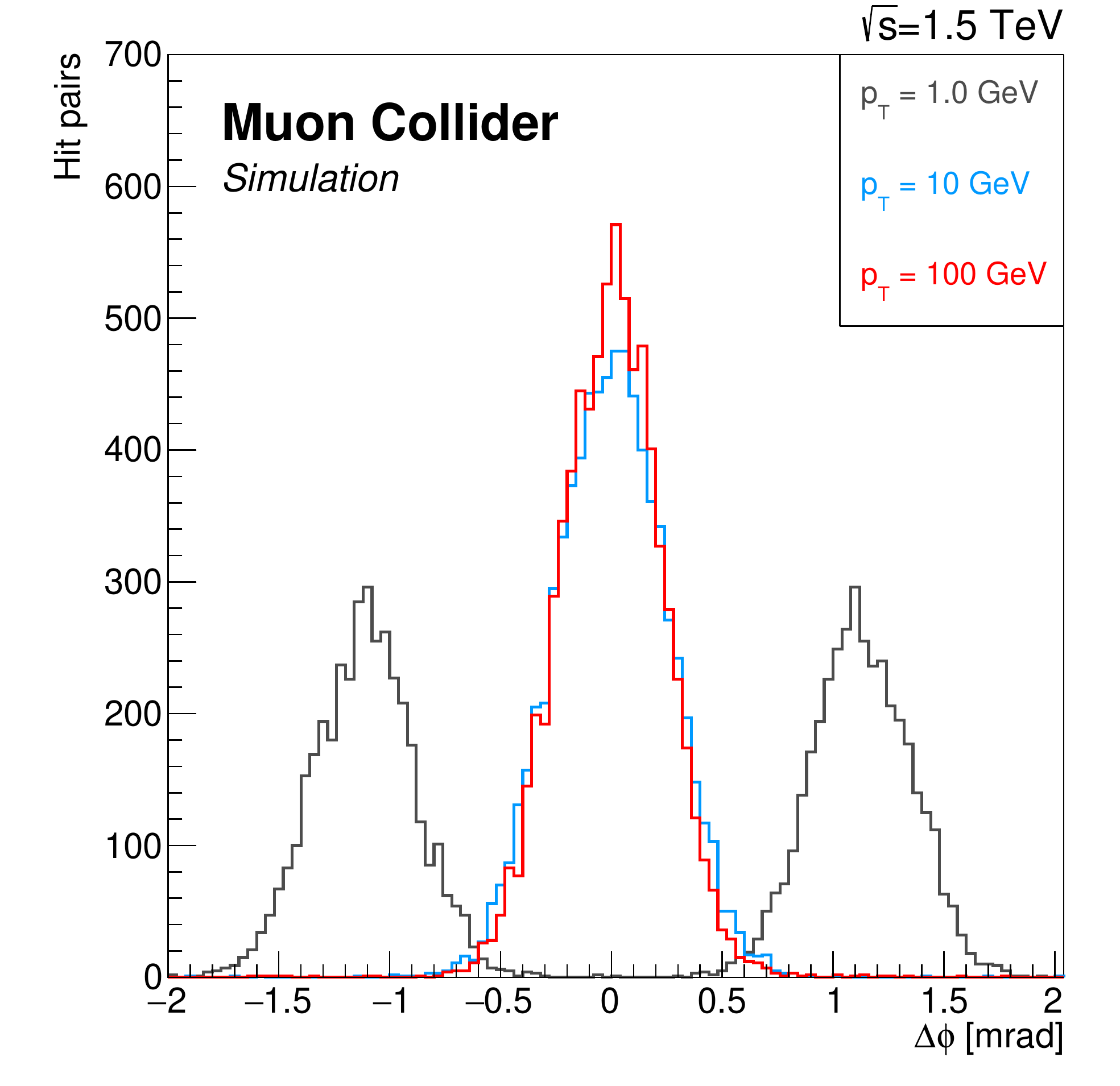} \hfill
    \includegraphics[width=0.47\textwidth]{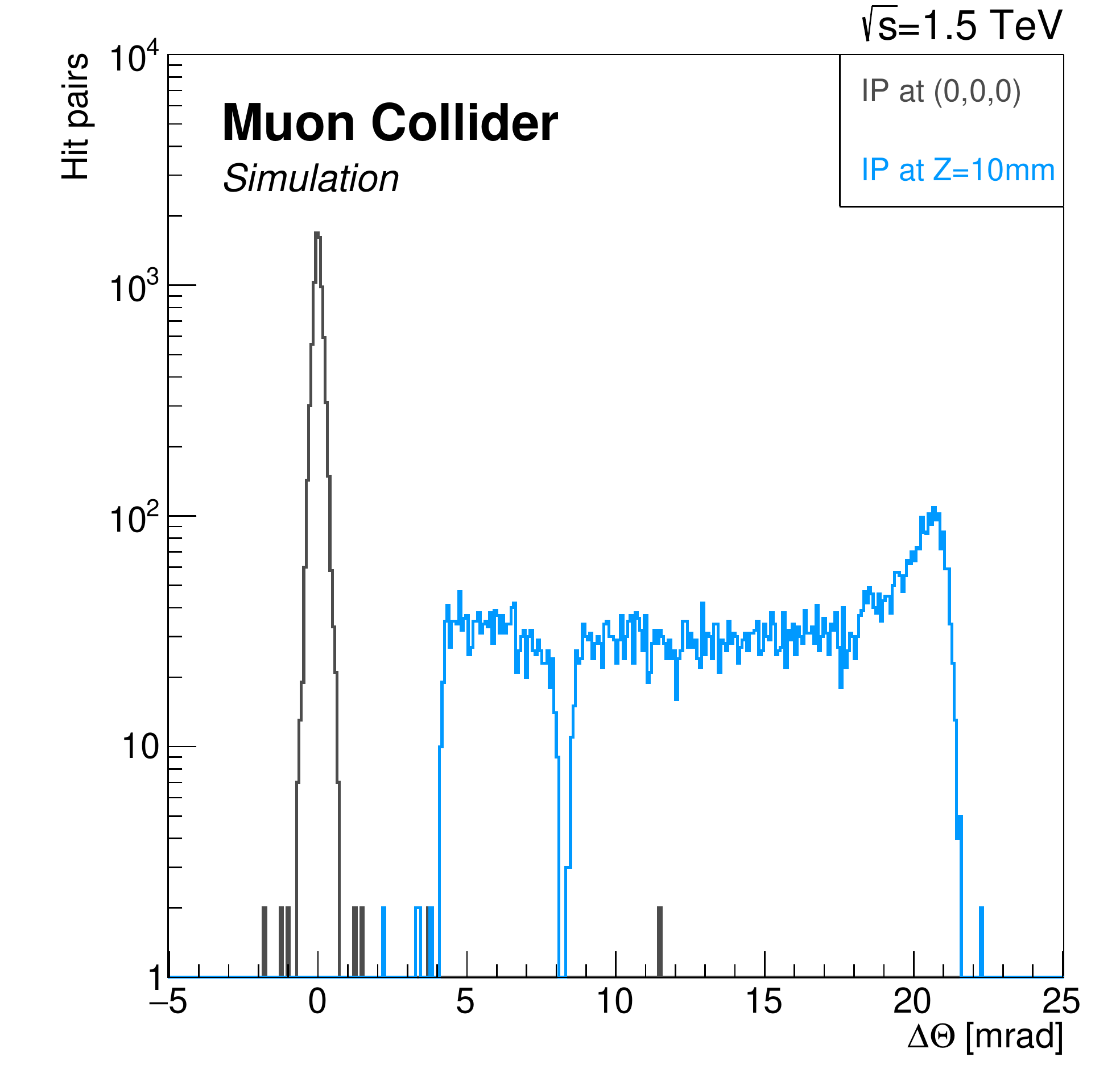}
    \caption{Left: dependence of $\Delta\phi$ in hit doublets on the transverse momentum of muon tracks. Two separate peaks become visible for low-\pt tracks corresponding to $\mu^+$ and $\mu^-$. Right: dependence of $\Delta\theta$ distribution from $\pt = \qty{1}{\GeV}$ muon tracks on the longitudinal displacement of the interaction point by \qty{10}{\milli\metre}. Both plots are obtained from hits in the innermost double-layer of the Vertex Detector in the barrel region.}
    \label{fig:tracking:perf-dl-realisticBS}
\end{figure}

For simplicity the two variables used for filtering are the polar ($\Delta\theta$) and azimuthal ($\Delta\phi$) angular distances. Fig.~\ref{fig:tracking:perf-dl-realisticBS} shows the distributions of $\Delta\theta$ and $\Delta\phi$ in single-muon events with a realistic beamspot in the first layer of the vertex detector. The distribution in other double-layers for $p=\qty{1}{\GeV}$ muons are shown in Fig.~\ref{fig:tracking:perf-dl-noBS}. The bi-modal nature of the $\Delta\phi$ distribution for low energy muons is the result of the circular path that charged particle take in the transverse plane. This biases DL filtering toward charge particles with higher $\pT$ unless very loose cuts are used.

\begin{figure}
    \centering
    \includegraphics[width=0.45\textwidth]{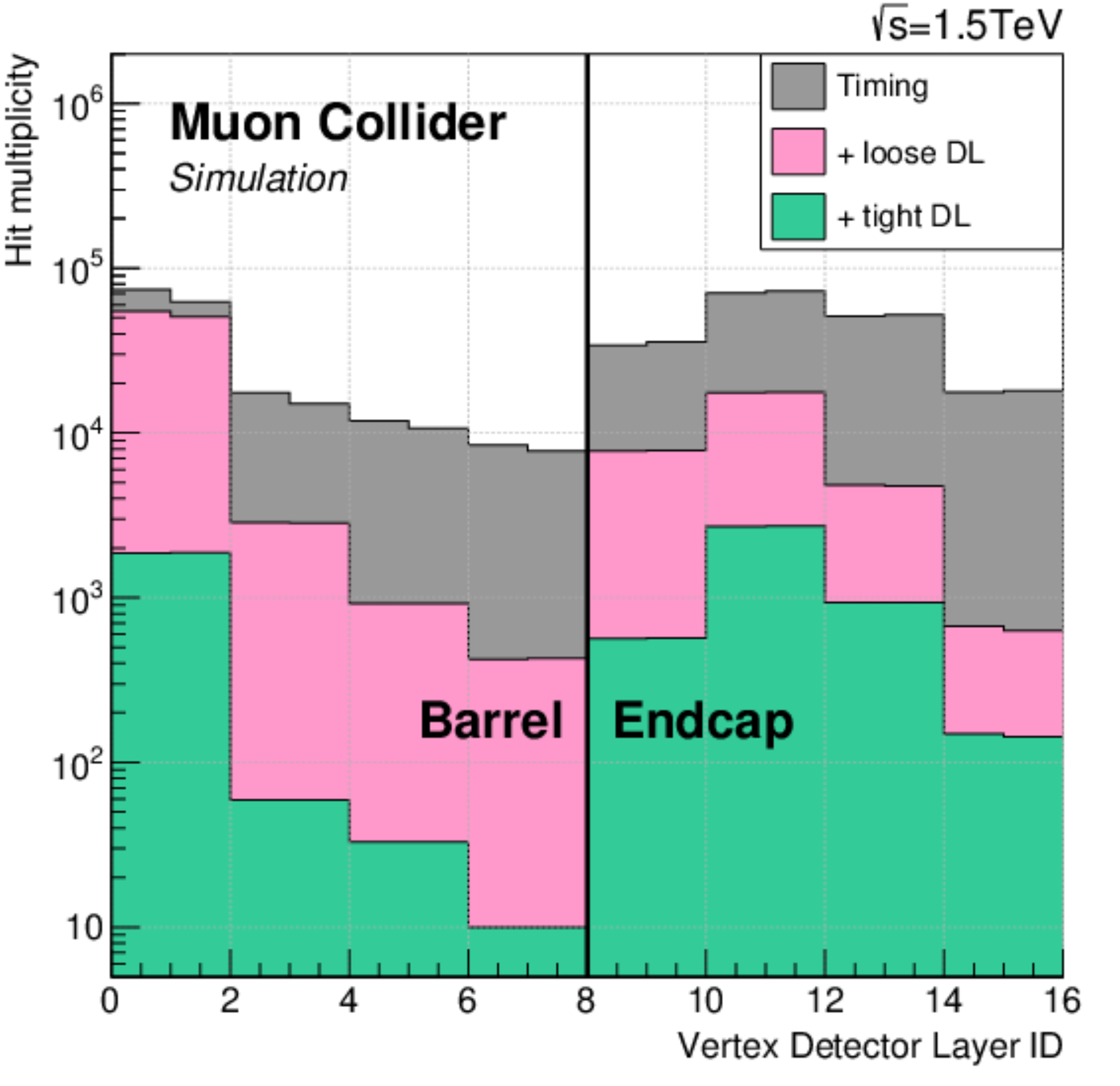}
    \caption{Expected reduction of hit multiplicity in the Vertex Detector achieved by applying the loose or tight double-layer filtering.}
    \label{fig:tracking:perf-dl-hits-reduction}
\end{figure}

Table~\ref{tab:tracking:dl-rejection} lists the two operating points ({\em loose} and {\em tight}) used to filter hits for the CT algorithm in two stages. Fig.~\ref{fig:tracking:perf-dl-hits-reduction} shows the hit multiplicity (mostly BIB-hits) as a function of VXD layer. 

The {\em loose} working point targets high efficiency reconstruction of $p>\qty{1}{\GeV}$ muons with the realistic beamspot size (Fig.~\ref{fig:tracking:perf-dl-realisticBS}). It reduces the number of hits in the innermost double-layer by a factor of two. This reduces the CT reconstruction time from $\approx\qty{1}{week/event}$ to $\approx\qty{2}{days/event}$. While significant, this is still not practical for sample production. Instead the {\em loose} working point is used with a special CT configuration as a first stage of a two-stage reconstruction process. The first stage reconstructs high-\pT~tracks to precisely determine the IP position, which can be used for

The {\em tight} working point is optimized for the scenario when the interaction point is precisely known (Fig.~\ref{fig:tracking:perf-dl-noBS}). It has a hit survival rate of $\approx{}2\%$ in the inner-most layer. This reduces the hit multiplicity enough for the CT algorithm to complete in $\approx\qty{2}{min/event}$. It is used as the second stage of the track reconstruction algorithm, with the IP determined from the first-stage.

The two-stage doublet-layer filtering provides a computationally viable method to track reconstruction in the presence of the BIB. However its efficiency is very limited for reconstructing non-prompt particles, including $b$-meson decays.

\begin{table}[htb]
    \centering

    \begin{tabular}{c|l||c|c|c|c||c|c|c|c}
    \multicolumn{2}{r||}{}       & \multicolumn{4}{c||}{\textbf{Barrel}} & \multicolumn{4}{c}{\textbf{Endcap}}\\
    \multicolumn{1}{c}{} & Layer IDs & 0,1 & 2,3 & 4,5 & 6,7 & 0,1 & 2,3 & 4,5 & 6,7\\
    \hline\hline
    \multirow{3}{.75in}{\centering \textbf{Loose DL} selections} 
    &Max. $\Delta\phi$ (mrad)    & 2.8 & 2.0 & 1.7 & 1.5 & 2.1 & 1.7 & 1.6 & 1.5 \\
    &Max. $\Delta\theta$ (mrad)  & 35 & 18 & 10 & 6.5 & 3.5 & 1.5 & 0.7 & 0.5  \\
    &Hit surival fraction        &\multicolumn{4}{c||}{55\%} &\multicolumn{4}{c}{18\%}\\
    \hline
    \multirow{3}{.75in}{\centering \textbf{Tight DL} selections} 
    &Max. $\Delta\phi$ (mrad)    & 3.0 & 2.0 & 1.6 & 1.5 & 2.2 & 1.8 & 1.7 & 1.6\\
    &Max. $\Delta\theta$ (mrad)  & 0.5 & 0.4 & 0.3 & 0.25 & 0.2 & 0.18 & 0.12 & 0.1 \\
    &Hit survival fraction       &\multicolumn{4}{c||}{2\%} & \multicolumn{4}{c}{2\%}\\
    \end{tabular}
    
    \caption{Angular selection on double-layers for hit suppression. A loose (left) and tight (right) selection is shown. The hit survival rate represents the fraction of hits (mostly from BIB) surviving the selections.}
    \label{tab:tracking:dl-rejection}
\end{table}

\subsection{Combinatorial Kalman Filter}
\label{sec:trk-ckf}
The Combinatorial Kalman Filter (CKF) algorithm seeded using hit triplets was implemented using the ACTS library v13.0.0. The implementation is a Marlin Processor that serves as a drop-in replacement for the existing tracking Processors. In addition to providing an alternate algorithm designed for large hit multiplicity, ACTS also provides a heavily optimized code for fast computation. The triplet seeding + CKF used in this section reconstructs tracks at the rate of \qty{4}{\min/event}. It provides the first practical and comprehensive tracking solution for the MCD.

The seeds for the CKF algorithm are formed from hit triplets in the four layers of the Vertex Detector. Only hits in the outer half of doublet layer are considered. Several heuristics are use to determine if each triplet is compatible with a track and can be used as a seed. The seeding algorithm is configured using the ACTS default values, with the exception of
\begin{itemize}
    \item radial distance between hits is between \qty{5}{\mm} and \qty{80}{\mm},
    \item minimum estimated \pT~is \qty{500}{\MeV},
    \item maximum forward angle of \qty{80}{\degree},
    \item extrapolated collision region is within \qty{1}{mm} of detector center,
    \item average radiation length per seed is 0.1,
    \item allowed amount of scattering is \qty{50}{\sigma},
    \item the middle hits in each seed are unique.
\end{itemize} 

This configuration has not been fully optimized. For example, the size of the collision region is smaller than the expected beam-spot size. Also one needs to explore the use of the Outer Tracker for seeding. Its lower BIB-hit multiplicity can allow the loosening of some requirements.

\begin{figure}
    \centering
    \includegraphics[width=0.49\textwidth]{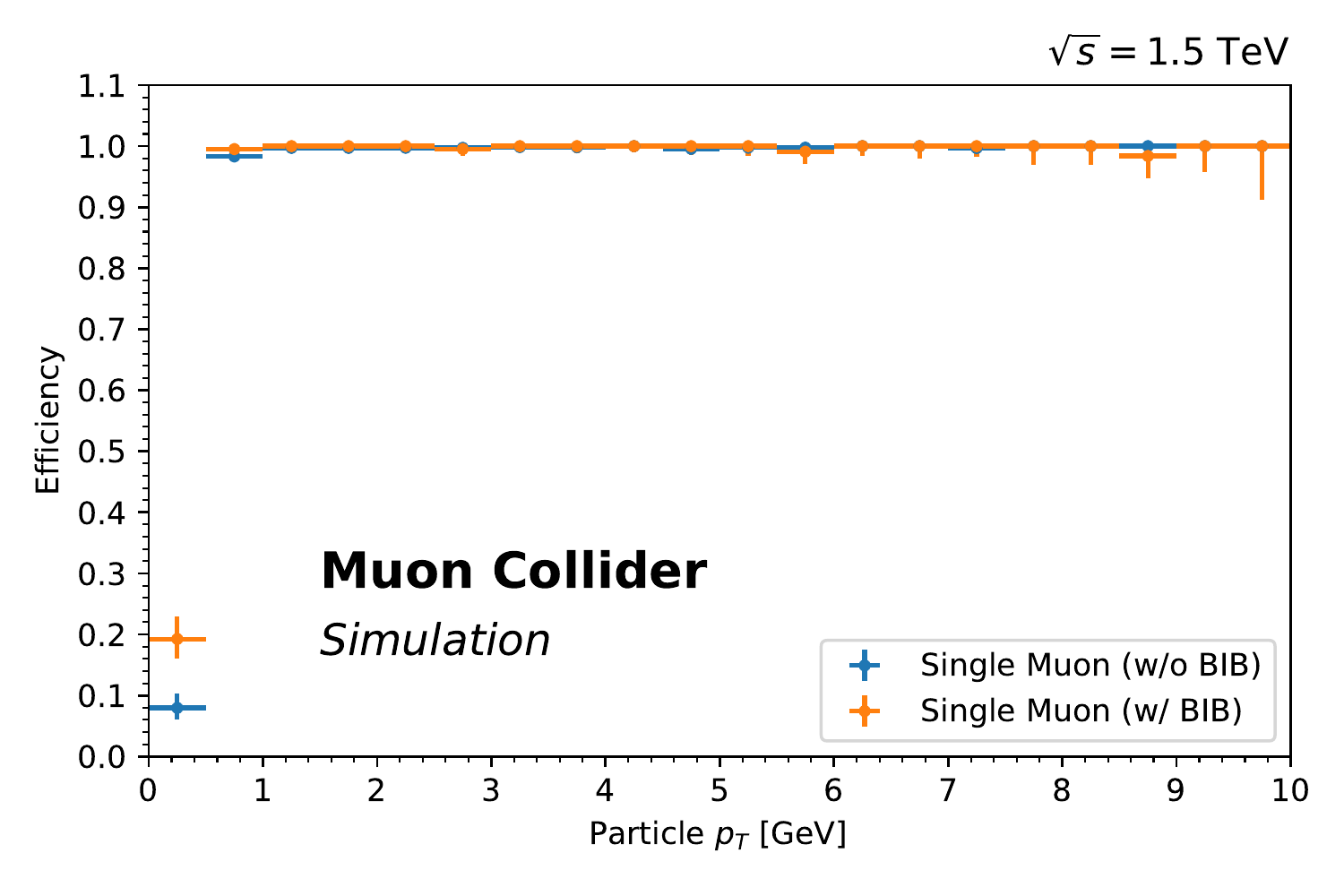}
    \includegraphics[width=0.49\textwidth]{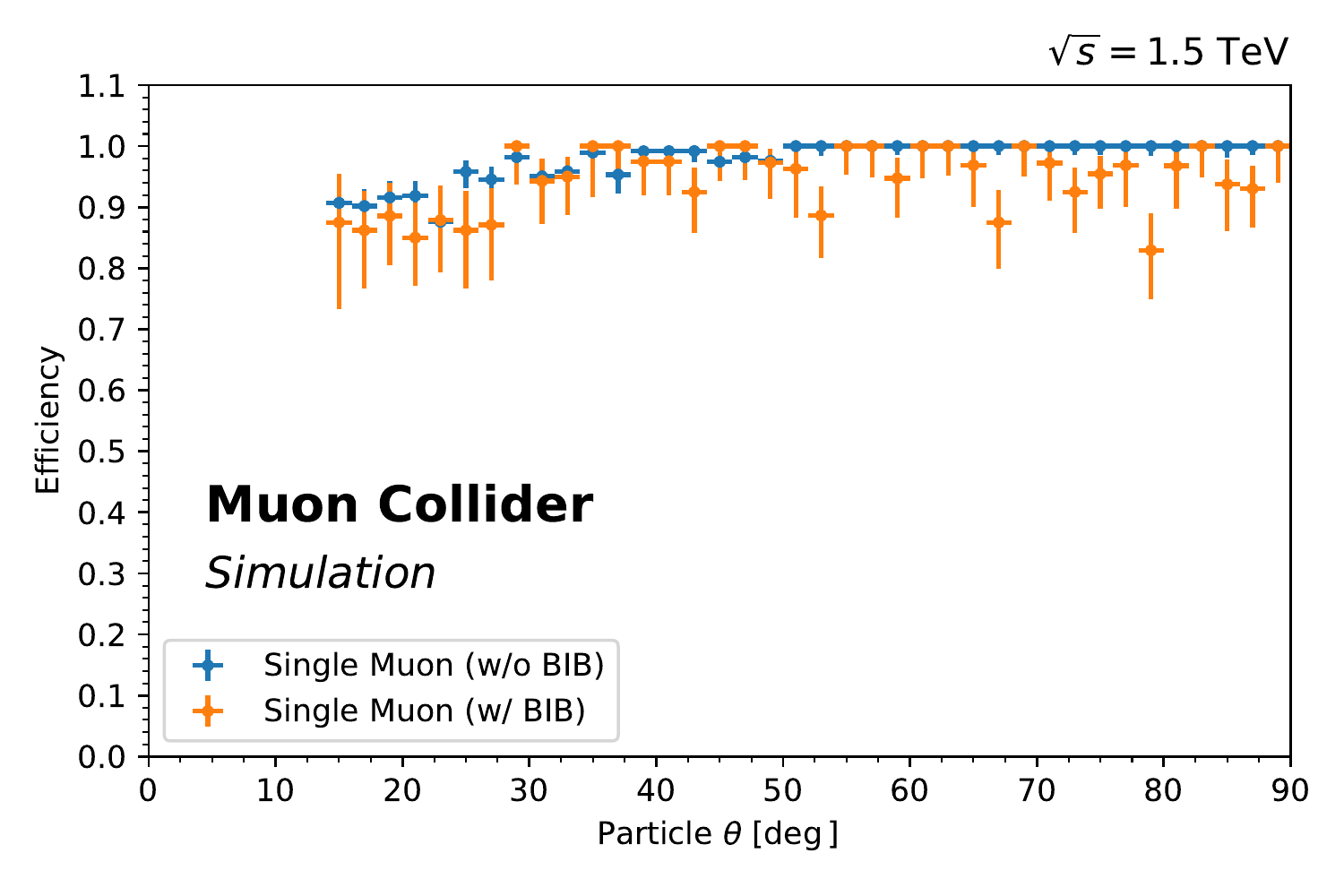}
    \caption{Seed finding efficiency for events containing a single muon with BIB overlay for several definition collision region ("rough $z_0$") as a function of truth \pT (left) and $\lambda$ (right). \note{update with final style} \note{remove alt seeding configurations}}
    \label{fig:trk-seeds}
\end{figure}

Around 150,000 seeds are found per event. The efficiency of the seeding algorithm is shown in Fig.~\ref{fig:trk-seeds}. Seeds are found for more than 90\% of the muons with \pT>\qty{2}{GeV}. Loosening the collision region definition increases the amount of fake seeds and reduces the seed finding efficiency due to the seed overlap removal. The latter can be addressed, at a cost in run time, by allowing multiple seeds to share the same middle hit.

The CKF is run inside-out, meaning that the track extrapolation starts from the inner-most seed hit and continues outward in the radial direction. The initial track parameters are estimated from the seed. The CKF algorithm has only two tunable parameters that are set as follows;
\begin{itemize}
    \item maximum one (closest) hit added at each layer,
    \item hit search window at each layer of \qty{10}{\chi^2}.
\end{itemize}

As with the seeding algorithm, these values are not optimized. The consideration of only a single hit at each layer means that the CKF algorithm will not branch to consider multiple track candidates for a single seed. This seems very tight, however a good tracking reconstruction efficiency is still seen. This underlines the difference between BIB and pile-up; the BIB-hits are a "random" background that is not compatible with the trajectory of a track.

\begin{figure}
    \centering
    \includegraphics[width=0.49\textwidth]{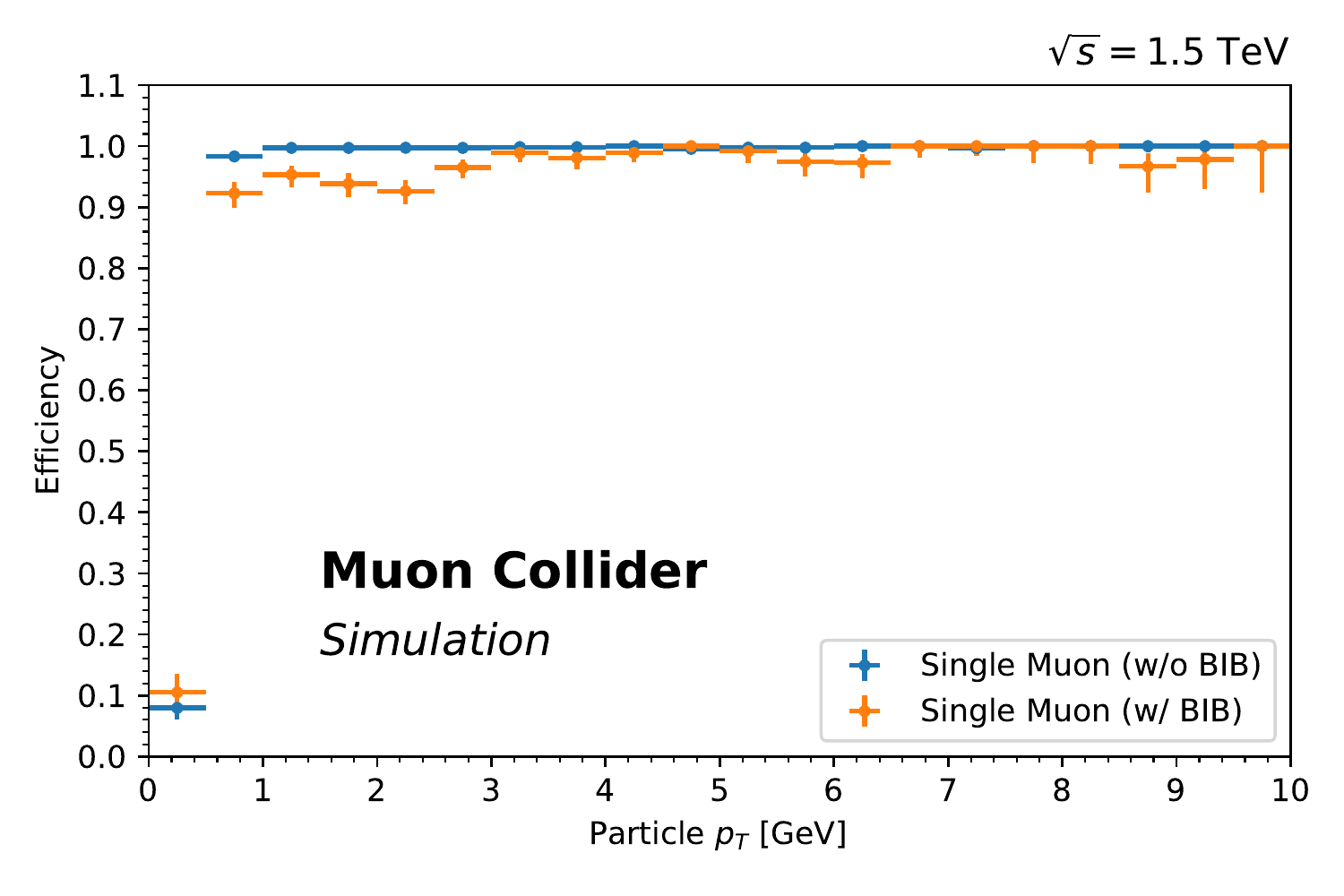}
    \includegraphics[width=0.49\textwidth]{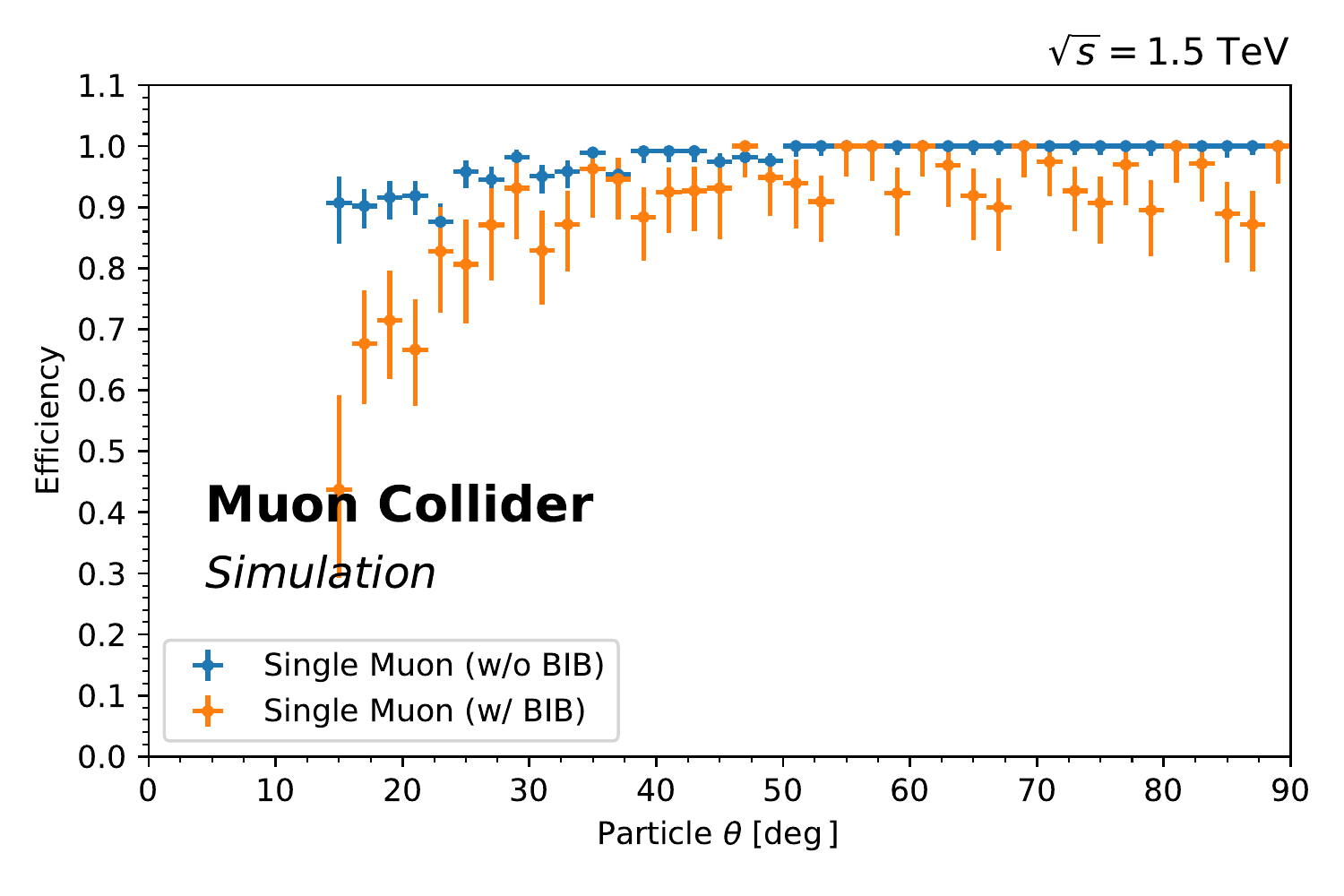}
    \caption{Track reconstruction efficiency for events containing a single muon with (blue) and without (orange) BIB overlay as a function of truth \pT (left) and $\theta$ (right).}
    \label{fig:trk-ckf}
\end{figure}

\begin{figure}
    \centering
    \includegraphics[width=0.49\textwidth]{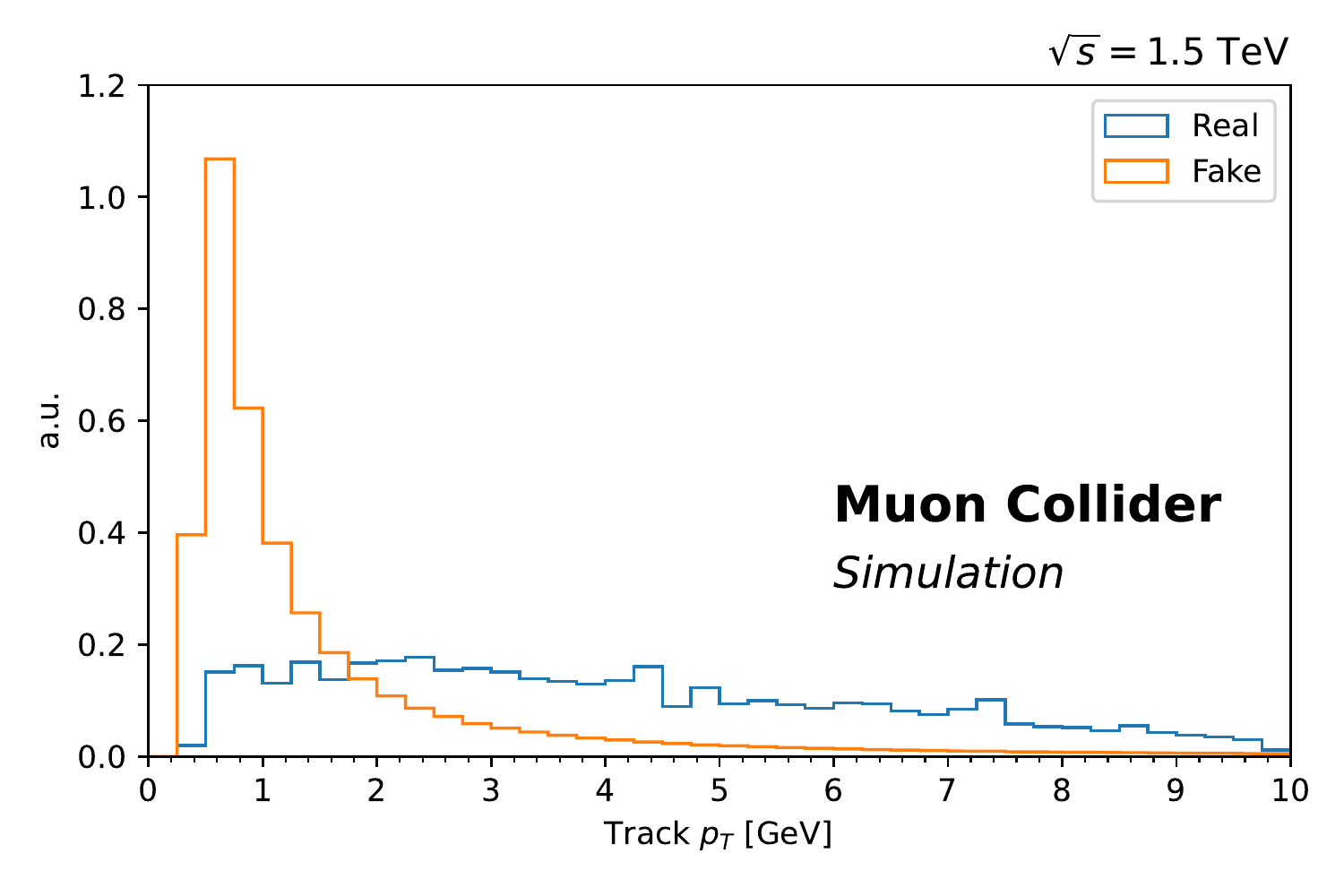}
    \includegraphics[width=0.49\textwidth]{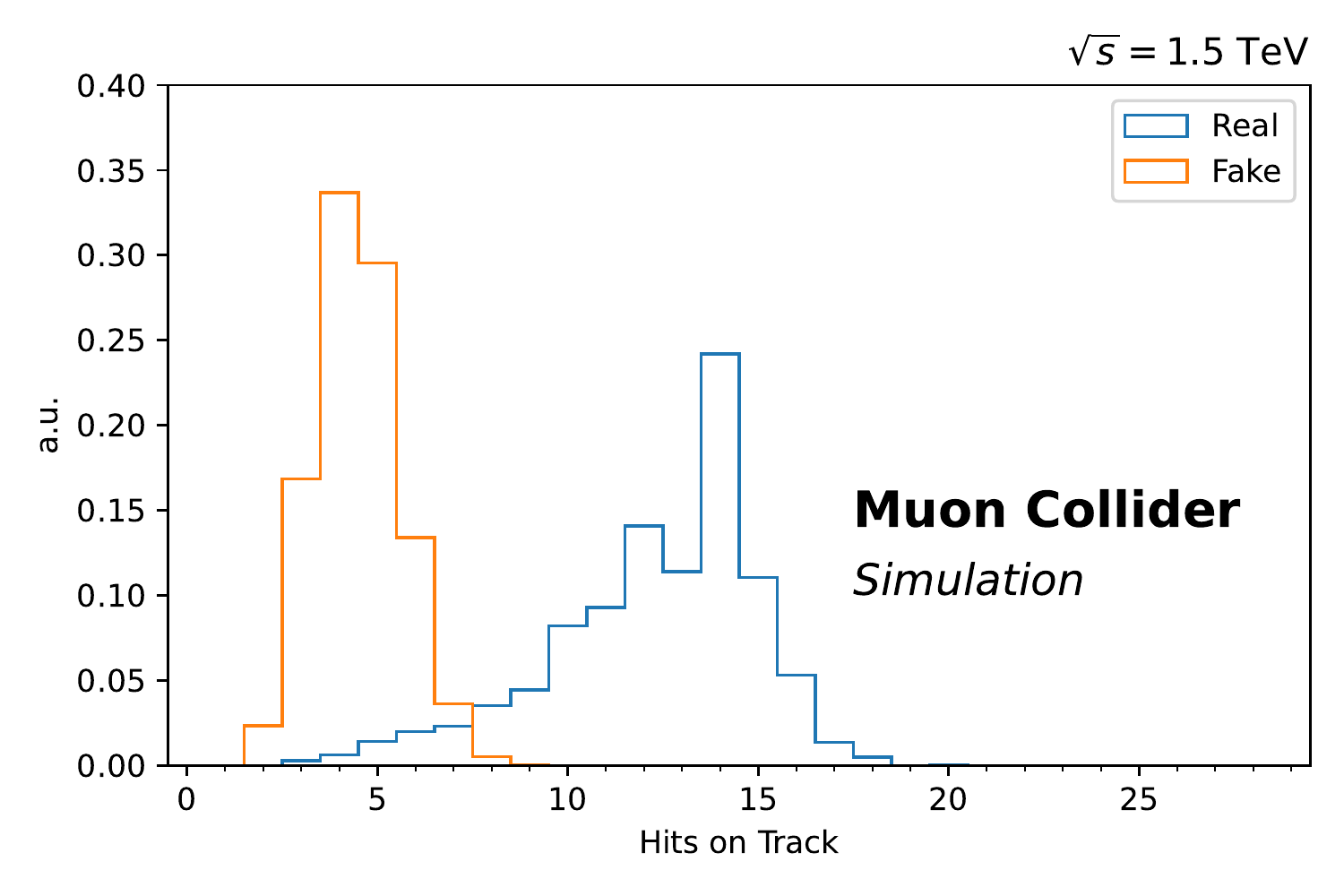}
    \caption{Track \pT (left) and $N_{\text{hit}}$ (right) distributions between real (blue) and fake (orange) tracks in single muon events. \note{Need to update with latest plots}}
    \label{fig:trk-ckf-fake}
\end{figure}

Fig.~\ref{fig:trk-ckf} shows the track reconstruction efficiency as a function of particle \pT~and $\theta$ for single muon events with $p=\qty{10}{\GeV}$. Muons with $\pt>\qty{2}{\GeV}$ are reconstructed with 90\% efficiency or greater even in the presence of BIB. There is a considerable amount of fake tracks ($\approx100,000$ per event). Fig.~\ref{fig:trk-ckf-fake} compares the \pT~distribution and the associated number of hits between real and fake tracks. Fake tracks are mostly at low \pT and are associated with a significantly smaller number of hits. The latter further underlines the randomness of the BIB-hits and provides a handle for reducing the fake rate.

\begin{figure}
    \centering
    \includegraphics[width=0.49\textwidth]{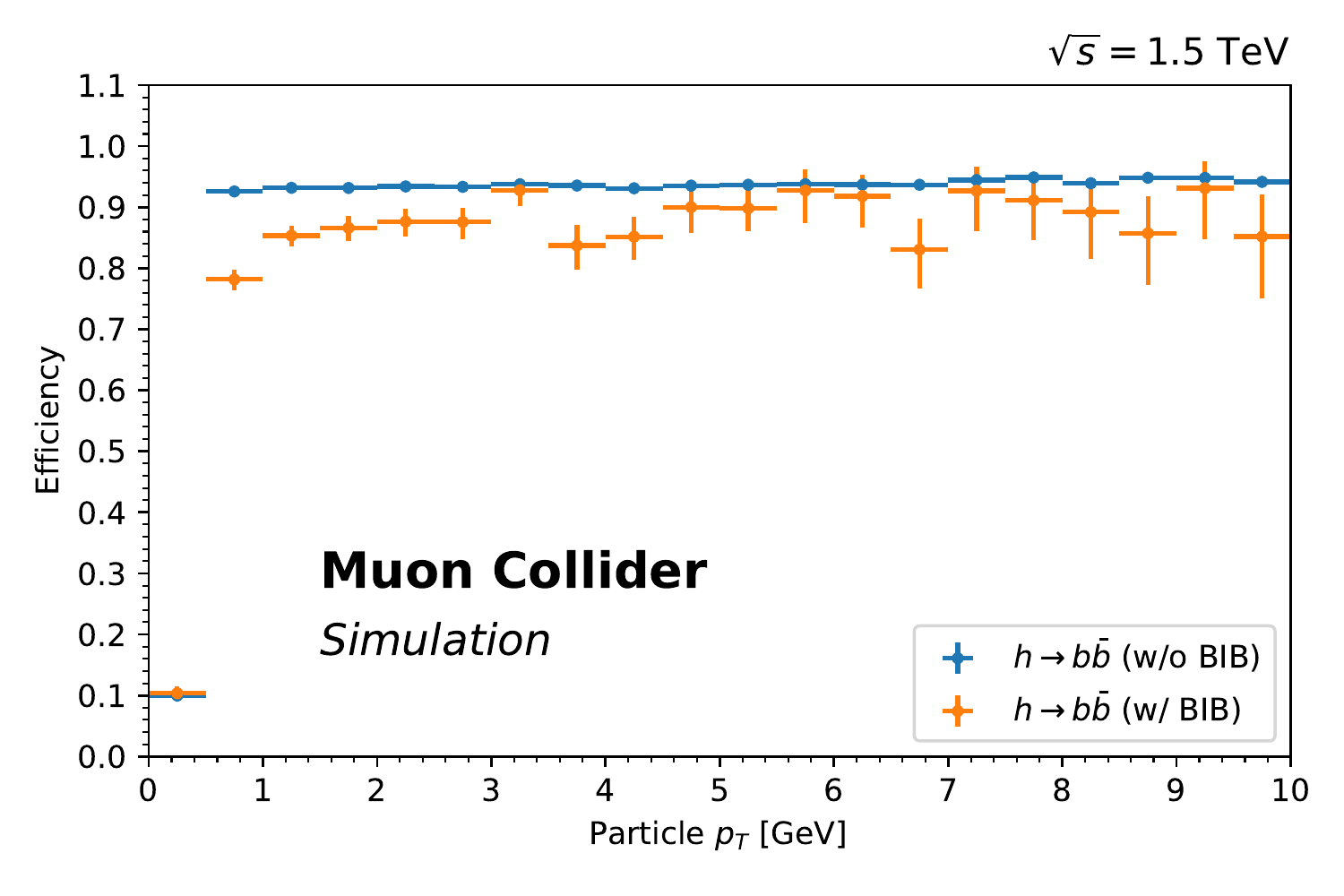}
    \includegraphics[width=0.49\textwidth]{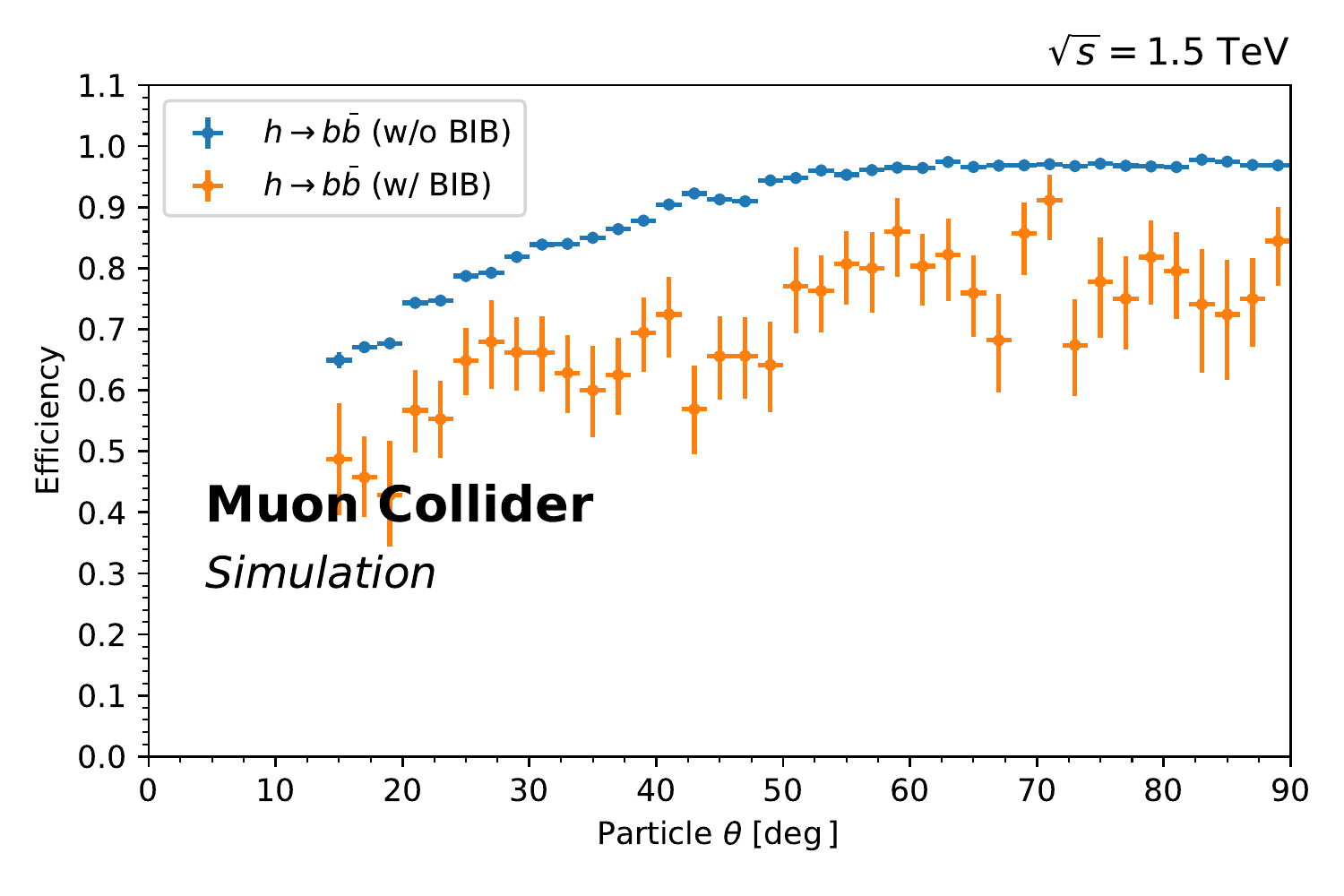} \\
    \includegraphics[width=0.49\textwidth]{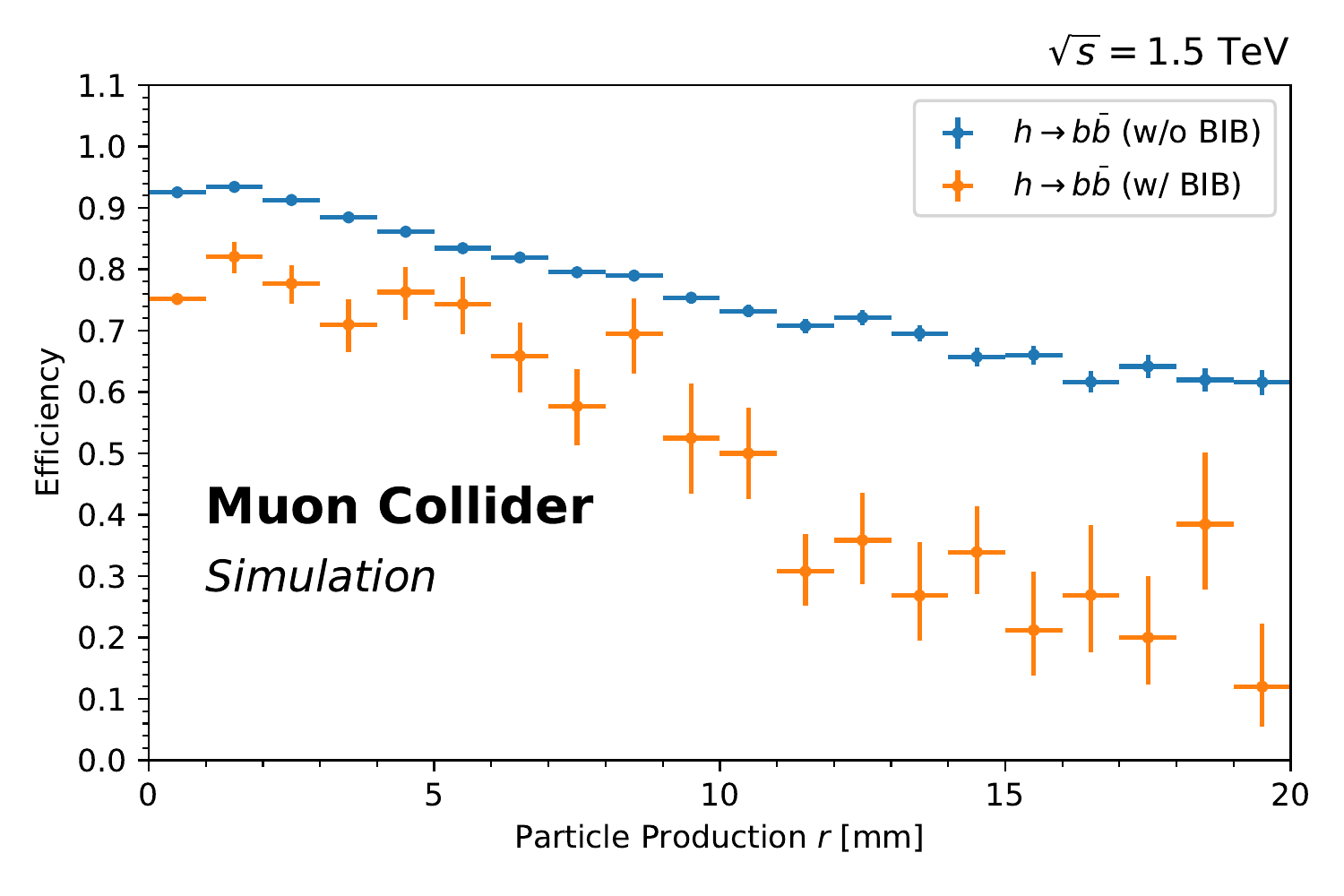}
    \caption{Track reconstruction efficiency for events containing a Higgs boson decaying to two $b$-quarks with (blue) and without (orange) BIB overlay as a function of truth \pt (left), $\theta$ (right) and production radius (bottom).}
    \label{fig:trk-ckf-higgs}
\end{figure}

Fig.~\ref{fig:trk-ckf-higgs} shows the track reconstruction efficiency in a sample containing Higgs bosons decaying to two $b$-quarks. This is a practical demonstration CKF performance on a realistic event. Maintaining a good track reconstruction efficiency as a function of production radius is important for identifying secondary vertices of $b$-mesons. The tight configuration maintains above 70\%  efficiency up to \qty{5}{\mm}. While not perfect, it is a significant improvement over the inefficiency of doublet-layer filtering required for CT.

%% file: jets.tex
Hadronic jets are fundamental objects in a Muon Collider experiment. As an example, the measurements of the tri-linear and quadri-linear Higgs couplings rely on the reconstruction and identification of $b$ jets from the Higgs boson decay, as the dominant decay signature of the Higgs boson is into $b$-quarks pair, with a branching ratio of $58 \%$. 
Jet reconstruction is one of the most difficult reconstruction tasks at a Muon Collider, since almost all sub-systems are involved, and the impact of the BIB is significant in all of them, with different features in different sub-detectors.
The jet reconstruction algorithm employed is described in this section, and its performance is discussed. The algorithm has been designed to reconstruct jets in the presence of the BIB, but it is far from being fully optimized, and further studies are needed in the future.
However we are going to demonstrate that even at this stage the jet reconstruction can achieve a decent performance. Given the high level of details of the full simulation (including the BIB impact) this give us the confidence that measurements with jets are possible at a Muon Collider, and further dedicated advancements on MDI, detectors and reconstruction algorithms would go in the direction of improving the physics reach.
The algorithm works as follows:
\begin{enumerate}
    \item tracks are reconstructed using the Combinatorial Kalman Filter algorithm and are filtered depending on the number of hits in the sub-systems;
    \item calorimeter hits are selected by requiring a hit time window and an energy threshold;
    \item tracks and calorimeter hits are used by the PandoraPFA algorithm to obtain reconstructed particles;
    \item the reconstructed particles are clustered into jets with the $k_t$ algorithm;
    \item requirements are applied to remove fake jets;
    \item a jet energy correction is applied.
\end{enumerate}
The jet performance has been evaluated on simulated samples of $b\bar{b}$, $c\bar{c}$ and $q\bar{q}$ dijets, where $q$ stands for a light quark ($u$,$d$ or $s$). These samples have been generated with an almost uniform dijet $p_T$ distribution from 120 to \qty{200}{\GeV}. Samples of $\mu^+ \mu^- \to H ( \to b\bar{b}) + X $ and $\mu^+ \mu^- \to Z ( \to b\bar{b}) + X $ at $\sqrt{s}=3$ TeV are also used to study the dijet invariant mass resolution. The BIB at $\sqrt{s}=1.5$ TeV has been overlayed with all the samples.

\subsection{Track filter}
\begin{figure}[htb]
\centering
\includegraphics[width=0.48\textwidth]{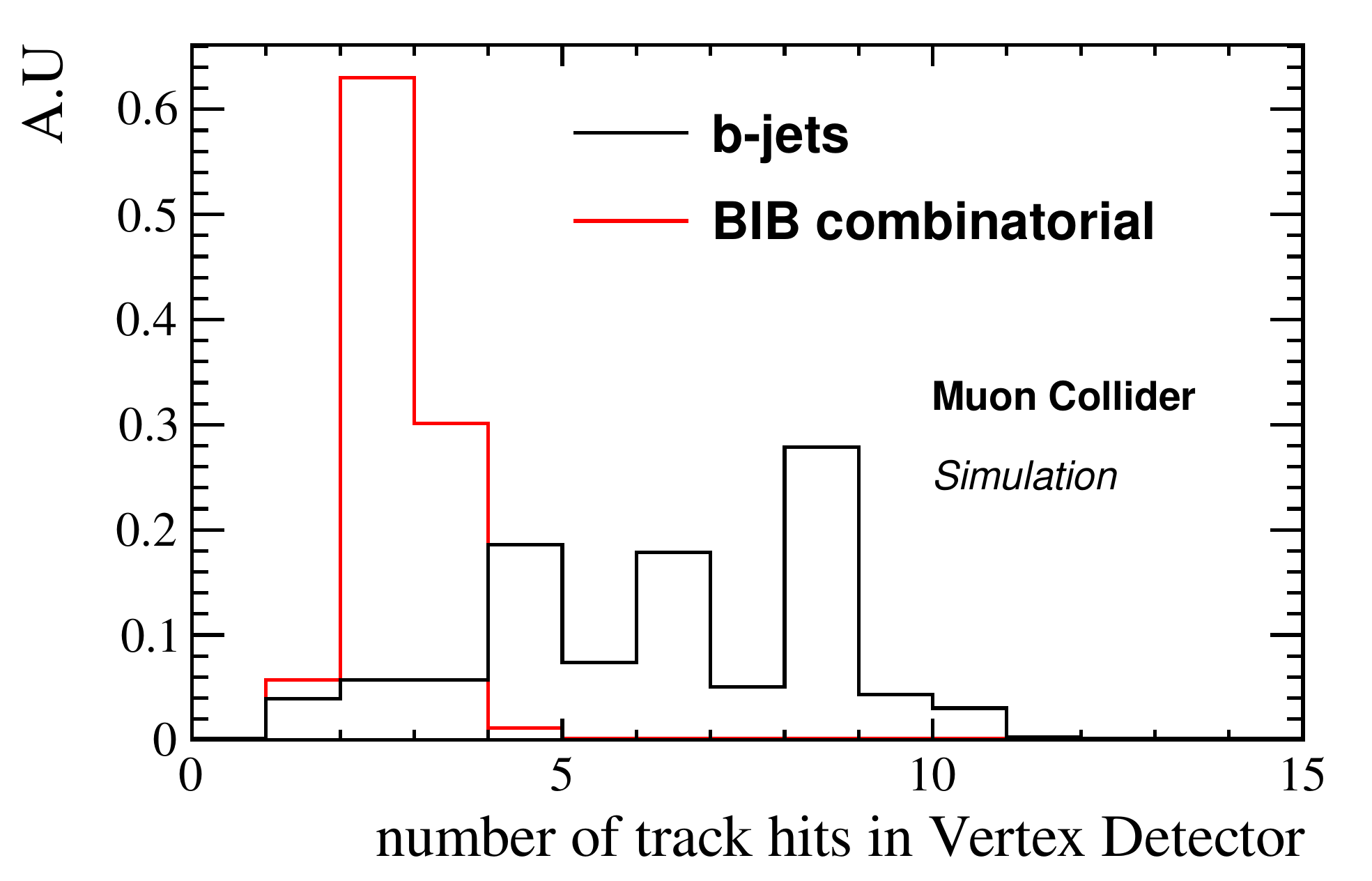}
\includegraphics[width=0.48\textwidth]{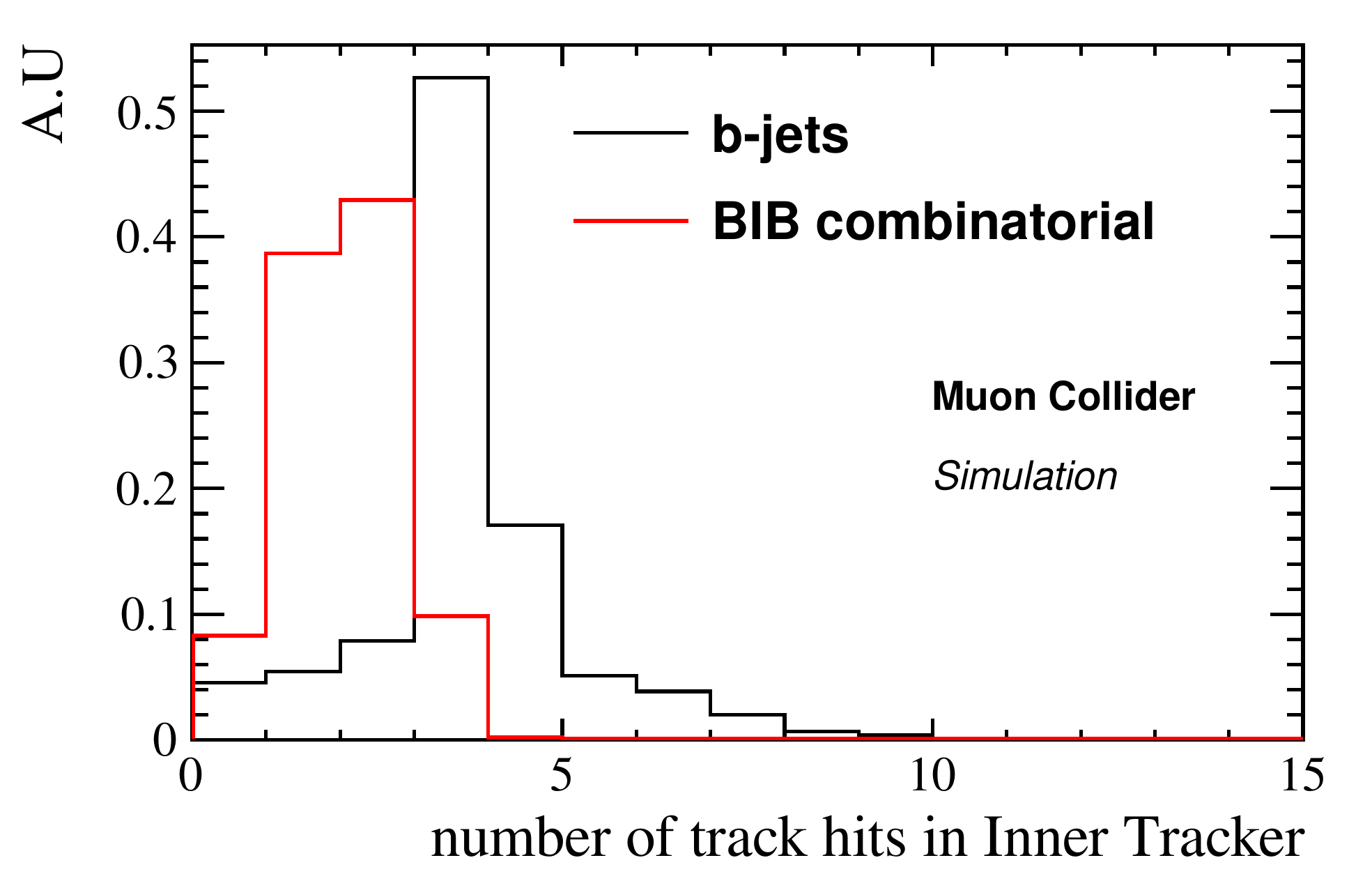}
\caption{Distributions of the number of track hits in Vertex Detector and Inner Tracker, for $b$-jet tracks and BIB combinatorial tracks. Distributions are normalized to the same area. The distribution in the Vertex Detector peaks at even numbers due to the double-layer structure of the detector.}
\label{fig:track_filter}
\end{figure}

Tracks are reconstructed using the Combinatorial Kalman Filter algorithm described in Section~\ref{sec:trk-ckf}. Since more than 100k fake tracks per event are found in this way, they are filtered depending on the number of hits in the Vertex Detector and Inner Tracker. In Fig.~\ref{fig:track_filter} hit distributions for tracks from $b$-jets and BIB are compared. 
A number of Vertex Detector hits greater than 3 and a number of Inner Tracker hits greater than 2 are required for each track, reducing the average number of BIB combinatorial tracks per event from more than 100k to less than 100.

\subsection{Calorimeter hit selection}

\begin{figure}[htb]
\centering
\includegraphics[width=0.6\textwidth]{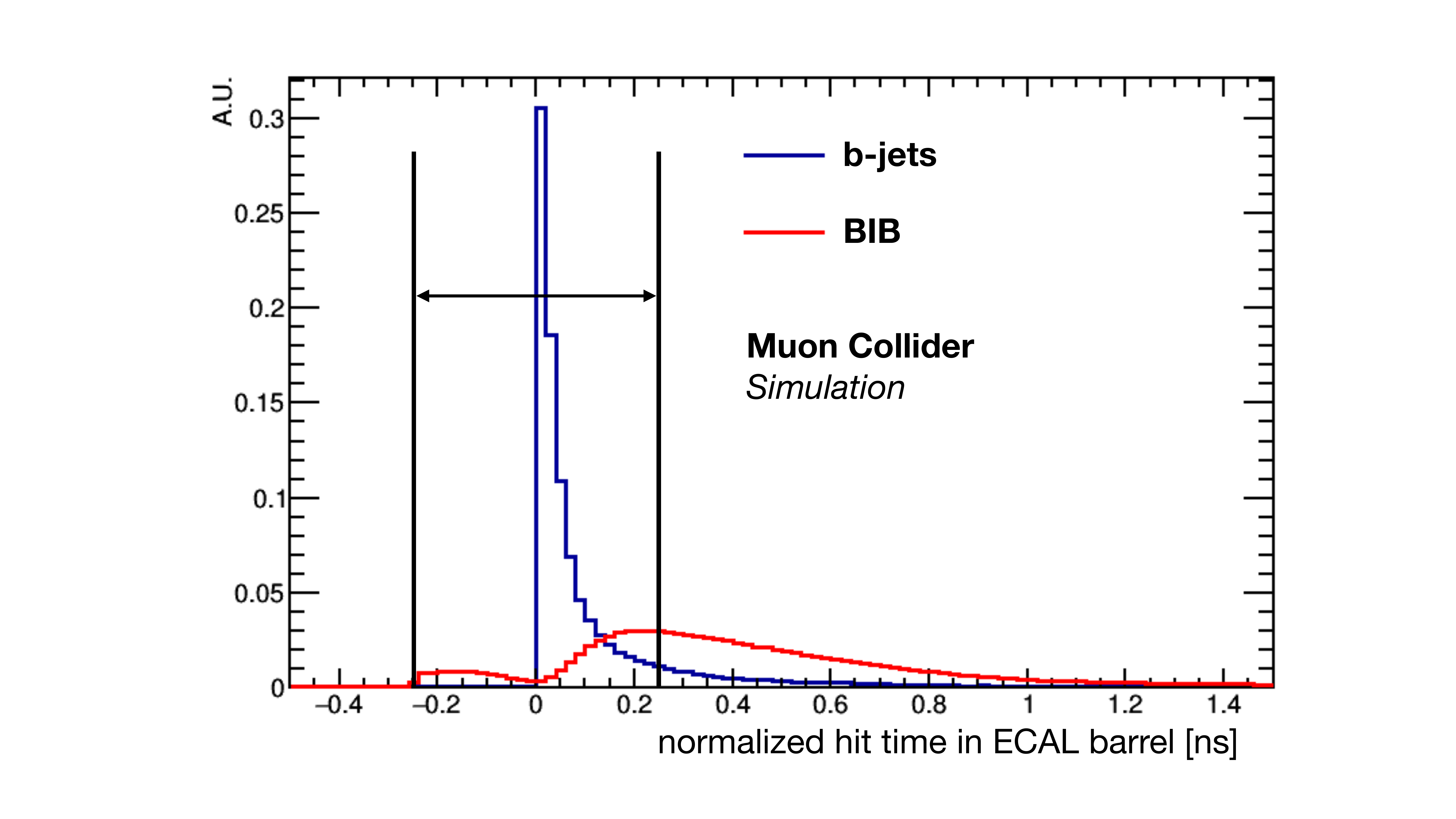}
\caption{Normalized hit time in ECAL barrel, for $b$-jets and BIB. Both distributions are normalized to the same area. Notice that this time is not smeared for the detector time resolution. The time window of \qty{\pm 250}{\pico\second} applied in the jet reconstruction is shown.}
\label{fig:calo_time}
\end{figure}

Calorimeter hits are filtered depending on the normalized hit time, defined as $t_N = t - t_0 - cD$, where $t$ is the absolute hit time, $t_0$ is the collision time, $c$ is the speed of light, and $D$ is the hit distance from the origin of the reference system.
A time window of $\pm 250$ ps is applied to remove most of the BIB hits but preserving the signal, as can be seen in Fig.~\ref{fig:calo_time}. 
As rule of thumb we can assume that a time window of width $\Delta$ is applicable if the Full Width at Half Maximum (FWHM) of the cell time resolution is below $\Delta/3$. Therefore, in this particular case, we assume a FWHM of at least 167 ps, that should be achievable by state-of-the-art calorimeter technologies.
    
Several calorimeter hit energy thresholds have been tested, in order to reduce hits produced by BIB. It has been found that a threshold of 2 MeV applied to both ECAL and HCAL is a compromise between jet performance and computing time.
In fact this requirement reduces the average number of ECAL Barrel hits from 1.5M to less than 10k.
The computing time of the jet algorithm exponentially grows with the number of calorimeter hits, therefore with the current resources we are not able to reduce the thresholds far below 2 MeV. However the impact of reducing the thresholds at the close value of 1 MeV is described in Section \ref{sec:future_jets}.
 
\subsection{Particle flow, jet clustering and fake jet removal} 
    
Calorimeter hits and tracks are given as input to the PandoraPFA algorithm, that produces as output reconstructed particles known as particle flow objects. The PandoraPFA algorithm is described in detail in \cite{THOMSON200925}.
Then particle flow objects are clustered into jets by the $k_T$ algorithm. A cone radius of $R=0.5$ is used.

\begin{figure}[htb]
\centering
\includegraphics[width=0.48\textwidth]{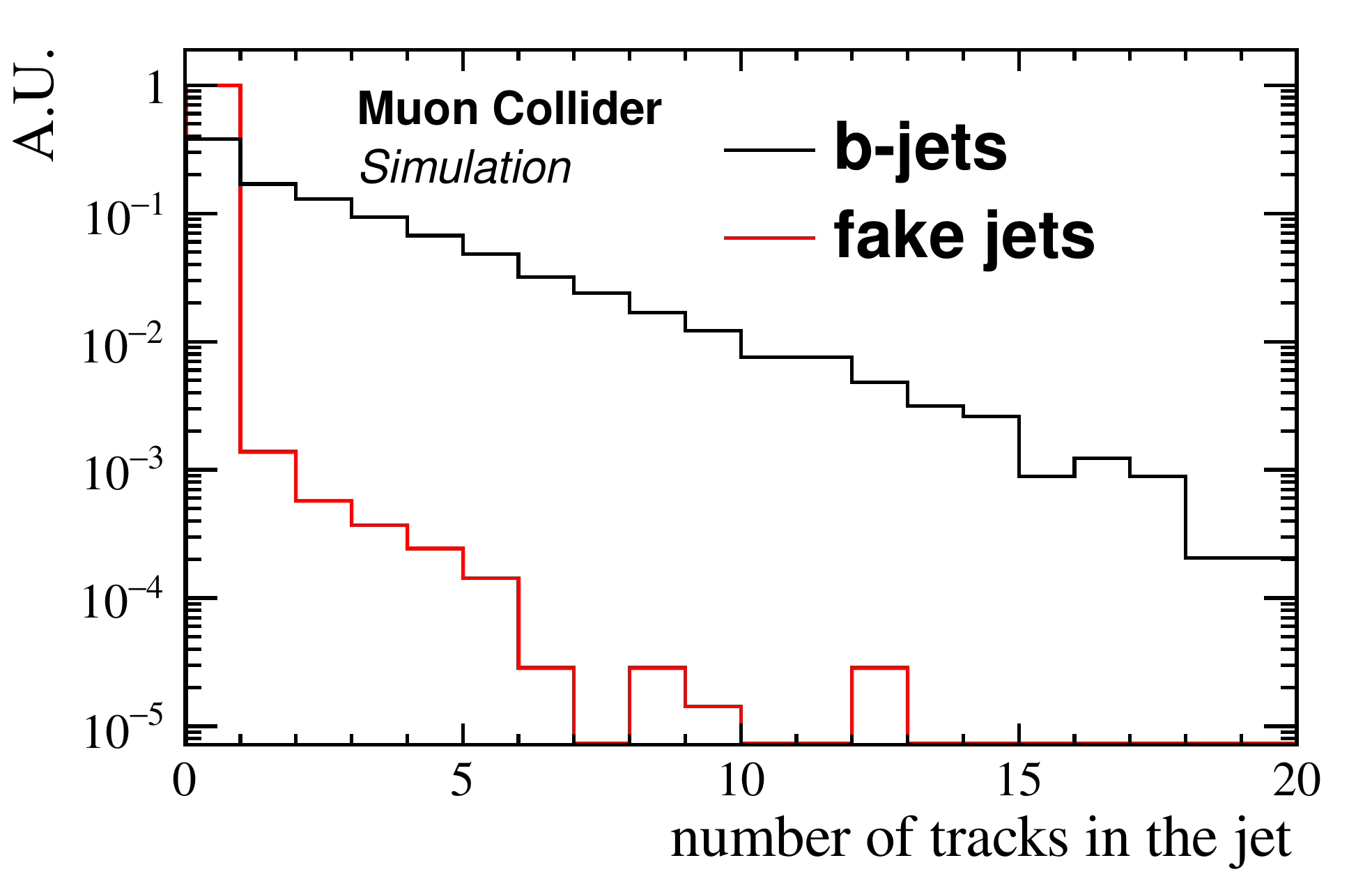}
\includegraphics[width=0.48\textwidth]{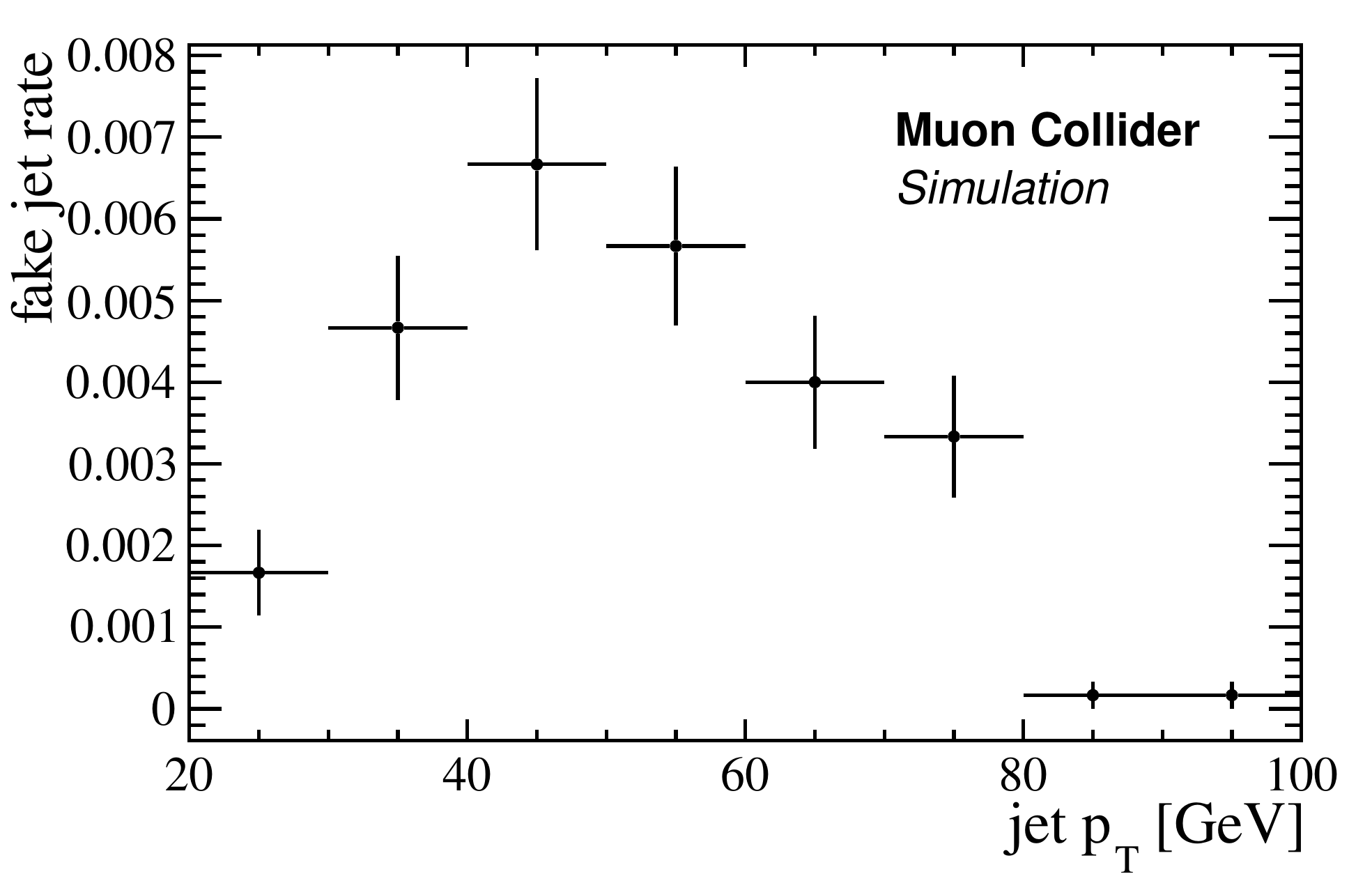}
\caption{Number of tracks associated to real $b$-jets and to fake jets from BIB (left). Jet fake rate as a function of the jet \pt (right), obtained after requiring at least one track associated to the jet.}
\label{fig:jet_ntracks}
\end{figure}

At this point an average number of 13 fake jets per event is reconstructed, which need to be removed by applying additional quality criteria. The number of tracks in the jet has been found to be the most discriminating feature between $b$-jets and fake jets, as shown in Fig.\ref{fig:jet_ntracks} (left).

It can be seen that most of the fake jets from BIB have no tracks associated to them. Therefore, requiring at least one track allows to greatly reduce the rate of fake jets.
In Fig.~\ref{fig:jet_ntracks} (right) the fake jet rate as a function of jet \pt is presented. It has been obtained using a sample of $b\bar{b}$ dijets and defined as the average number of reconstructed fake jets per event, which is well below 1\%.

\subsection{Jet momentum correction}

The jet 4-momentum is defined by the sum of 4-momenta of particles that belong to the jet. The jet axis is identified by the jet momentum direction. We define the fiducial region of the jet reconstruction by selecting jets with pseudorapidity $|\eta|<2.5$, where $\eta = - \mathrm{log}[\mathrm{tan}(\theta/2)]$, with $\theta$ angle between the jet axis and beam axis.

In order to recover the energy lost by reconstruction inefficiencies, as well as to take into account BIB contamination, a correction to the jet 4-momentum is applied. This correction has been determined by comparing the reconstructed jet $p_T$ with the corresponding truth-level jet $p_T$. Truth-level jets are defined as jet clustered by applying the $k_t$ algorithm to Monte Carlo particles. Reconstructed and truth-level jets are matched if their $\Delta R$ distance is below 0.5, \emph{i.e.}:
\begin{displaymath}
\Delta R = \sqrt{(\Delta \eta)^2 + (\Delta \phi)^2} < 0.5,
\end{displaymath}
where $\Delta \eta$ and $\Delta \phi$ are respectively the pseudo-rapidity the azimuthal angle differences between the reconstructed-level jet axis and truth-level jet axis. 
If more than one reconstructed jet is matched to the same truth-level jet than the one with lower $\Delta R$ is chosen. 

\begin{figure}[htb]
    \centering
    \includegraphics[width=0.48\textwidth]{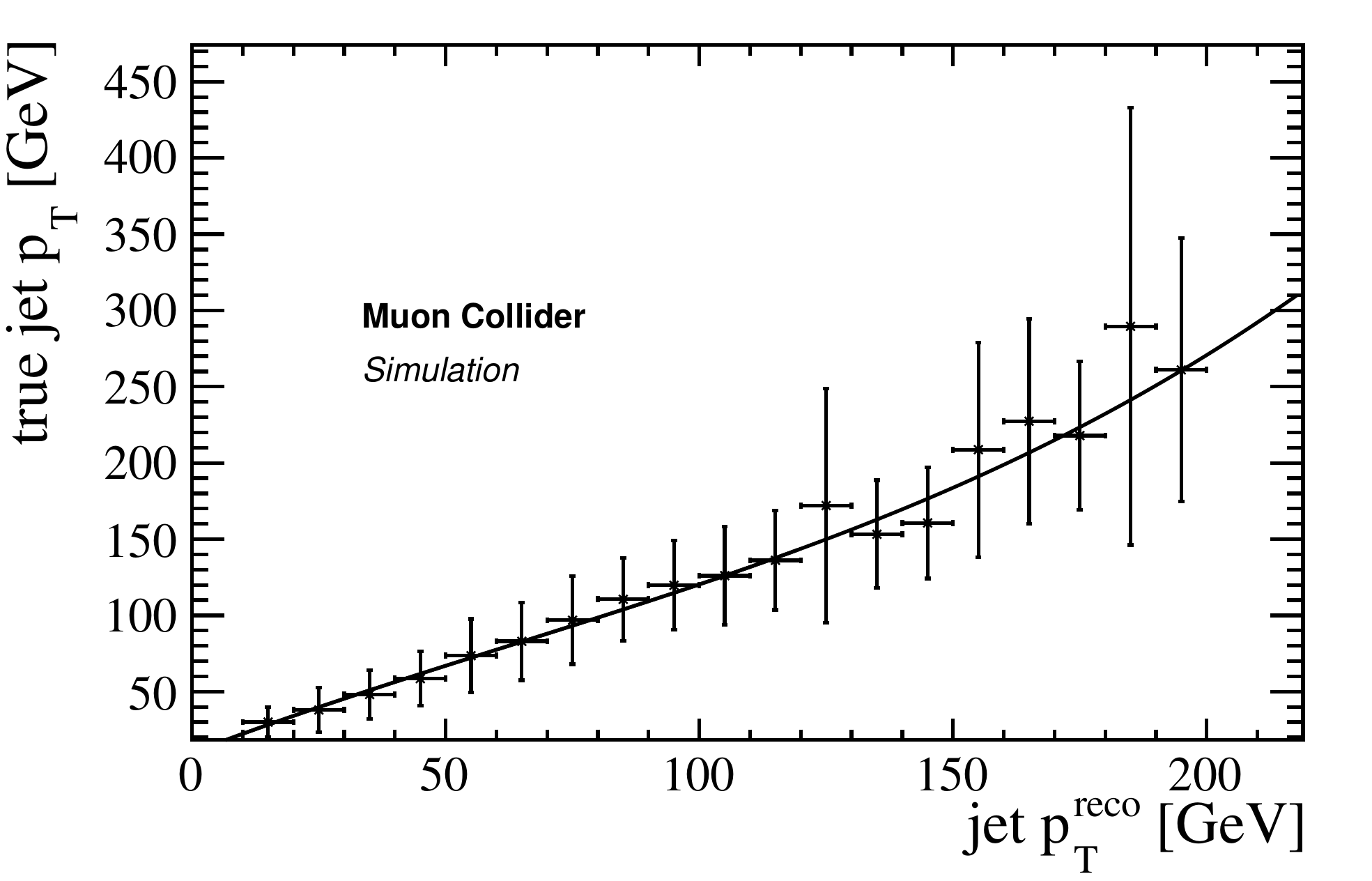}
    \includegraphics[width=0.48\textwidth]{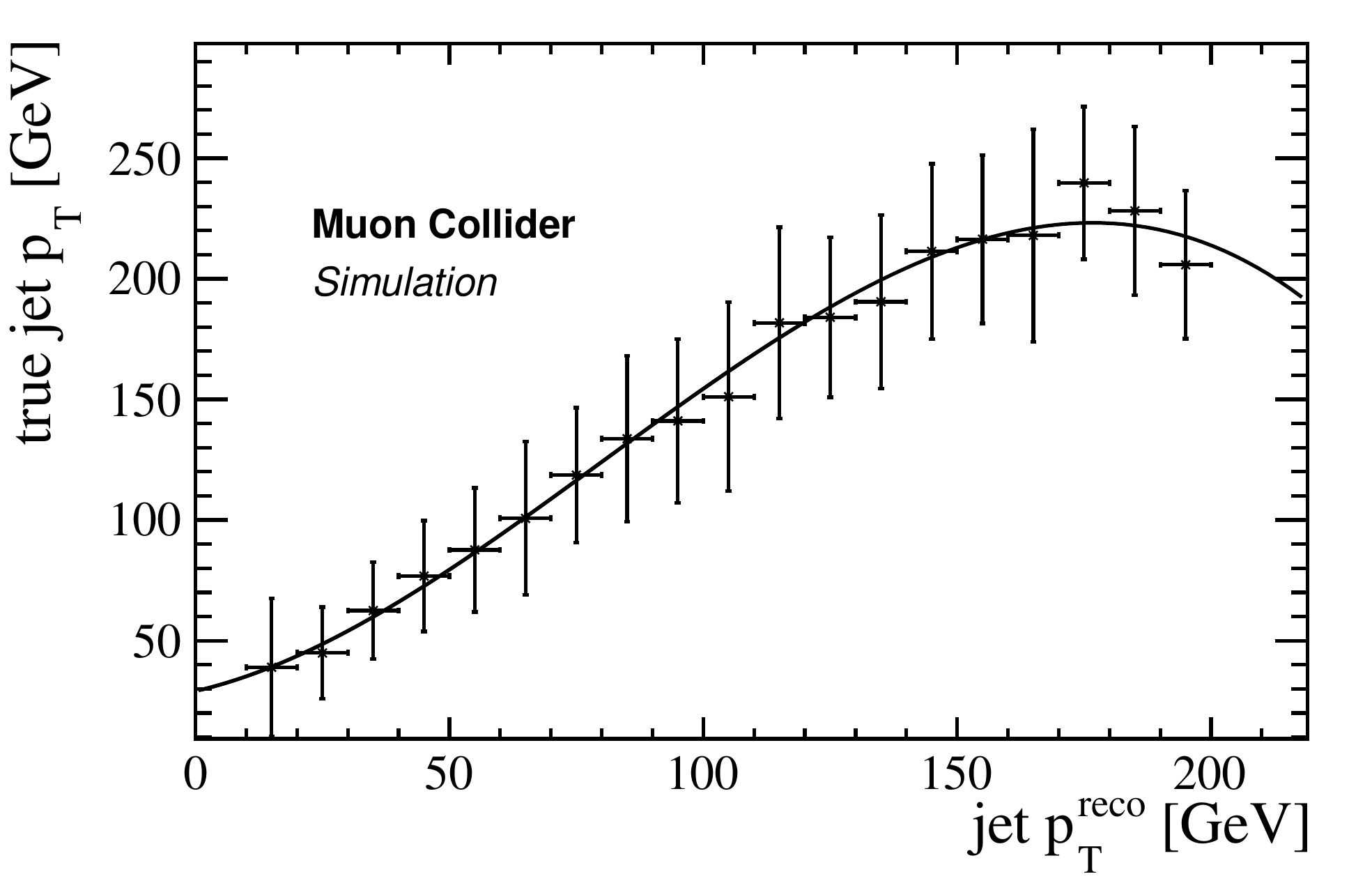}
    \caption{Transfer functions used for jet momentum corrections. The average truth-level jet $p_T$ as a function of the reconstructed jet $p_T$ is shown, for $0<|\eta|<0.5$ (left) and $1.5<|\eta|<2.0$ (right). Error bars represent the standard deviation of the truth-level jet \pT in each interval.}
    \label{fig:jec}
\end{figure}

The correction is evaluated in five equal-width intervals of reconstructed $|\eta|$ between 0 and 2.5. Each pseudo-rapidity interval is further divided in 19 equal-width intervals of reconstructed $p_T$ between 10 and 200 GeV. For each interval the average and standard deviation of the truth-level jet $p_T$ distribution is calculated. Transfer functions as then obtained in each $\eta$ interval by fitting the average truth-level jet $p_T$ as a function of reconstructed jet $p_T$. Examples of transfer functions are shown in Fig.~\ref{fig:jec}. These functions are then used obtain the scale factor that is applied to each component of the reconstructed 4-momentum.

\subsection{Jet reconstruction performance}

\begin{figure}[htb]
\centering
\includegraphics[width=0.48\textwidth]{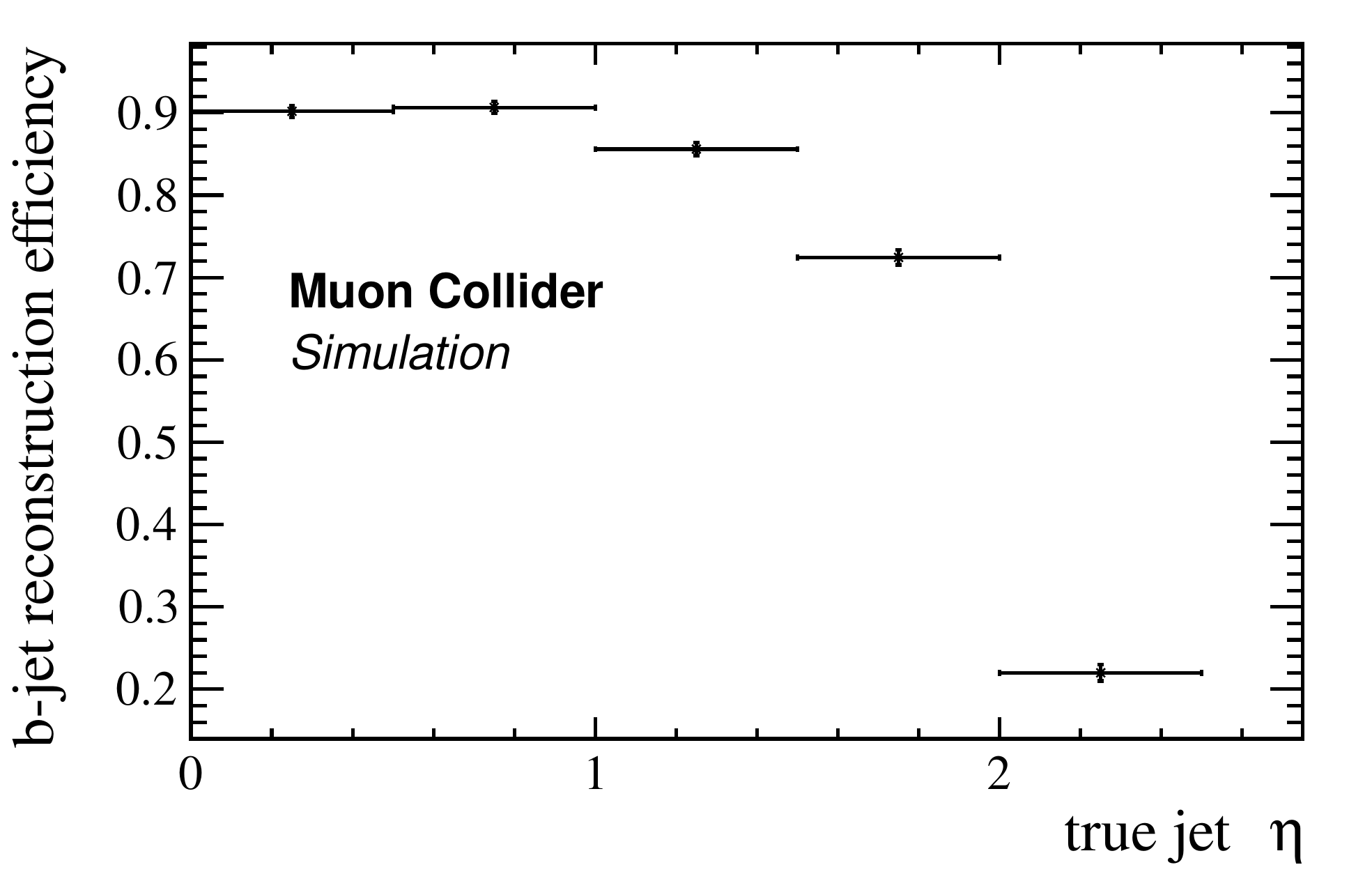}
\includegraphics[width=0.48\textwidth]{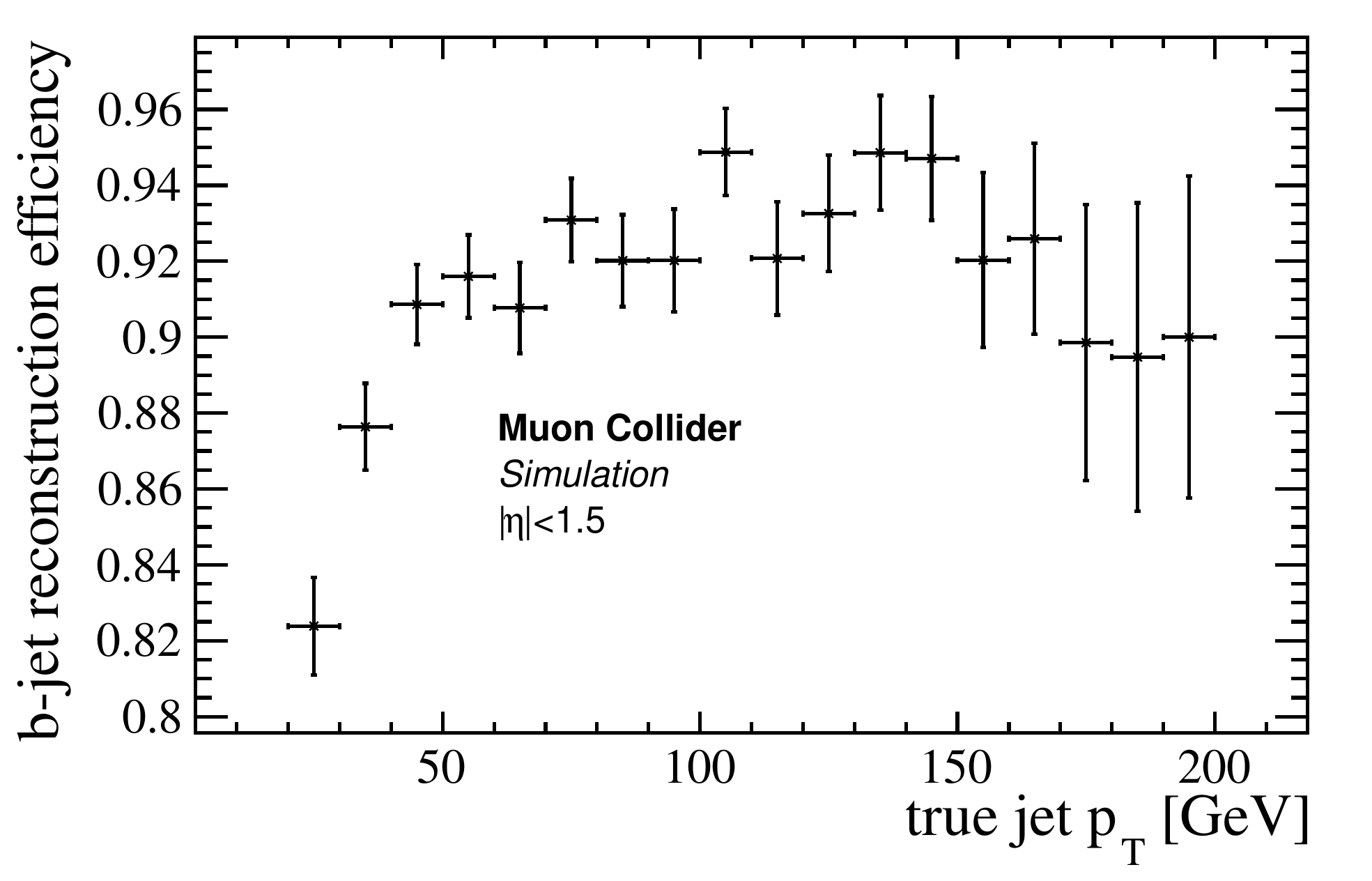}
\caption{Efficiency of $b$-jet reconstruction as a function of truth-level jet $\eta$ (left) and as a function of the truth-level jet $p_T$ (right, for $|\eta|<1.5$).}
\label{fig:bjet_eff}
\end{figure}

The $b$-jet reconstruction performance is first evaluated on the simulated sample of $b\bar{b}$ dijets. In Fig.~\ref{fig:bjet_eff} (left) the $b$-jet efficiency as a function of the true $b$-jet $\eta$ is shown. It is worth noting that this is only the jet reconstruction efficiency, while the flavor tagging performance is studied in Section~\ref{sec:jets:heavyflavour}. The efficiency is around 90$\%$ in the central region, but there is a drop for $|\eta|>2$, where the efficiency is 20$\%$. It has been verified that this drop is mainly due to the fake jet removal requirement, since many jets without reconstructed tracks are found in the forward region. In Fig.~\ref{fig:bjet_eff} (right) the $b$-jet efficiency as a function of the true $b$-jet $p_T$ in the region $|\eta|<1.5$ is presented, and it goes from 82$\%$ at low $p_T$ to 95$\%$ at higher $p_T$.

\begin{figure}[htb]
\centering
\includegraphics[width=0.48\textwidth]{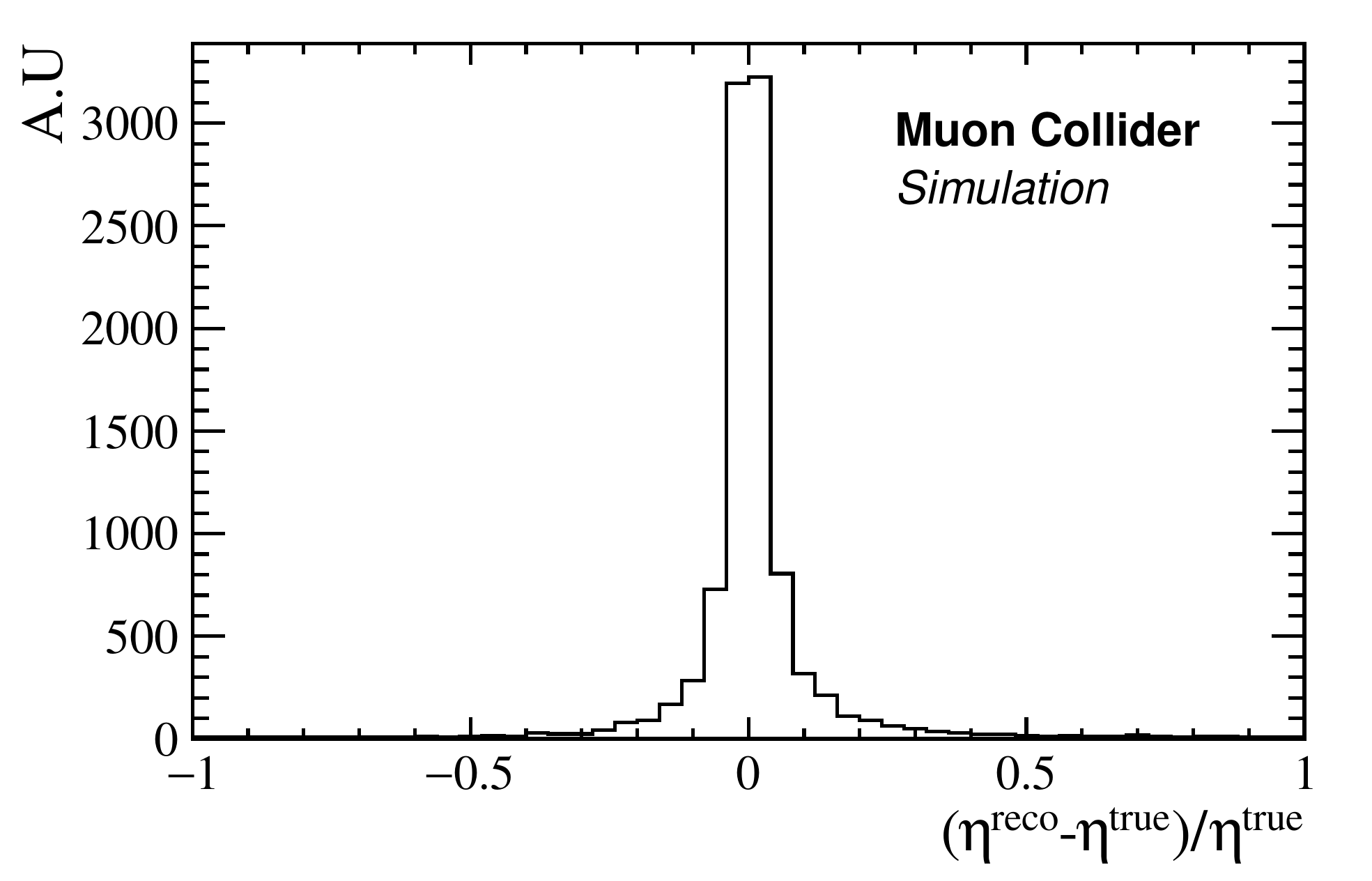}
\includegraphics[width=0.48\textwidth]{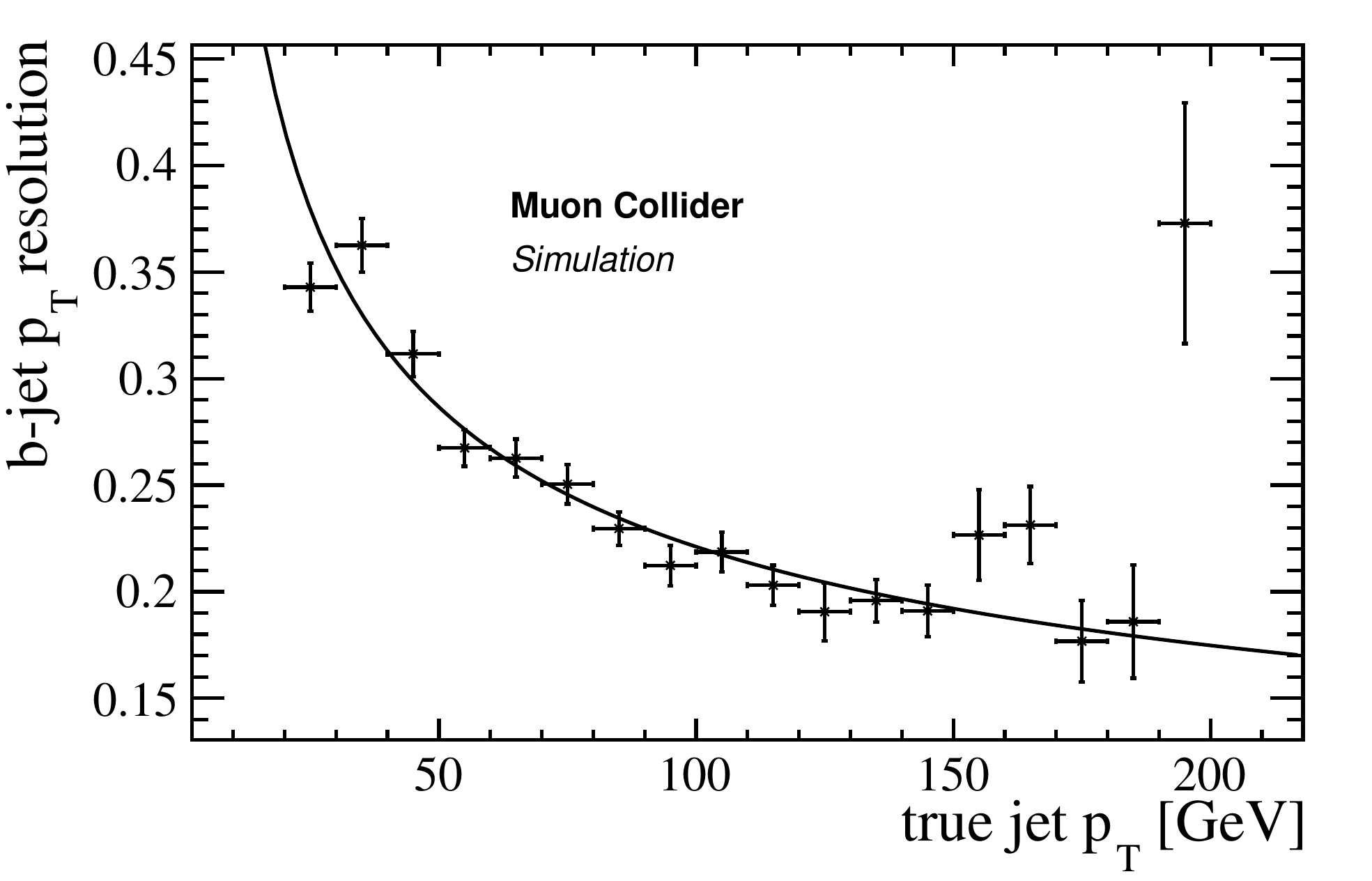}
\caption{Left: relative difference between reconstructed and true jet pseudo-rapidity. Right: $b$-jet $p_T$ resolution as a function of the jet $p_T$.}
\label{fig:bjet_reso}
\end{figure}

The relative difference between reconstructed and true $b$-jet $\eta$ is shown in Fig.~\ref{fig:bjet_reso} (right): the standard deviation of this distribution, directly related to the jet-axis angular resolution, is 14$\%$.
The $b$-jet $p_T$ resolution as a function of the true $b$-jet $p_T$ is shown in Fig.~\ref{fig:bjet_reso} (right).
The resolution goes from 35$\%$ for jet $p_T$ around 20 GeV to 20$\%$ for high jet $p_T$.

\begin{figure}[htb]
\centering
\includegraphics[width=0.48\textwidth]{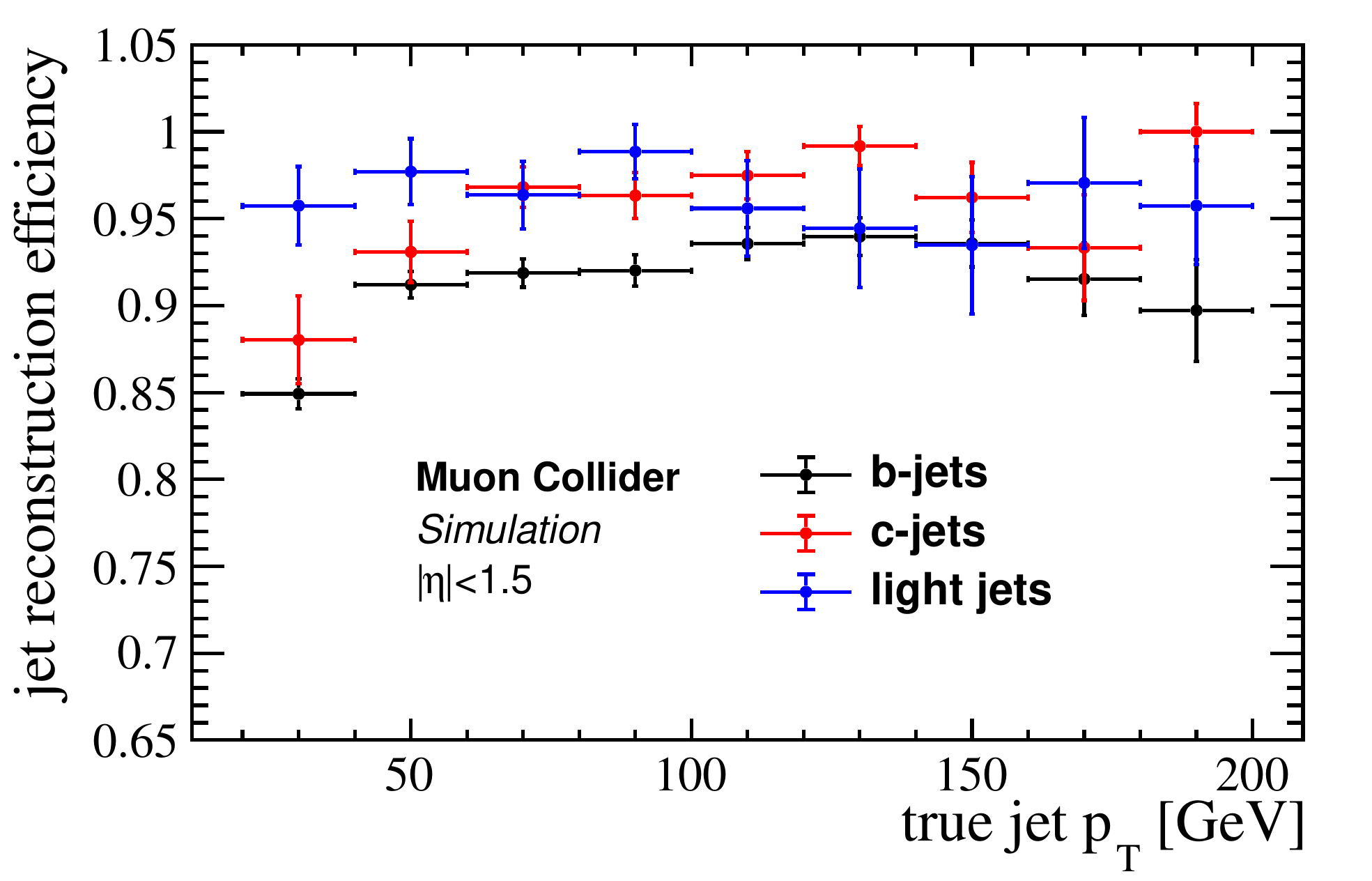}
\includegraphics[width=0.48\textwidth]{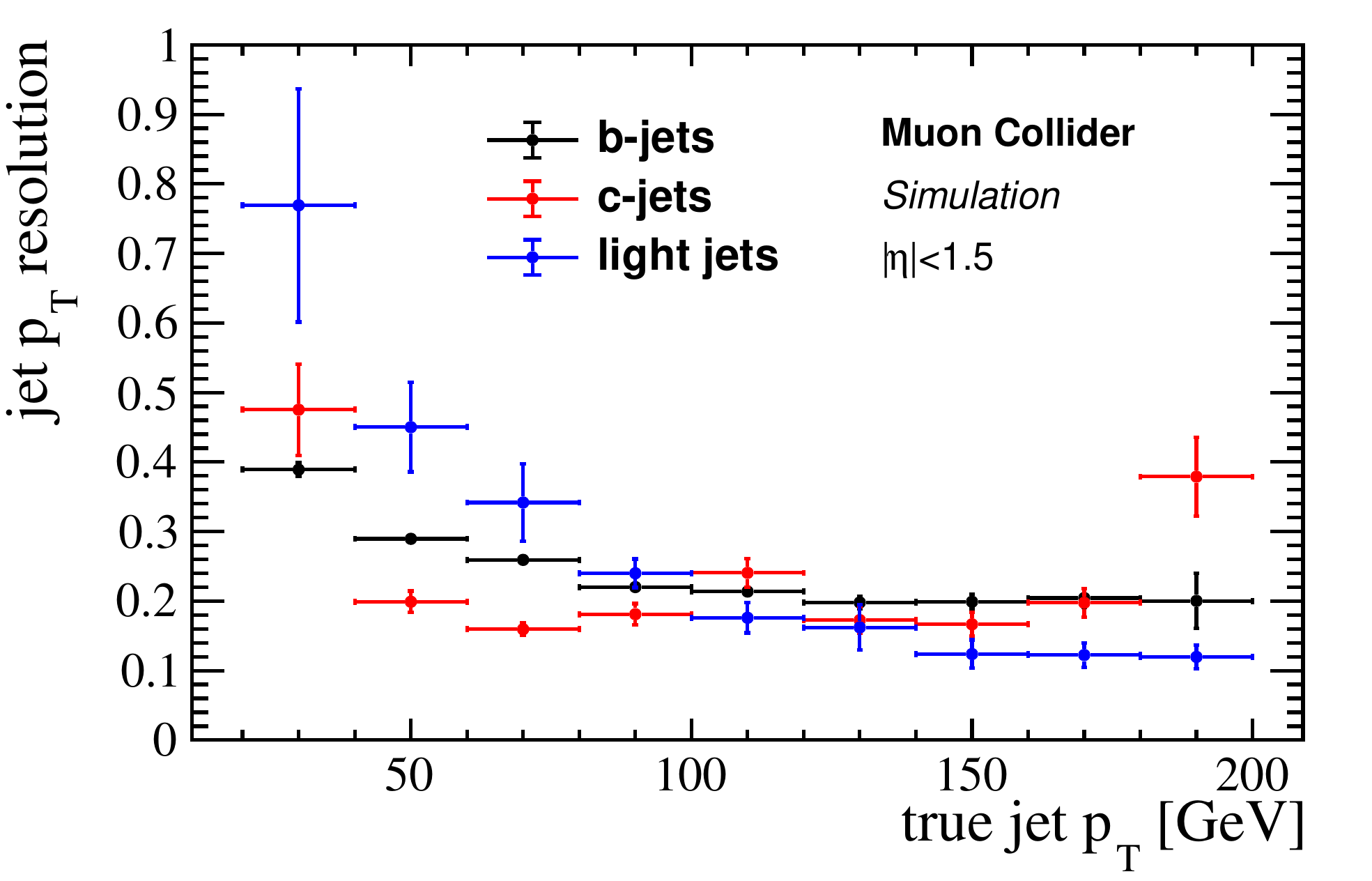}
\caption{Left: Jet reconstruction efficiency (left) and jet $p_T$ resolution (right) as a function jet $p_T$ for $b$-jets, $c$-jets and light jets in the central region $|\eta|<1.5$. It has been checked that differences between the jet flavours are mainly due to different jet $\eta$ distributions in the three samples.}
\label{fig:jet_flavour}
\end{figure}

The jet reconstruction performances obtained separately in  the $b\bar{b}$, $c\bar{c}$ and $q\bar{q}$ samples and in the central region ($|\eta|<1.5$) are compared in Fig.~\ref{fig:jet_flavour}. It can be seen that both the efficiencies and $p_T$ resolutions are of the same order, however some differences exist between the jet flavours: it has been checked that these are mainly due to different jet $\eta$ distributions in the three samples. 

\subsection{Invariant mass reconstruction of Higgs and Z bosons to b-quarks}

\begin{figure}[htb]
\centering
\includegraphics[width=0.48\textwidth]{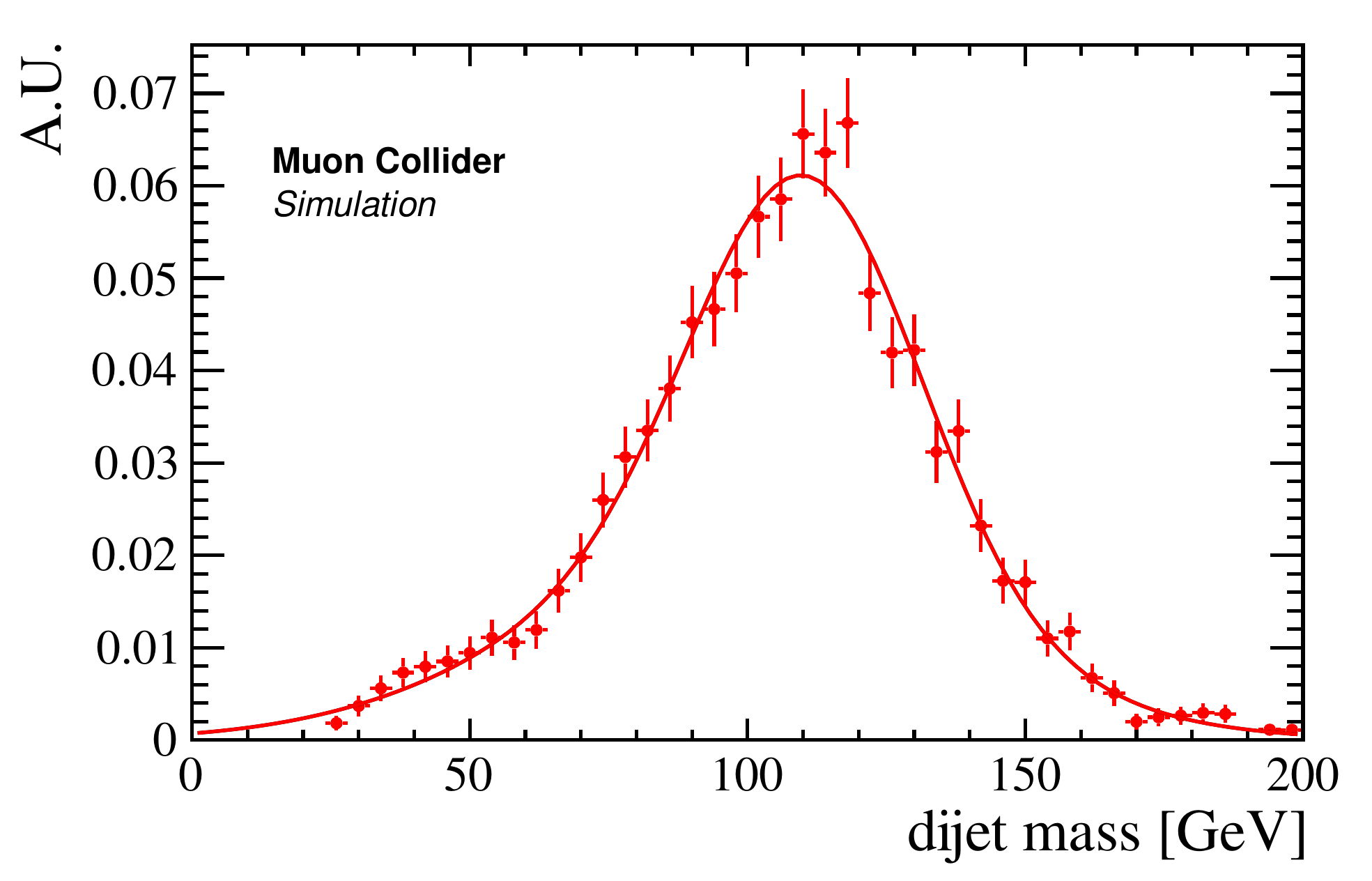}
\includegraphics[width=0.48\textwidth]{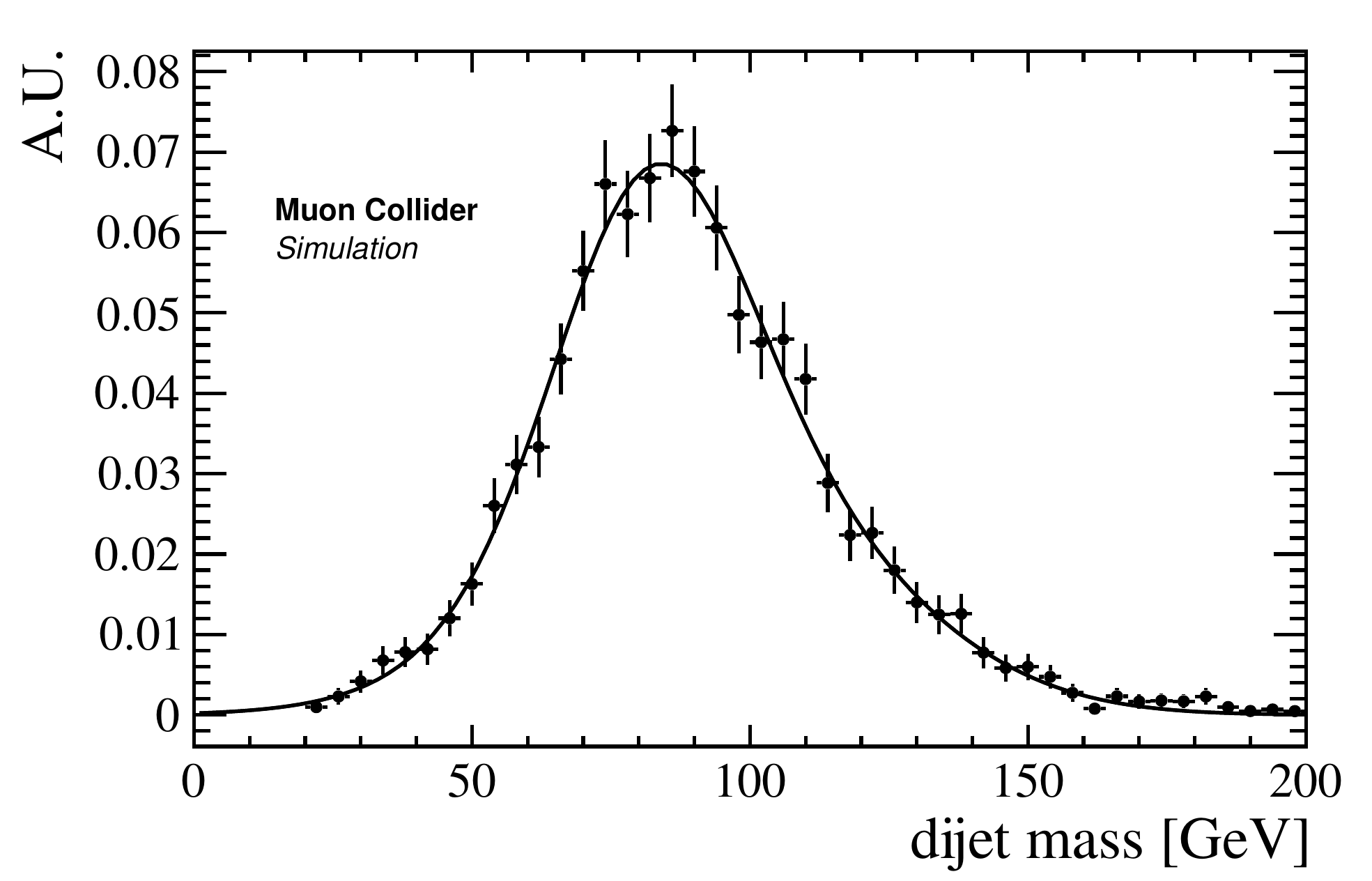}
\includegraphics[width=0.48\textwidth]{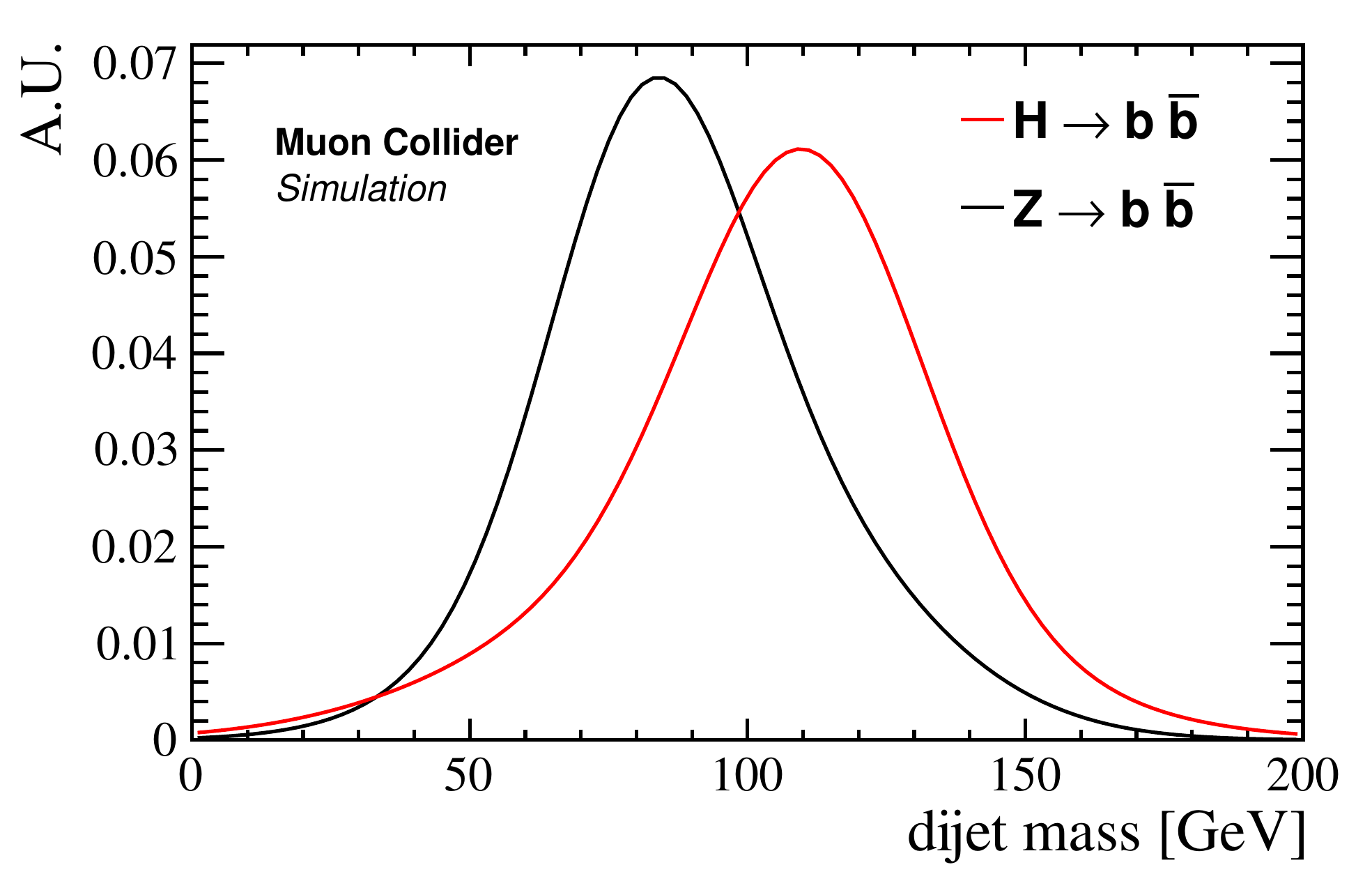}
\caption{Dijet invariant mass distributions for $H \rightarrow b \bar{b}$ (top left) and $Z \rightarrow b \bar{b}$ (top right) are shown. Distributions are normalized to the same area, and are fitted with double gaussian functions. The shapes are compared is the bottom plot.}
\label{fig:hbb}
\end{figure}

The jet reconstruction is applied to the simulated samples of $H \rightarrow b \bar{b}$ and $Z \rightarrow b \bar{b}$ to probe the dijet invariant mass reconstruction.
The invariant mass separation between these two processes is of paramount importance for physics measurements at the muon collider. In this study both jets are required to have $p_T>40$ GeV and $|\eta|<2.5$. In Fig.~\ref{fig:hbb} the dijet invariant mass distributions for $H \rightarrow b \bar{b}$ and $Z \rightarrow b \bar{b}$ are shown. The distributions are fitted with double gaussian functions, and the shapes are compared. A relative width, defined as the standard deviation divided by the average value of the mass distribution, of 27$\%$(29$\%$) for $H \rightarrow b \bar{b}$($Z \rightarrow b \bar{b}$) is found. It can be seen that a significant separation is achieved.
\subsection{Future prospects on jet reconstruction}
\label{sec:future_jets}
\begin{figure}[htb]
    \centering
    \includegraphics[width=0.48\textwidth]{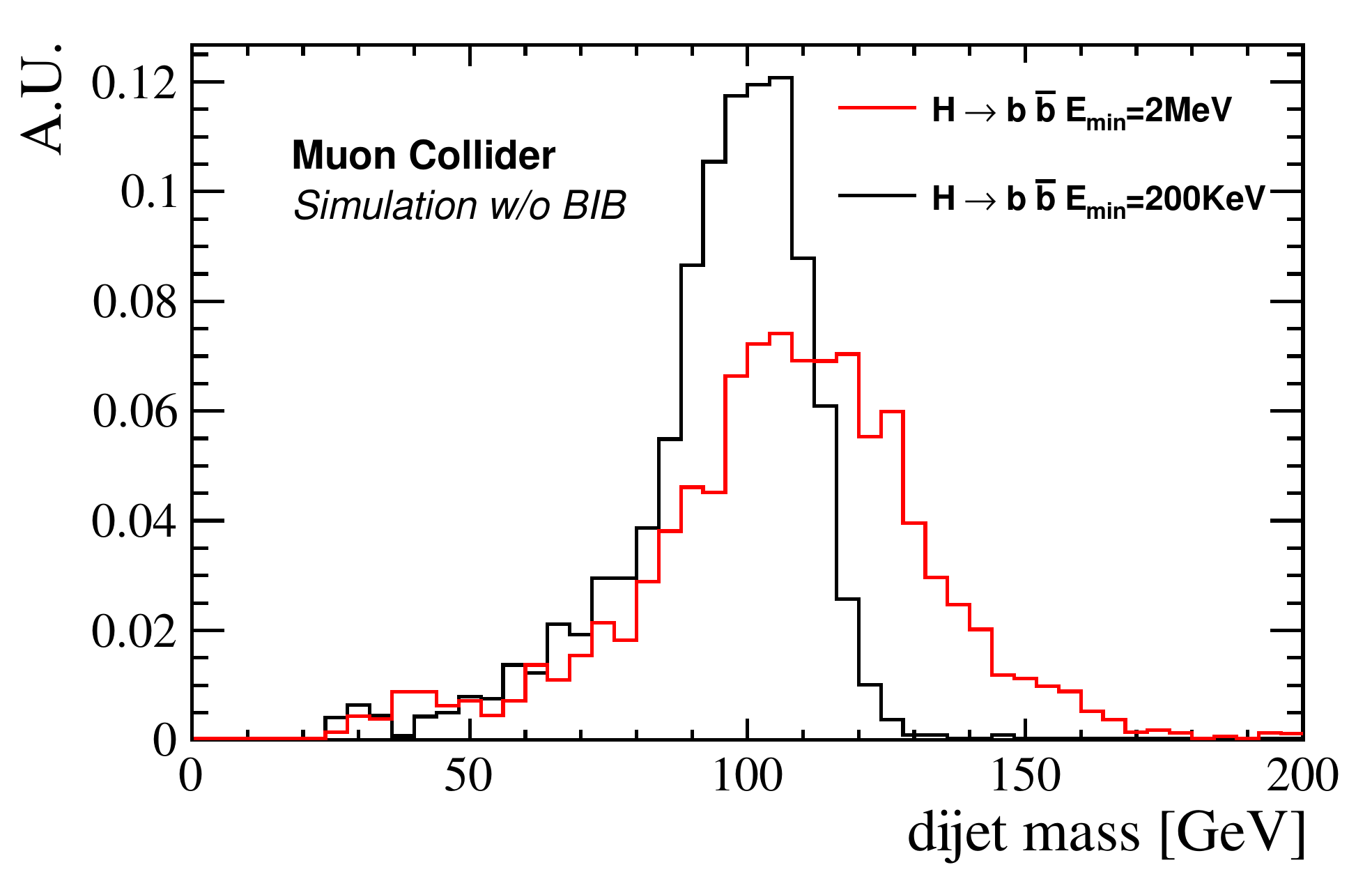}
    \includegraphics[width=0.48\textwidth]{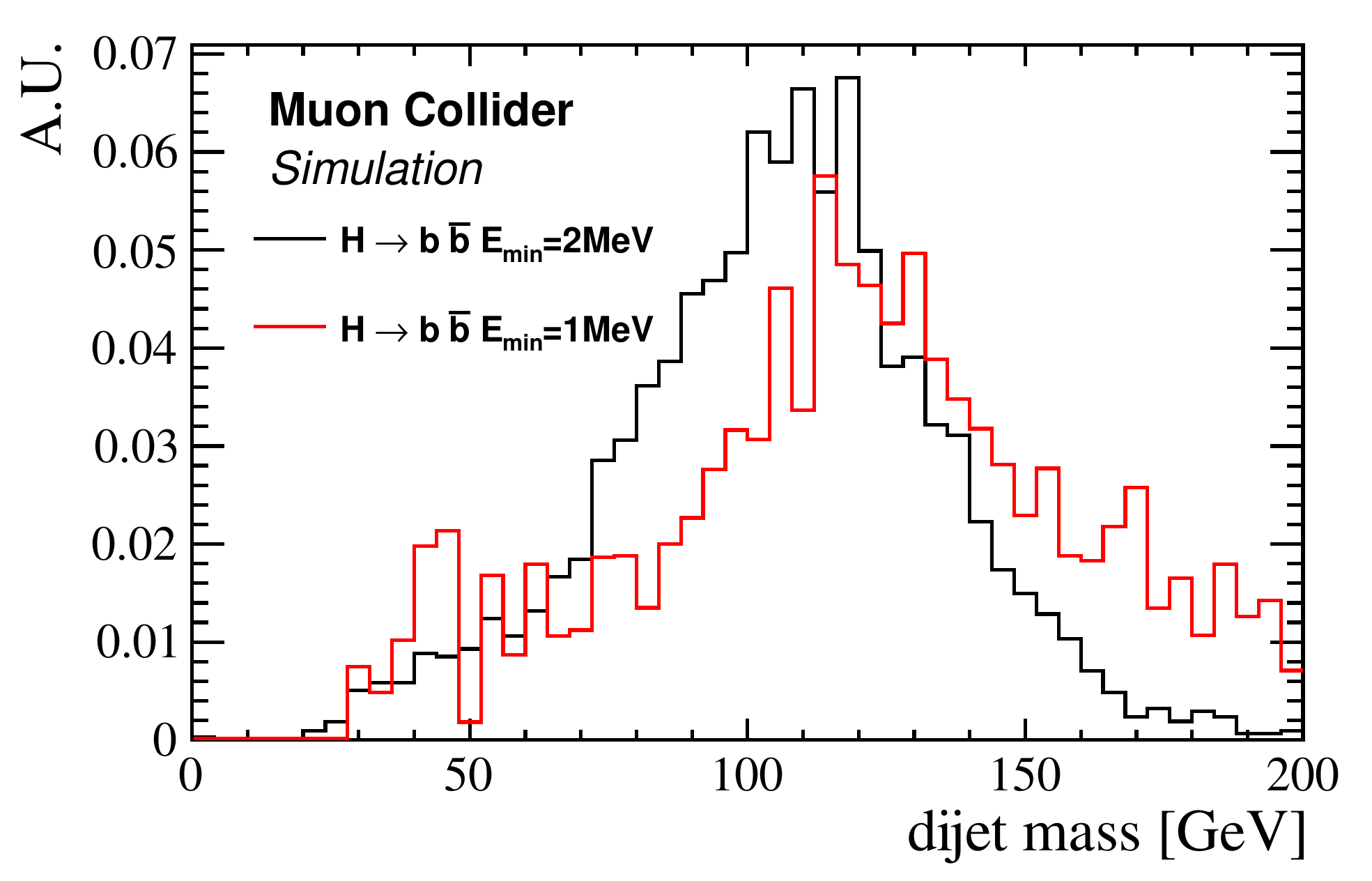}
    \caption{Left: $H \rightarrow b \bar{b}$ dijet invariant mass, reconstructed without the presence of the BIB and with 2 MeV and 200 KeV calorimeter hit energy thresholds. Right: $H \rightarrow b \bar{b}$ dijet invariant mass reconstructed with 2 MeV and 1 MeV thresholds. Distributions are normalized to the same area.}
    \label{fig:h_2_vs_1}
\end{figure}

\begin{figure}[htb]
    \centering
    \includegraphics[width=0.6\textwidth]{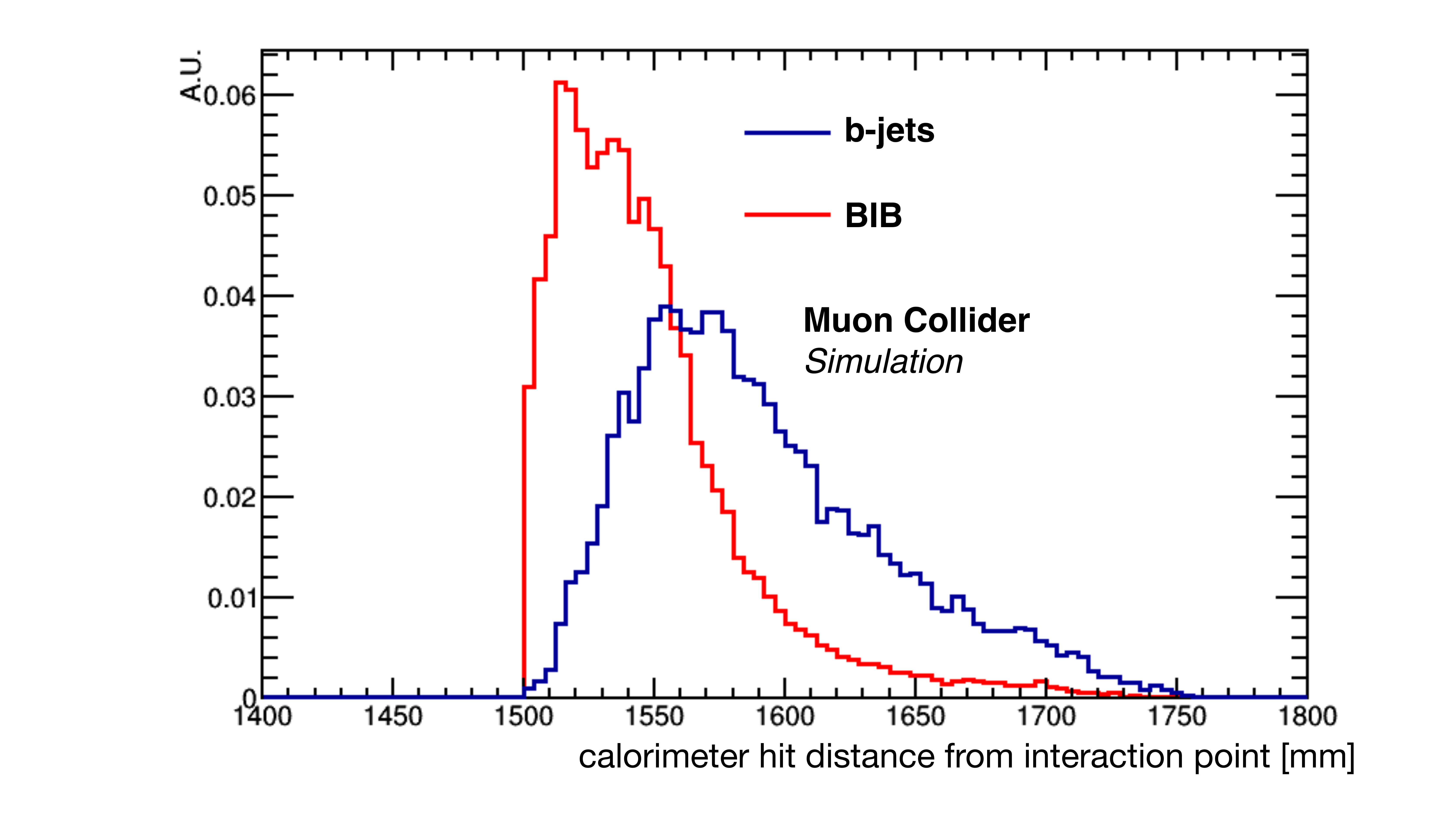}
    \caption{Distribution of the ECAL barrel hits distance from the interaction point (weighted for the hit energy), for both $b$-jets and BIB. Both distributions are normalized to the same area.}
    \label{fig:calo_hit}
\end{figure}

Before discussing the heavy-flavour jet identification, we notice that, at this stage, the jet reconstruction algorithm can be improved in several ways. In this Section some guidelines are given:
\begin{itemize}
    \item \textbf{track filter}: it has been verified that the track filter has a different impact in the central and the forward region, in particular the efficiency in the forward region is lower. An optimized selection should be defined,
    \item \textbf{calorimeters threshold}: the hit energy threshold has been set to the relatively high value of 2 MeV, as a compromise between computing time and jet reconstruction performance. This is a major limitation in the jet performance as can be seen in Fig.~\ref{fig:h_2_vs_1} (left), where the $H \rightarrow b \bar{b}$ dijet invariant mass, reconstructed without the presence of the BIB, is compared between 2 MeV and 200 KeV thresholds. However reducing this threshold is not an easy task, given the large number of calorimeter hits selected from the BIB that contaminate the jet reconstruction. This is shown in Fig.~\ref{fig:h_2_vs_1} (right), where can be clearly seen that the performance at 1 MeV threshold is degraded with respect to 2 MeV. To tackle this problem an optimized algorithm should be developed: as an example thresholds that depend on the sensor depth could by applied, since the longitudinal energy distribution released by the BIB is different from the signal jets as shown in Fig.~\ref{fig:calo_hit} (right). A generalization of this idea could be the application of a multivariate-algorithm trained to select signal hits and reject BIB hits,
    \item \textbf{fake jet removal}: the fake jet removal applied in this study has an impact in reducing the jet efficiency in the forward region. Moreover this issue is highly dependent from the calorimeter thresholds. A fake removal tool based on machine learning and with jet sub-structure observables as input should be developed to solve this task.
\end{itemize}
    
%% file: jets_heavyflavour.tex
The $b$-jet identification algorithm described in this section relies on the reconstruction of the secondary vertices inside the jet cones, that are compatible with the decay of the heavy-flavour hadron.
 

  For the secondary vertices reconstruction, tracks are reconstructed with the Conformal Tracking algorithm described in Section \ref{sec:trk-roi}, where the region of interests are defined by cones with $\Delta R=0.7$ around the jet axes. In order to reduce the computational time, the Double Layer filter (see Section~\ref{sec:trk-dl}) is applied to reduce the number of hits due to BIB. The effect of this filter on the secondary vertex reconstruction is then corrected to obtain the efficiency prior to its application, as will be explained in Section \ref{sv_tagging_eff}.

   \subsubsection{Secondary vertex tagging}\label{sv_tag_algo}

  The vertexing algorithm employed for the primary and secondary vertex reconstruction is described in detail in \cite{SUEHARA2016109}. In order to reduce the amount of combinatorial tracks due to BIB, cuts are applied to the reconstructed tracks given as input to the algorithm. The main steps of the algorithm and cuts to the input tracks are here summarized:
  \begin{enumerate}
      \item \textbf{primary vertex finding}: tracks with $|D_{0}| \leq \qty{0.1}{\mm}$ and $|Z_{0}| \leq \qty{0.1}{\mm}$ are selected and given in input to the primary vertex fitter. Furthermore, to reduce the number of combinatorial short BIB tracks, each track is required to have at least four hits in the vertex detector;
      \item \textbf{tracks selection for secondary vertex finder}: tracks not used to reconstruct the primary vertex are given in input to the secondary vertex fitter. Fig.~\ref{fig:b_c_vs_bib} shows the distributions of the total number of hits in the vertex detector for combinatorial BIB tracks and for tracks matched at Monte Carlo level with particles generated by $b$ or $c$ hadrons decays. A minimum number of 4 hits in the vertex detector is required in order to keep most of the the tracks from $b$ and $c$ hadrons decay, while rejecting most of BIB tracks. 
      \begin{figure}[htb]
          \centering
          \includegraphics[width=0.48\textwidth]{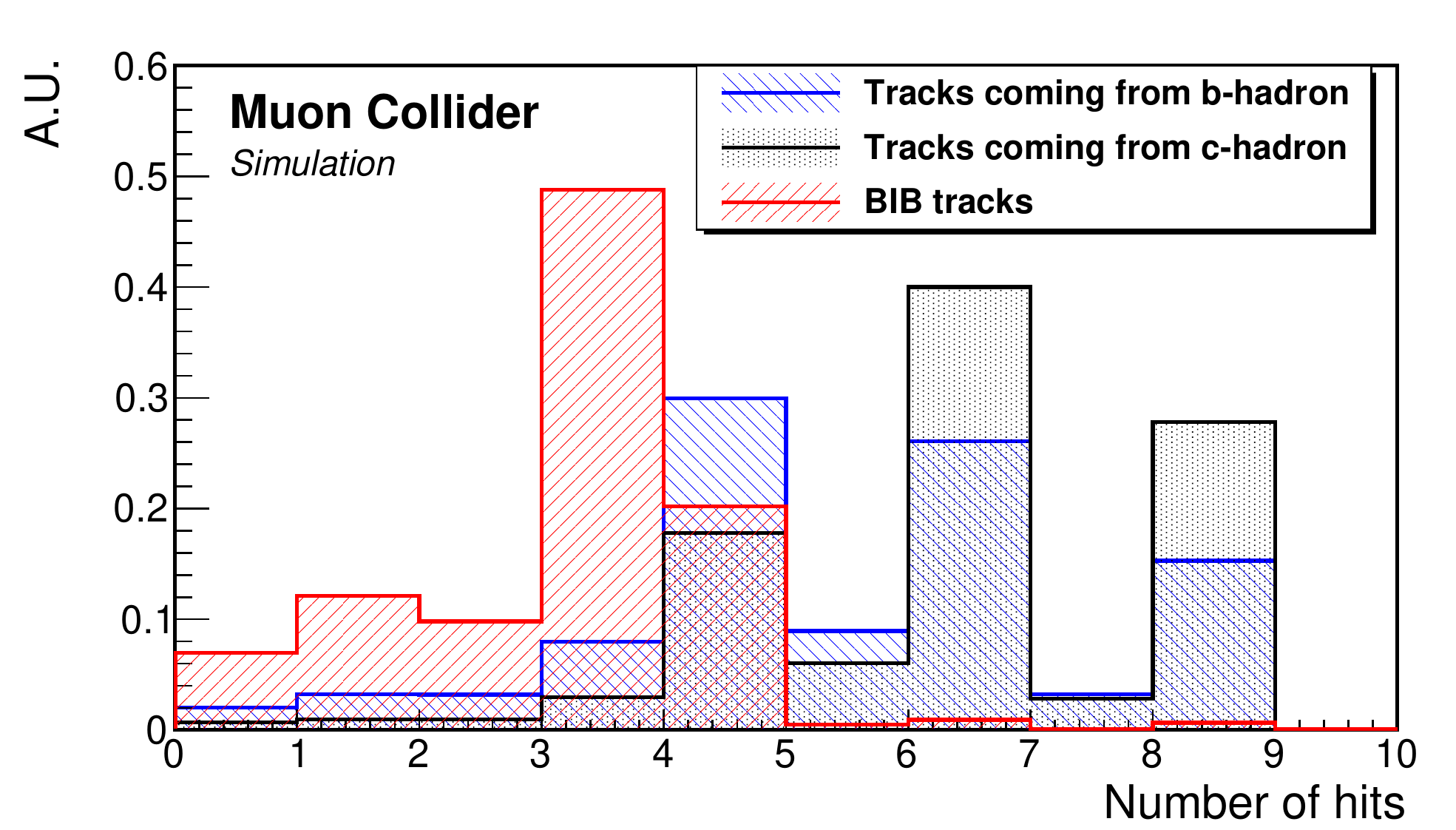}
          \caption{ Number of hits in the vertex detector of BIB combinatorial tracks (red) and of tracks matched with Monte Carlo truth particles coming from $b$ or $c$ hadrons (blue and black respectively) decay.  Distributions are normalized to the unity area.}
          \label{fig:b_c_vs_bib}
      \end{figure}
      Other cuts on the track \pT~, the maximum track $Z_0$ (longitudinal impact parameter) and $D_0$ (transverse impact parameter), and the $D_0$ and $Z_0$ errors are applied in order to further reduce the amount of BIB tracks. As an example, Fig.~\ref{fig:b_c_vs_bib_pt} (left) shows the $p_T$ distributions of BIB combinatorial tracks and tracks from $b$- or $c$-hadrons decays. By requiring $p_T > 0.8~\GeV$ about 80 $\%$ of the BIB tracks are rejected, while about 85-90 $\%$ of the tracks from $b$- or $c$-hadrons are kept. The $Z_0$ distributions of tracks coming from $b$- and $c$-hadrons and combinatorial BIB tracks are shown in Fig.~\ref{fig:b_c_vs_bib_pt} (right). The selection of tracks with $|Z_0| \leq \qty{5}{mm}$ is applied to reject the tails at large $Z_0$ that characterize BIB tracks;
      \begin{figure}[hbt]
          \centering
          \includegraphics[width=0.48\textwidth]{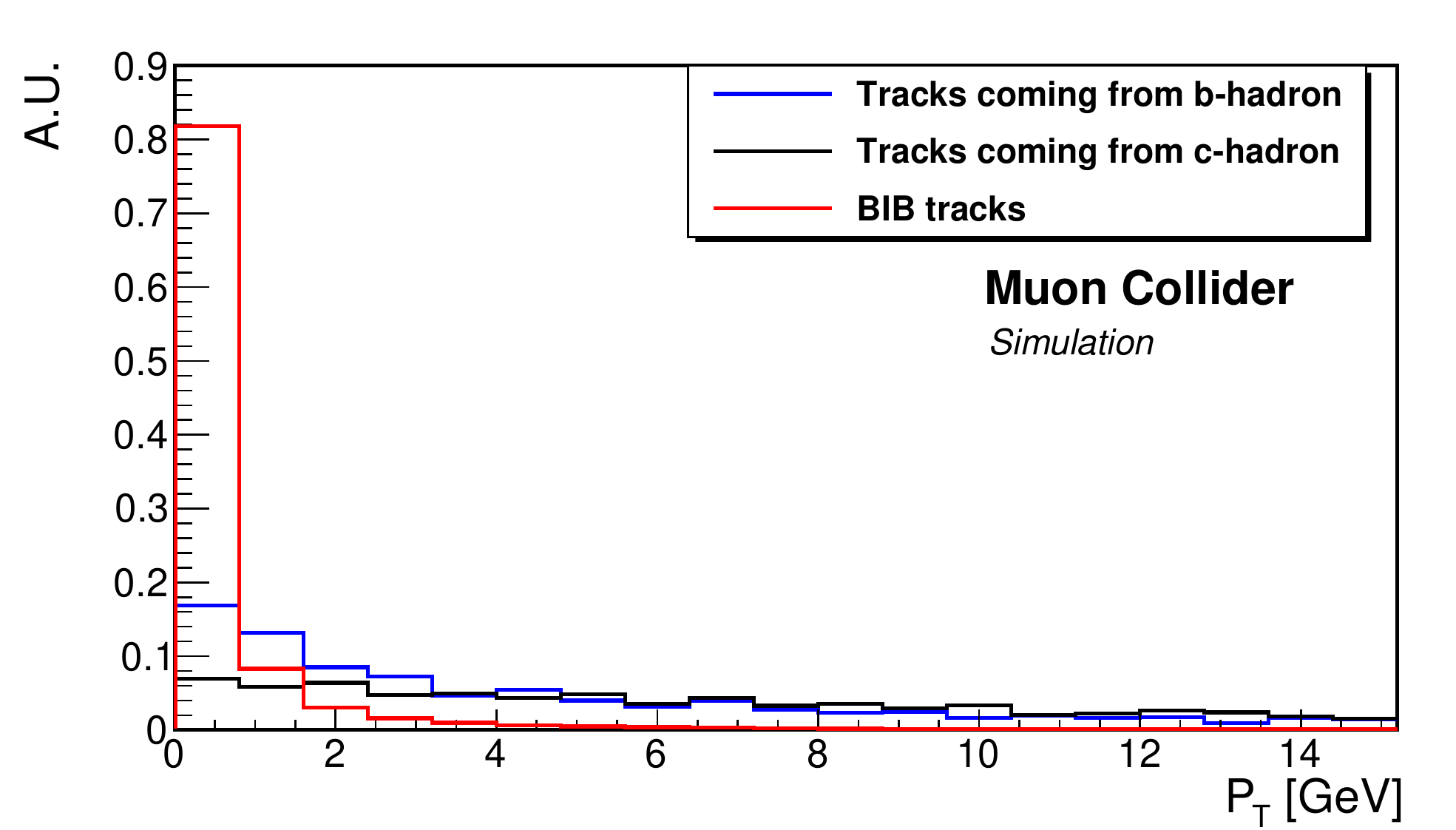}
        \includegraphics[width=0.48\textwidth]{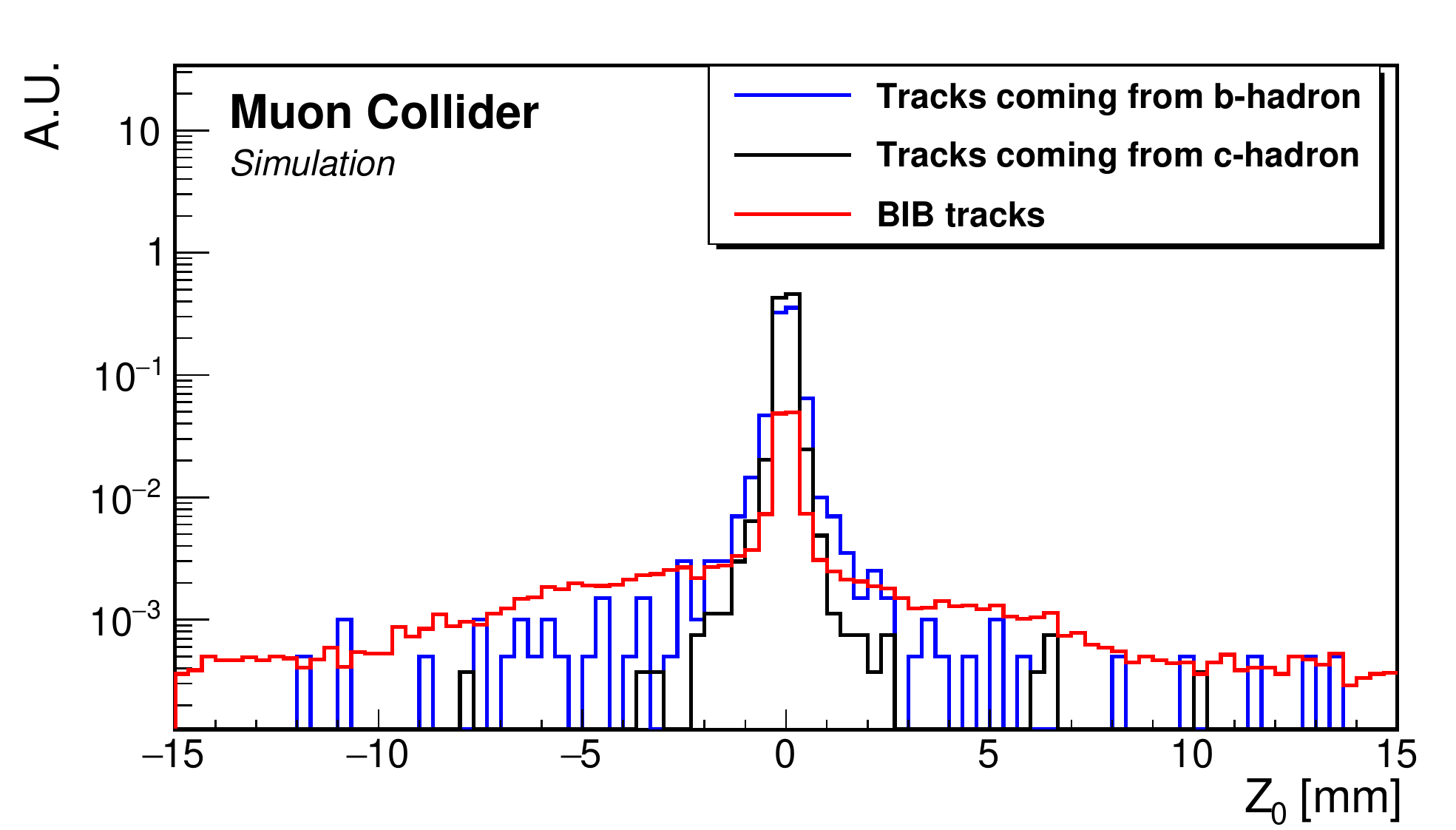}
 \caption{ Left: $p_T$ distribution of tracks coming from $b$ (blue) or $c$ (black) hadrons decay and combinatorial BIB (red) tracks. Right: $Z_0$ distribution of tracks coming from $b$ (blue) or $c$ (black) hadrons decay and combinatorial BIB (red) tracks. Distributions are normalized to the number of tracks.}
          \label{fig:b_c_vs_bib_pt}
      \end{figure}
      \item \textbf{secondary vertex finding}: tracks passing the requirements are used to build two-tracks vertex candidates, that must satisfy the following requirements: the invariant mass must be below 10 GeV and must be smaller than the energy of each track, the position with respect to the primary vertex must lie in the same side of the sum of the tracks momenta, and the $\chi^2$ of the tracks with respect the secondary vertex position must be below a threshold value, set to 5. Track pairs are also required not to be compatible with coming from the decay of neutral long lived particles.
      Additional tracks are iteratively added to the two-tracks vertices if they satisfy the above requirements. Tracks associated to more than one secondary vertex are assigned to the vertex with the lowest $\chi^2$ and removed from other vertices.
  \end{enumerate}

  \subsubsection{SV-tagging performance}\label{sv_tagging_eff}
In order to evaluate the $b$-tagging efficiency and mis-identification, a truth-level flavour is associated to the reconstructed jets with the following steps:
\begin{itemize}
   \item first, the truth-level jets are matched with the quarks from Monte Carlo to determine its flavour, requiring a distance $\Delta$R $<$ 0.5 between the truth-level jet axis and quark momentum. If more than one truth-level jet is found to match with the same quark, the one with the lowest $\Delta$R is chosen; 
    \item then, the flavour of the reconstructed jets is determined by matching them with the truth-level jets, by requiring a $\Delta$R distance between the jets axes below 0.5;
    \item if the reconstructed jet do not match with any truth-level jet, it is labeled as fake.
\end{itemize}

\begin{figure}[hbt]
          \centering
          \includegraphics[width=0.48\textwidth]{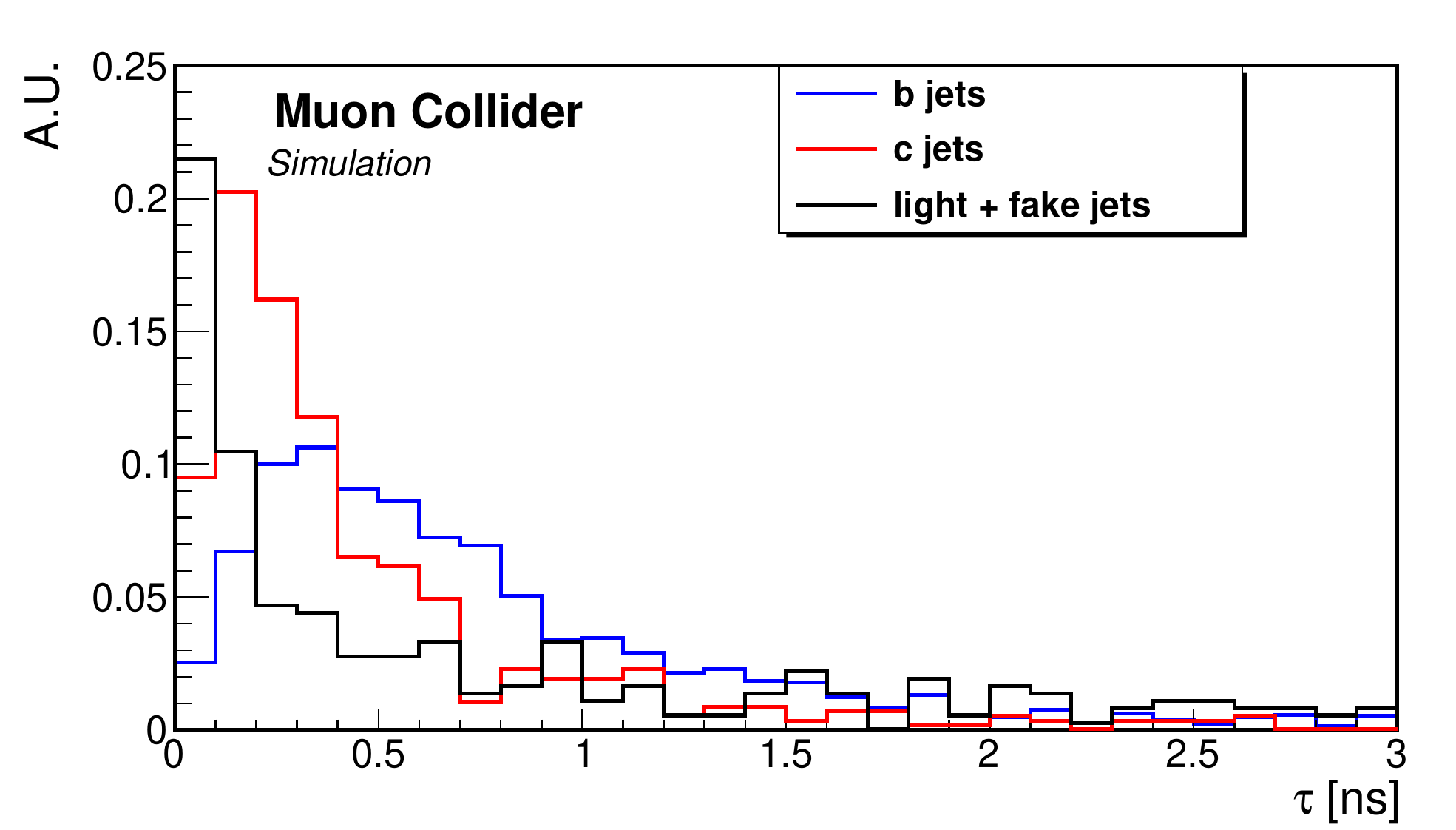}
         \caption{Distribution of the secondary vertex proper lifetime for $b$, $c$ and light jets tagged. Distributions are normalized to the unit area.}
          \label{fig:tau}
      \end{figure}

The characteristics of secondary vertices inside reconstructed jets have been studied in order to reduce the mis-identification of $c$, light and fake jets.
Fig.~\ref{fig:tau} shows the distribution of the secondary vertices proper lifetime ($\tau$) for $b$ jets, $c$ jets and light+fake jets.  A cut on $\tau>$ 0.2 ns rejects $\sim 30 \%$ of both $c$ and light+fake jets, while keeping $90 \%$ of $b$ jets. 
A reconstructed jet is identified (tagged) as $b$ jet, if at least one secondary vertex with $\tau>$ 0.2 ns is found inside its cone ($\Delta R <$ 0.5).

The $b$-tagging efficiency is defined as:
\begin{equation}
    \epsilon_b = \frac{N_{b,SV}}{N_{b}}
\end{equation}
where $N_{b,SV}$ is the number of tagged and truth-matched $b$ jets, while $N_{b}$ is the total number of truth-matched $b$ jets.
Then, the mistag on $c$ and light+fake jets is calculated as: 
\begin{equation}
    \mathrm{mistag}_{c,\mathrm{light}} = \frac{N_{c(\mathrm{light}),SV}}{N_{c(\mathrm{light})}}
\end{equation}
where $N_{c(\mathrm{light}),SV}$ is the number of tagged and truth-matched $c$(light+fake) jets, while $N_{c(\mathrm{light})}$ is the total number of truth-matched $c$(light+fake) jets.

\begin{figure}[hbt]
          \centering
          \includegraphics[width=0.48\textwidth]{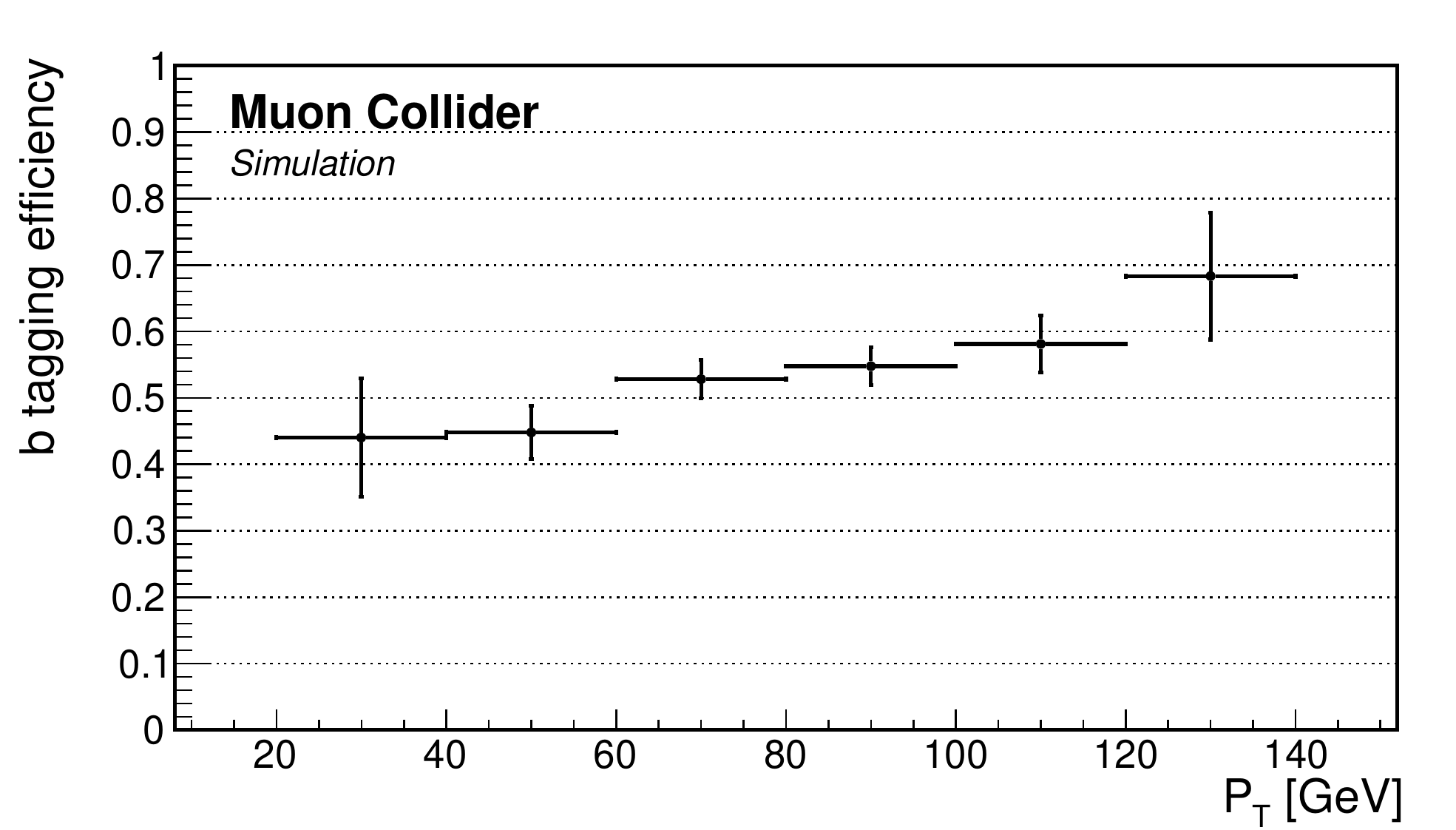}
          \includegraphics[width=0.48\textwidth]{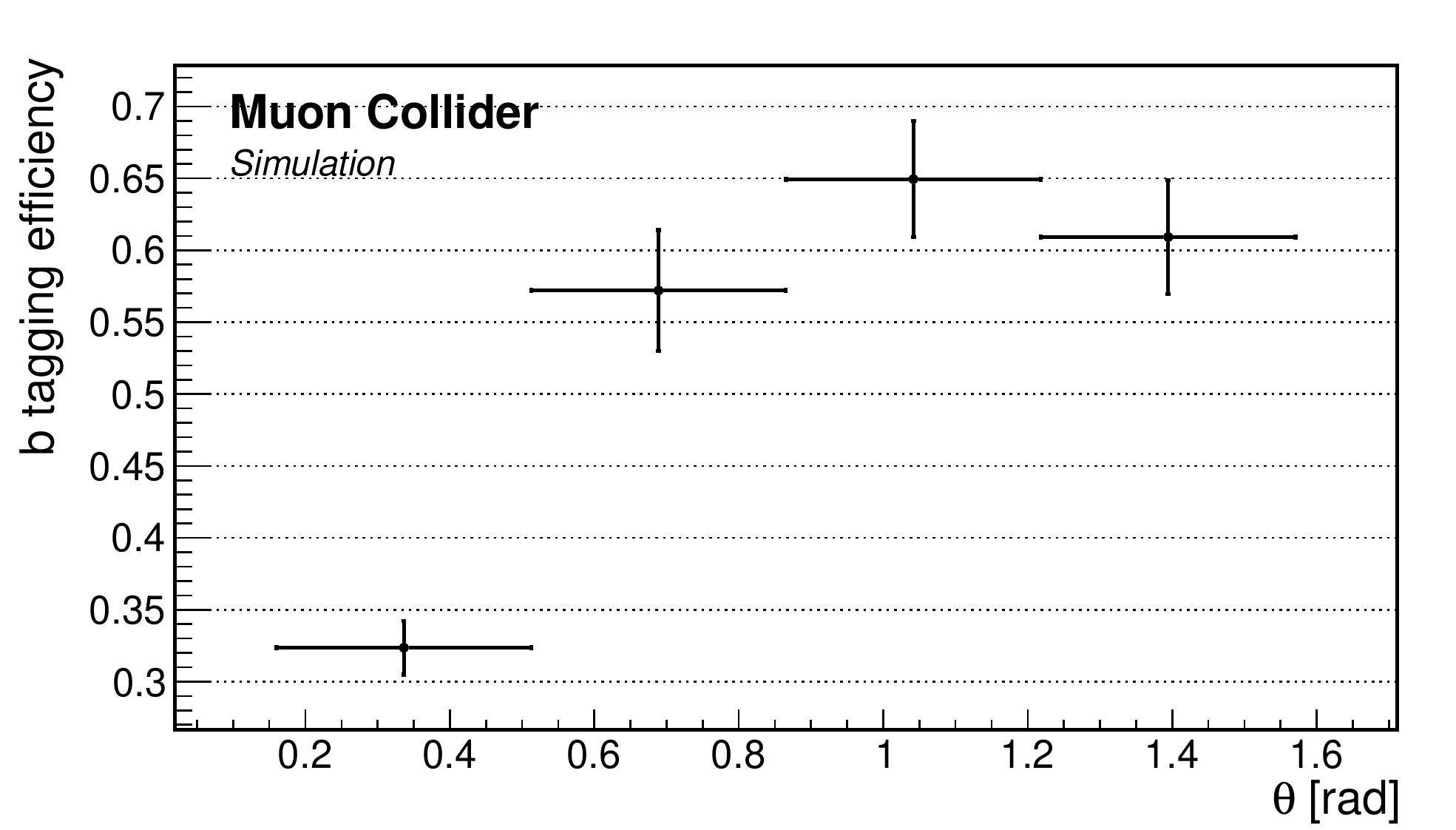}
          \caption{ Left: $b$-tagging efficiency as a function of \pT Right: $b$-tagging efficiency as a function of the angle between the jet and the beam axes. }
          \label{fig:b_eff_pt}
      \end{figure}

\begin{figure}[hbt]
          \centering
          \includegraphics[width=0.48\textwidth]{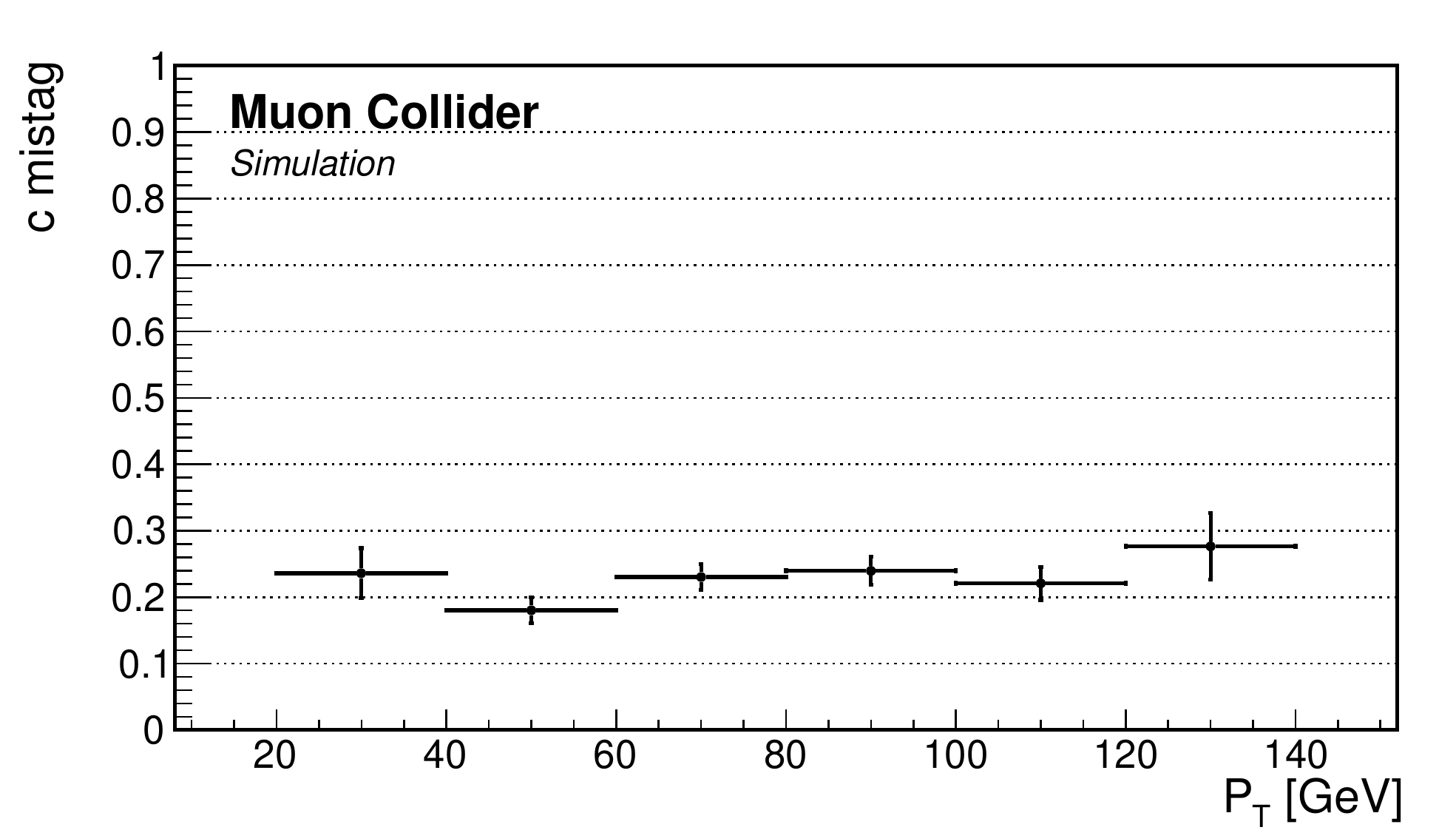}
          \includegraphics[width=0.48\textwidth]{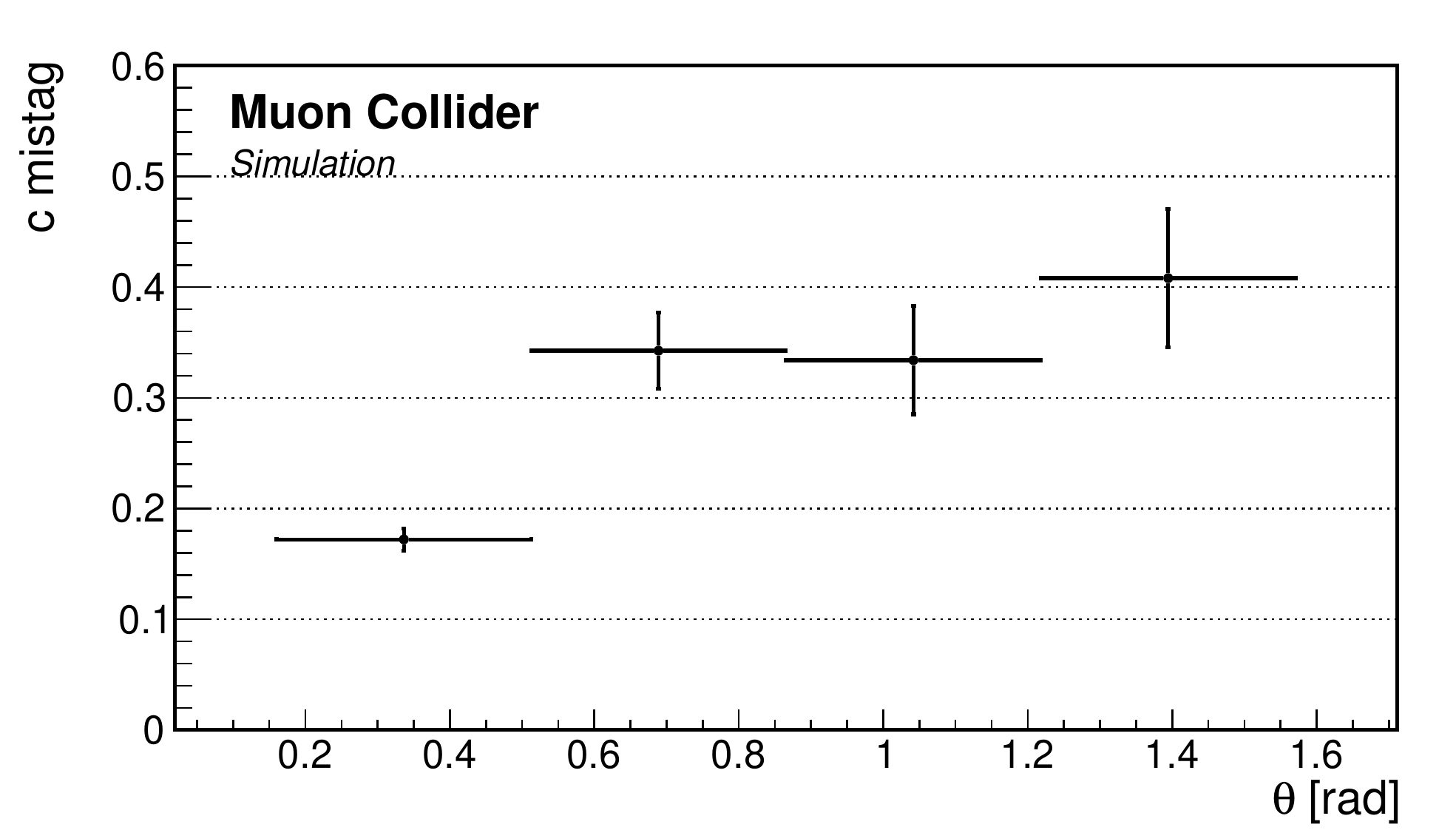}
          \caption{Mistag in $c \bar{c}$ dijet samples as a function of $p_T$ (left) and $\theta$ (right). }
          \label{fig:c_eff_pt}
      \end{figure}

\begin{figure}[hbt]
          \centering
          \includegraphics[width=0.48\textwidth]{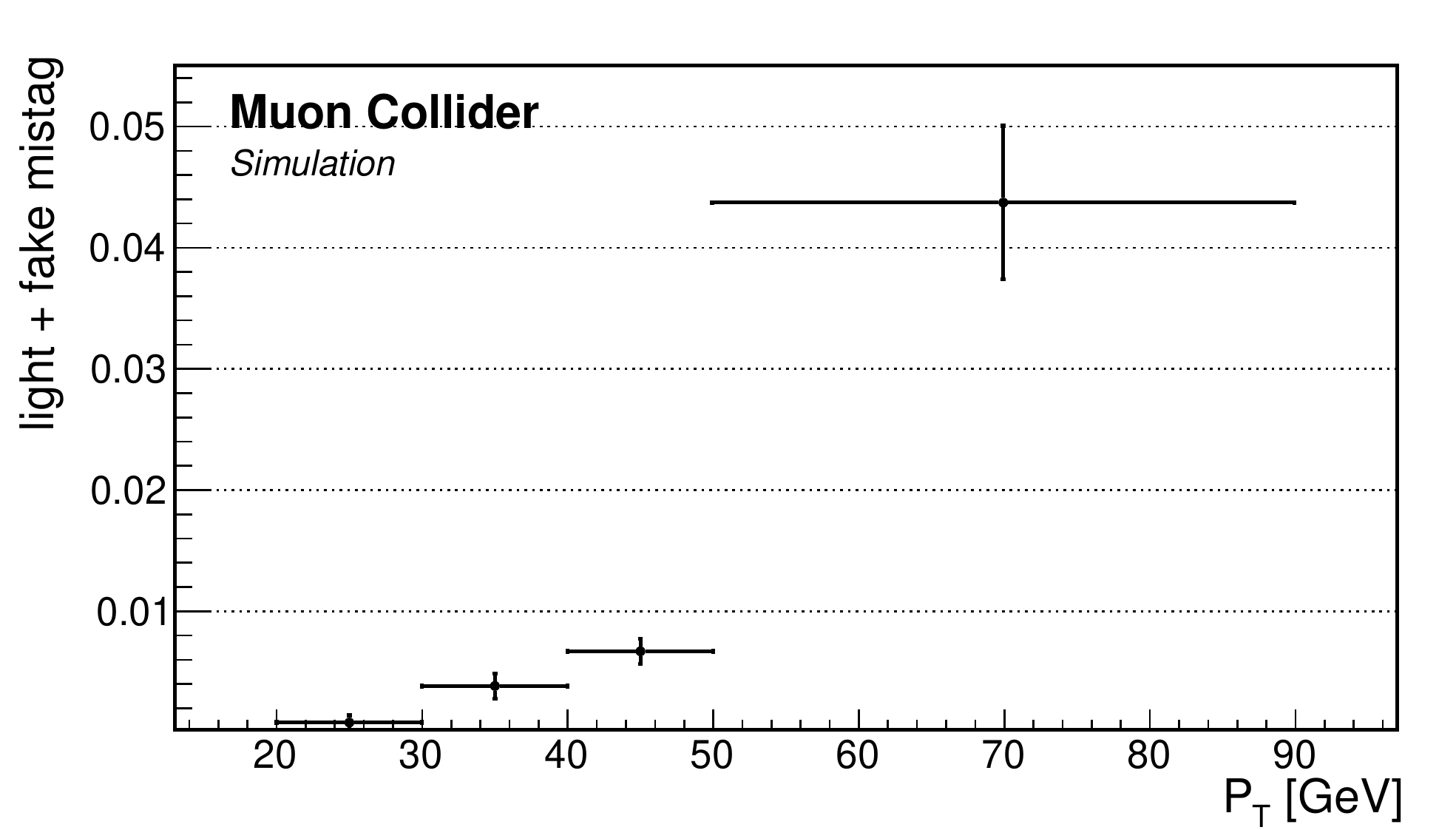}
          \includegraphics[width=0.48\textwidth]{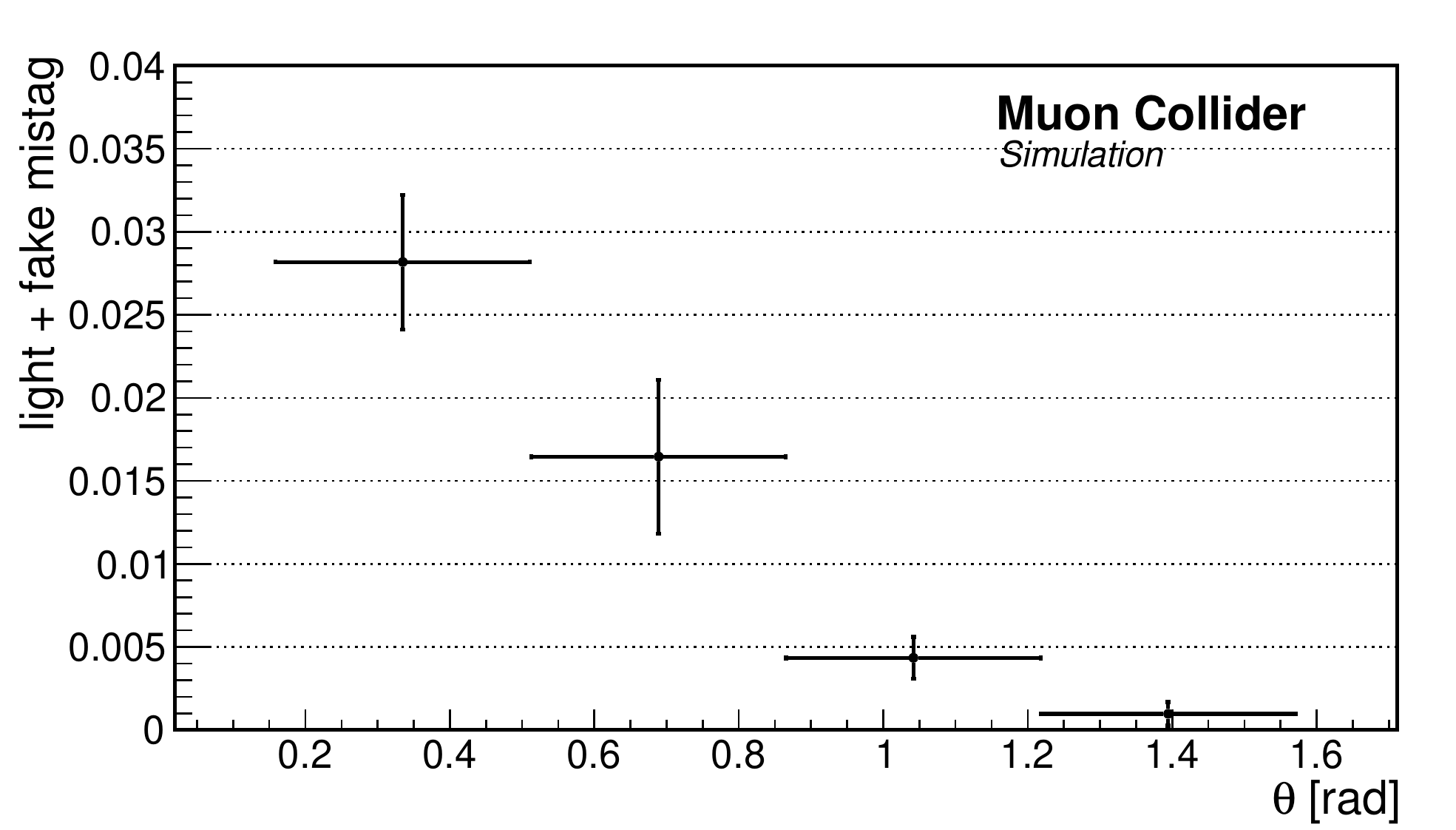}
          \caption{ Mistag in light dijet samples as a function of $p_T$ (light) and $\theta$ (right). }
          \label{fig:light_eff_pt}
      \end{figure}

The effect of the Double Layer Filter on the secondary vertex finding efficiency have been evaluated reconstructing $b\bar{b}$, $c\bar{c}$ and $q\bar{q}$ dijet samples without BIB, with and without the Double Layer filter. The following ratio has been calculated as a function of of the jet $p_T$ and of the angle of the jet axis with respect to the beam ($\theta$):
  \begin{equation}
      r (p_{T},\theta)= \frac{N_{noBIB, NoDL}}{N_{noBIB, DL}},
  \end{equation}
  where $N_{noBIB, NoDL}$ is the number of tagged jets without double layer filter and $N_{noBIB, DL}$ is the number of tagged jets with the double layer filter.
  The final tagging efficiencies are then corrected for this ratio, assuming that its value does not change in the presence of the BIB.
The $b$-tagging efficiency as a function of the jet $p_T$ and $\theta$ is shown in Fig.~\ref{fig:b_eff_pt}.
The efficiency is around $50 \%$ at low $p_T$ and increases up to $70-80 \%$ at high $p_T$.
At low $\theta$ the efficiency is lower ($\sim 30 \%$) probably due to the presence of the nozzles. 
The mistag for $c$ jets is shown in Fig.~\ref{fig:c_eff_pt} and is found to be around $20 \%$. As for $b$-jet efficiency, the $c$ mistag increases in the central region of the detector.
Fig.~\ref{fig:light_eff_pt} (left) shows the mistag for the light and fake jets versus jet $p_T$, up to 90 GeV, and is found to be lower than $1\%$ below 50 GeV, while increases to $5\%$ at higher jet $p_T$.
Fig.~\ref{fig:light_eff_pt} right shows the light+fake jet mistag as a function of $\theta$. 

As for jet reconstruction, the $b$-jet identification algorithm needs further improvements, but a solid starting point has been set. In particular it will take advantage of the advancements in the vertex detector and track reconstruction. Given the impact of the BIB, the features shown in this section are not sufficient to setup an effective $c$-tagging algorithm, therefore this case should be studied by defining a dedicated strategy for the muon collider environment.

\clearpage

    

%% file: photons.tex
The photon reconstruction and identification performance of the muon collider detector is assessed
in a sample of 100000 events with a single photon per event. 
Photons were generated in the nominal collision vertex at the center of the detector, uniformly
distributed in energy between 1 and 1500 GeV, in polar angle between $10^\circ$ and $170^\circ$, and in
the full azimuthal angle range. The sample was then processed with the detector simulation and fully reconstructed.
40000 events were also reconstructed with the beam-induced background overlaid.

The reconstruction procedure includes the CKF tracking and the calorimeter reconstruction.
Prior to track reconstruction, the tracker hits were processed with the Doublet-Layer filter~(see Section~\ref{sec:trk-dl}). 
Moreover, to get rid of most of the fake tracks due to the spurious hits from the background, 
a track quality selection is applied before the track refitting step, which requires at least 
three hits in the vertex detector and at least two hits in the inner tracker.
To reject part of the background hits in the calorimeters, an energy threshold of 2 MeV is applied 
to both the ECAL and HCAL hits.
Photons are reconstructed and identified with the Pandora Particle Flow algorithm. A detailed description of the package can be found in Ref.~\cite{THOMSON200925}.

The energy threshold of the calorimeter hits and the presence of the beam-induced background 
affect the energy scale of the reconstructed photons.
A correcting factor is applied to the reconstructed photon energy to make the detector response 
uniform as a function of the photon energy and the photon polar angle.
The correction was calculated from a comparison of the reconstructed photon energy with
the photon energy at generation level in an independent subsample of events.

\begin{figure}[t]
\centering
\includegraphics[width=0.48\textwidth]{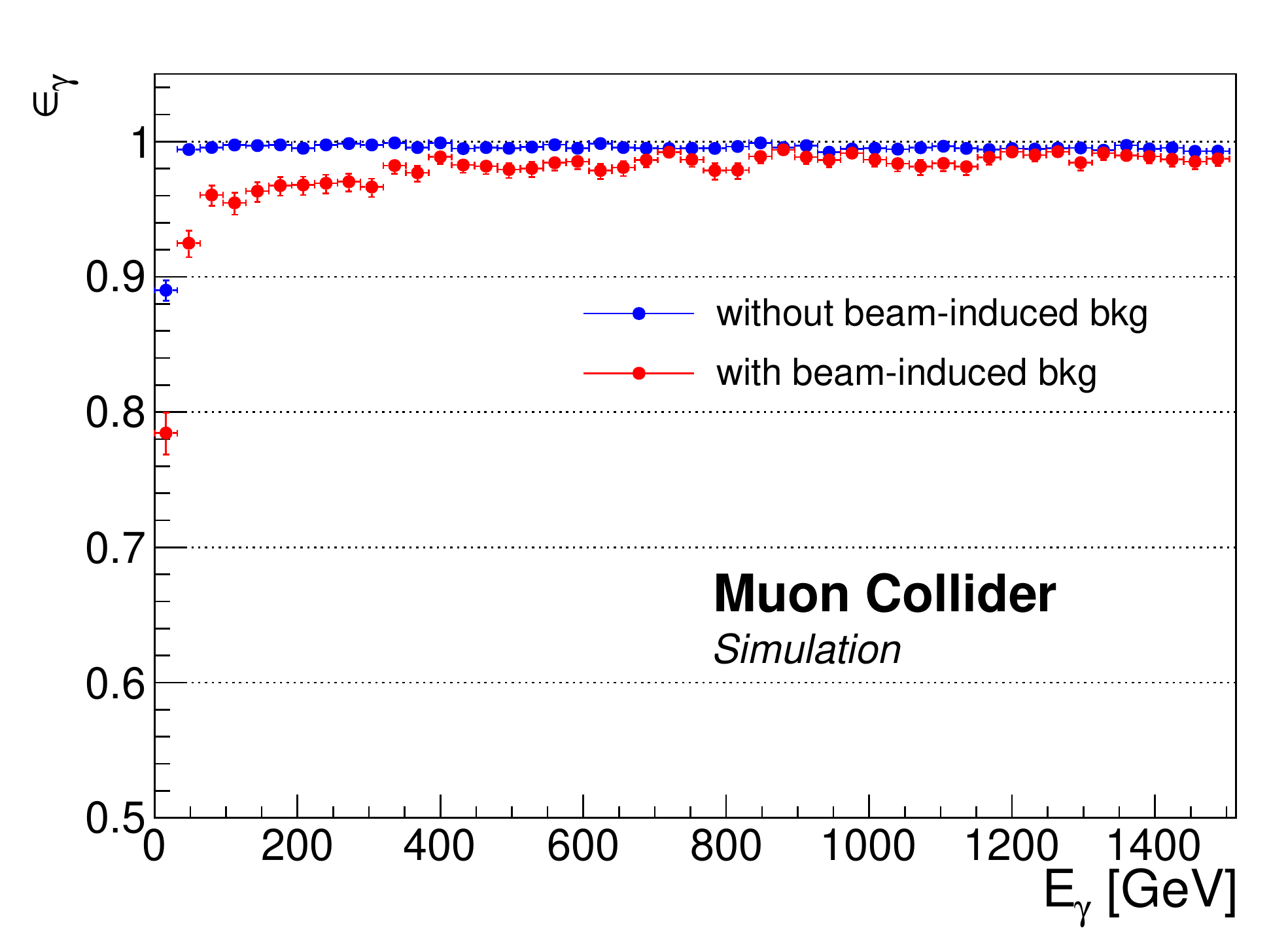}
\includegraphics[width=0.48\textwidth]{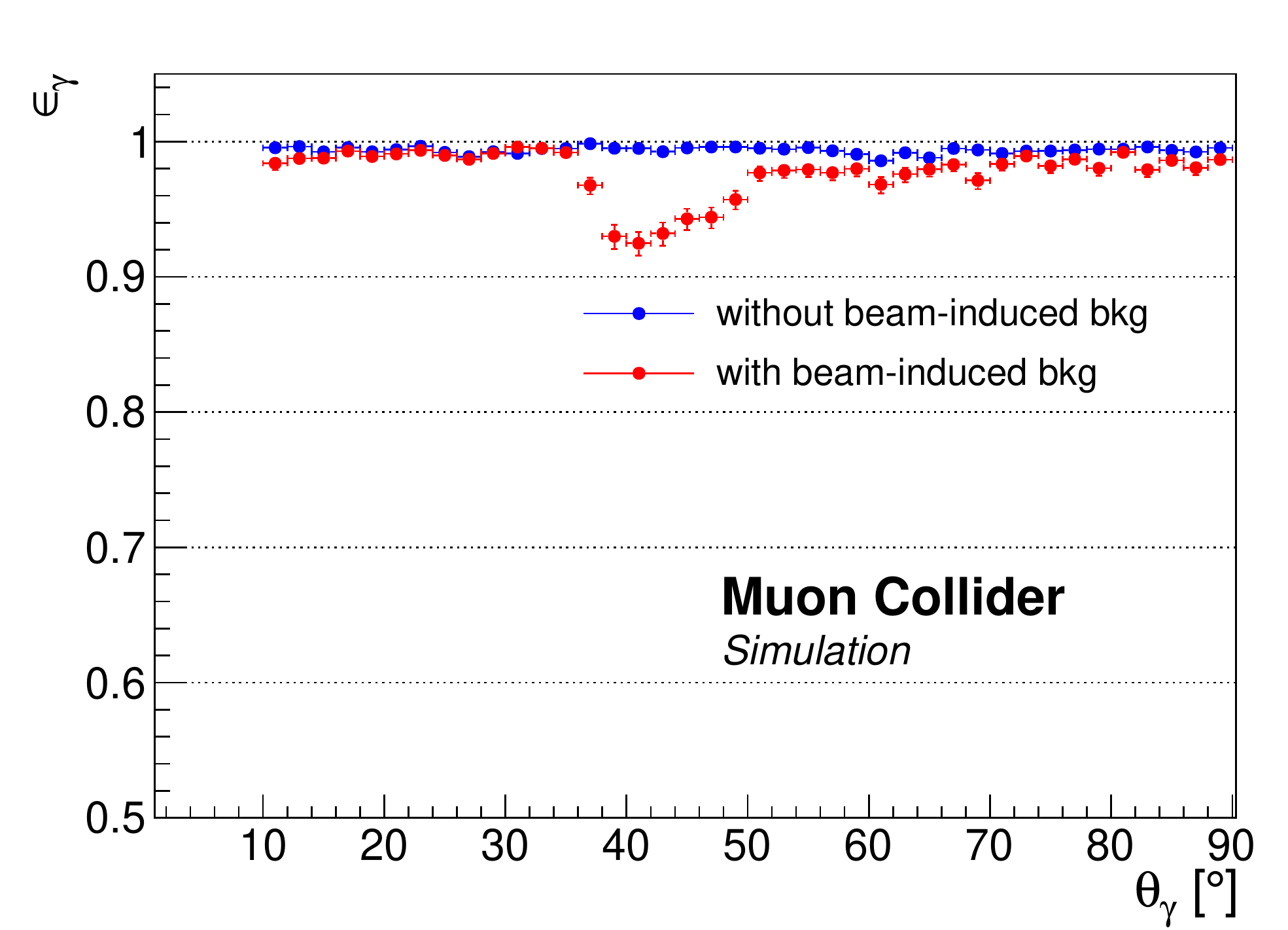}
\caption{Photon reconstruction efficiency as a function of the photon energy (left) and
    the photon polar angle (right).
\label{fig:ph_efficiency}}
\end{figure}

Fig.~\ref{fig:ph_efficiency} shows a comparison of the photon reconstruction efficiency as a 
function of the generated photon energy and polar angle $\theta$ with and without the BIB.
The efficiencies are defined as the fraction of generated photons in the range $10^\circ$ and 
$170^\circ$ that are matched to a reconstructed photon within a cone $R = \sqrt{(\Delta\phi)^2+(\Delta\theta)^2}<0.05$.

The dip in the region between the barrel and endcap calorimeters, which is reflected in the efficiency below \qty{400}{\GeV} \note{to be checked}.

effect of BIB on energy resolution
standard deviation $\sigma_{\Delta E}$ of a Gaussian fit to the distribution of
residuals $E_\gamma^{RECO}-E_\gamma$.
It may be seen that the BIB degrades significantly the photon reconstructed energy
in the region between the barrel and the endcap.

\begin{figure}[t]
\centering
\includegraphics[width=0.48\textwidth]{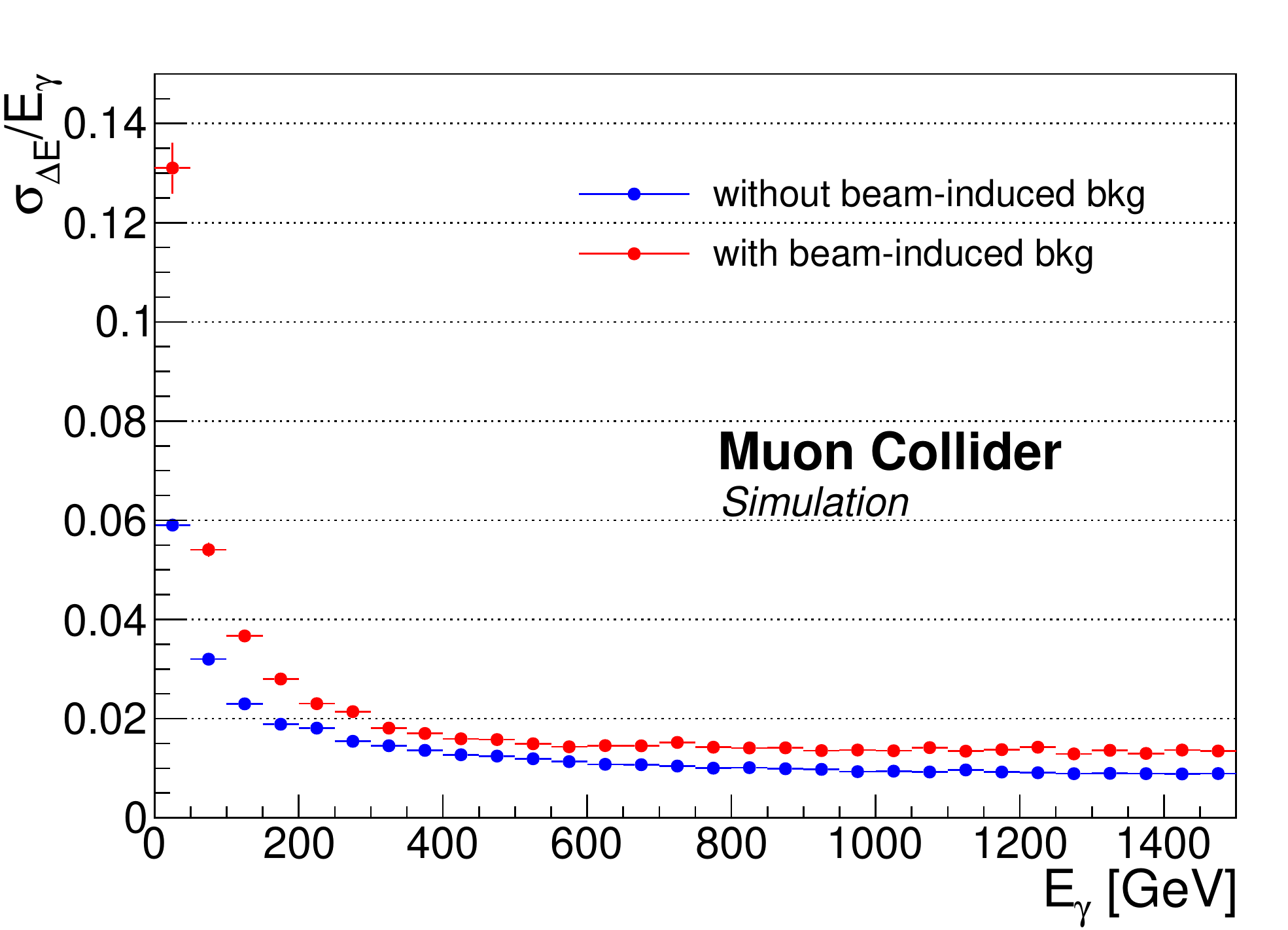}
\includegraphics[width=0.48\textwidth]{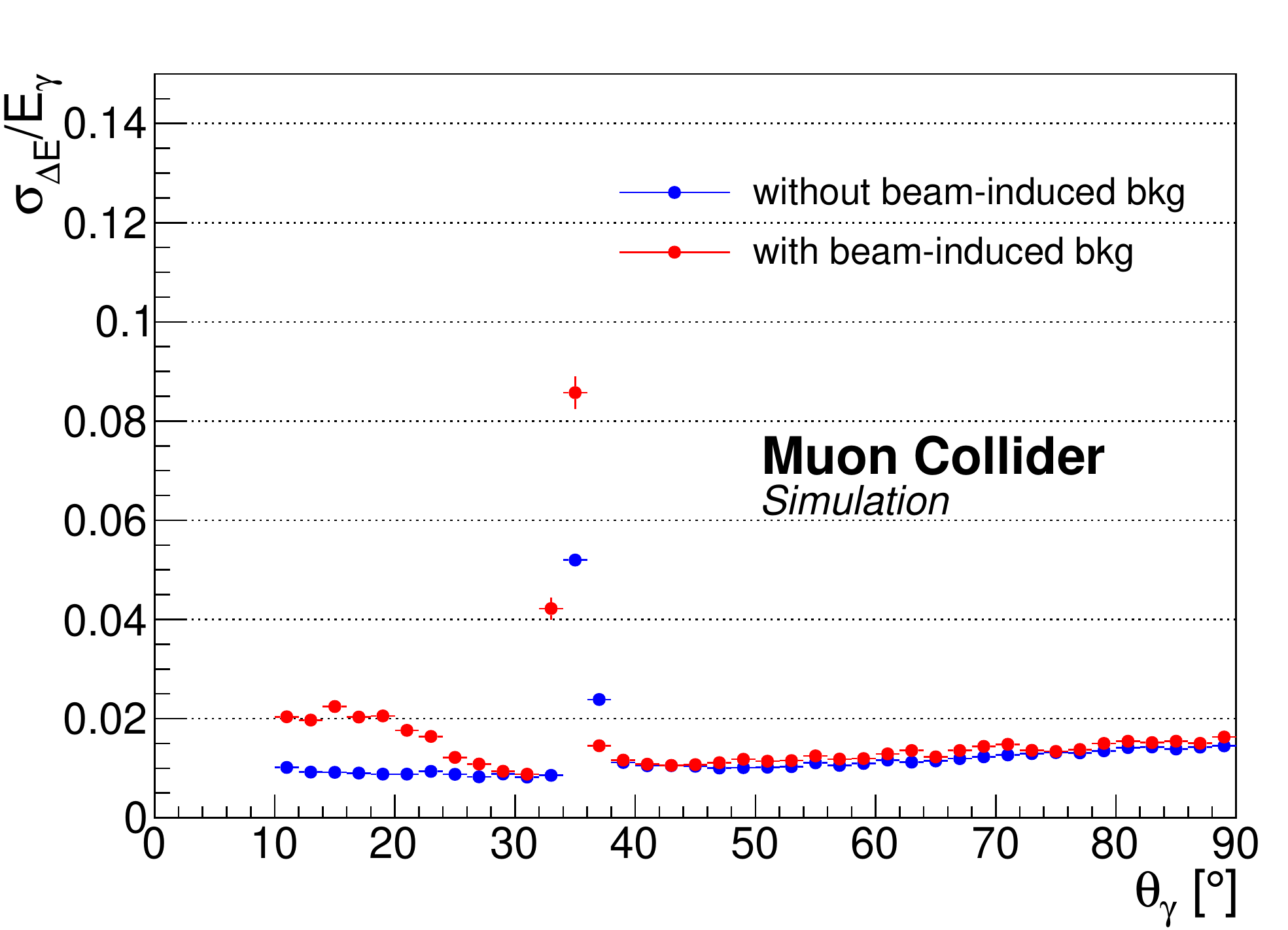}
\caption{Energy resolution of the reconstructed photond as a function of the photon
    energy (left) and the photon polar angle (right).
\label{fig:ph_eneres}}
\end{figure}    

\begin{figure}[t]
\centering
\includegraphics[width=0.48\textwidth]{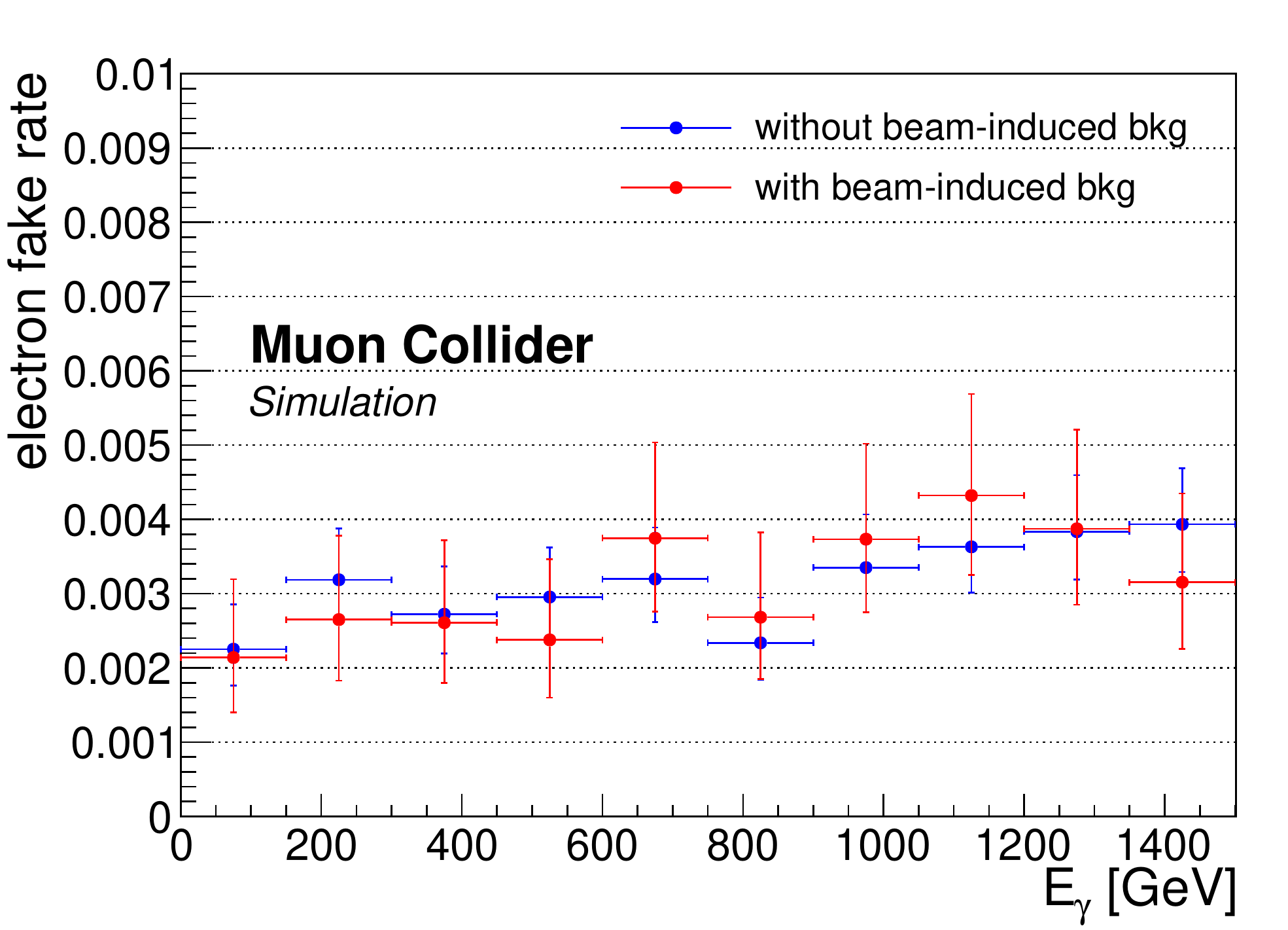}
\includegraphics[width=0.48\textwidth]{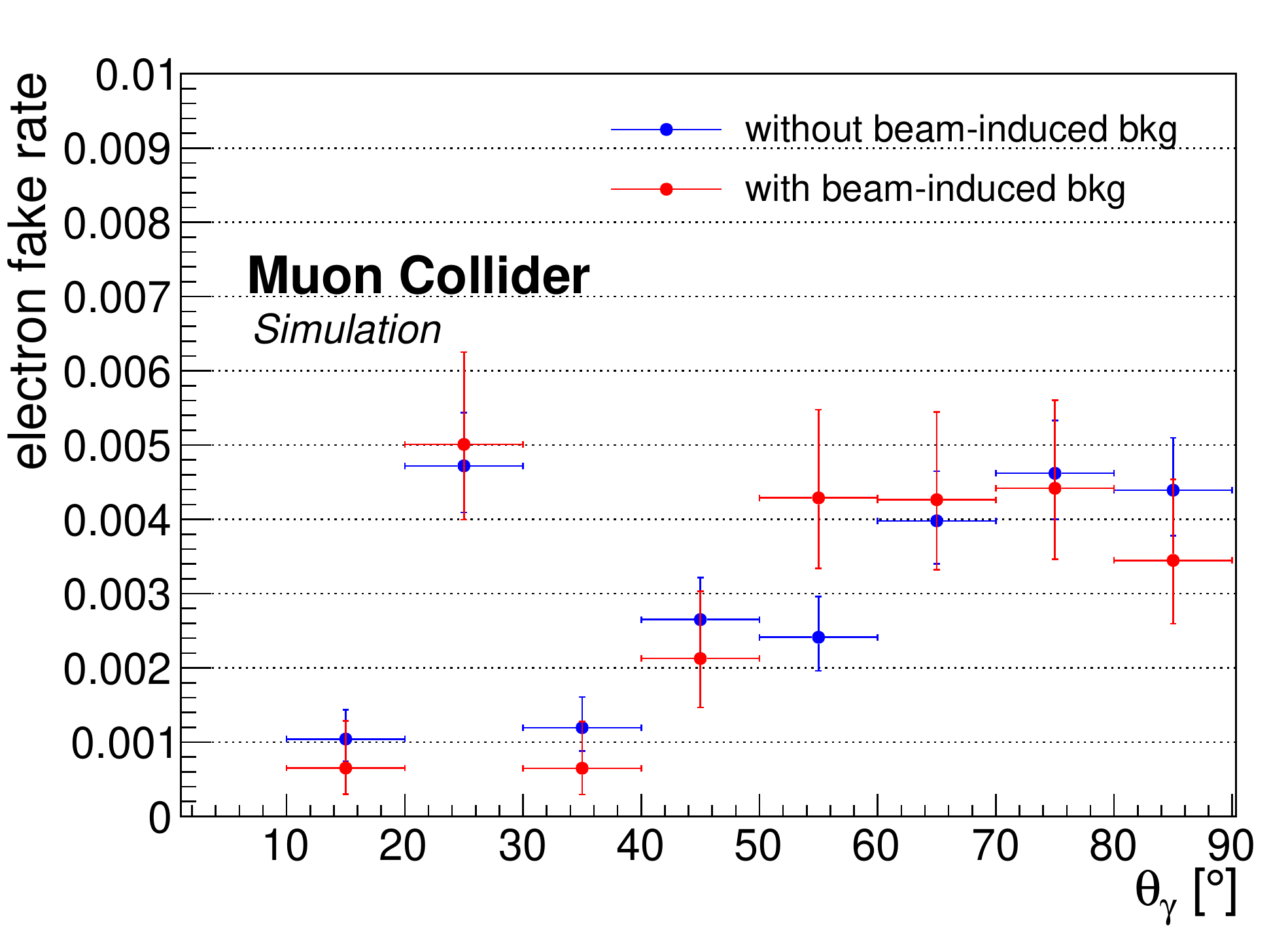}
\caption{Fraction of photons misidentified as electrons as a function of the photon energy (left)
    and the photon polar angle (right).\note{For future: we should check how many random photons we reconstruct just because of BIB as a function of photon energy/angle.}
\label{fig:ph_elefake}}
\end{figure}    

Fig.~\ref{fig:ph_elefake} reports the fraction of photons that are reconstructed
and identified as electons.

\note{Mention ML-based developments for HGCAL and their potential at the Muon Collider}.

%% file: electrons.tex
Performance for electron reconstruction and identification is obtained in a sample of 50000 events with a single electron per event produced at the nominal collision point. The generated electrons are uniformly distributed in energy between 1 and 1500 GeV, in polar angle between $10circ$ and $170circ$, and in azimuthal angle over the whole range.
The sample was processed with the detector GEANT4 simulation and the Marlin reconstruction algorithms. A subsample of 20000 events was also reconstructed with the beam-induced background overlaid.

Electrons are identified by means of an angular matching of the electromagnetic clusters, reconstructed in the same way as Section~\ref{sec:photons}, with tracks reconstructed with the CKF algorithm, as described in Section~\ref{sec:trk-ckf} in a $R=0.1$ cone.
The Double-Layer filter was used to clean the tracker hits upstream the track reconstruction and
tracks are required to have $\chi^2/\textrm{ndof} < 10$.
In the presence of the beam-induced background, the energy thresholds of the calorimeter hits play a dominant role for an efficient and precise cluster reconstruction.
In this study, a threshold of 5 MeV was set. Futher developments and more sophisticated algorithms are needed to improve the cluster reconstruction, particularly in the lower energy regime. 

\begin{figure}[t]
\centering
\includegraphics[width=0.48\textwidth]{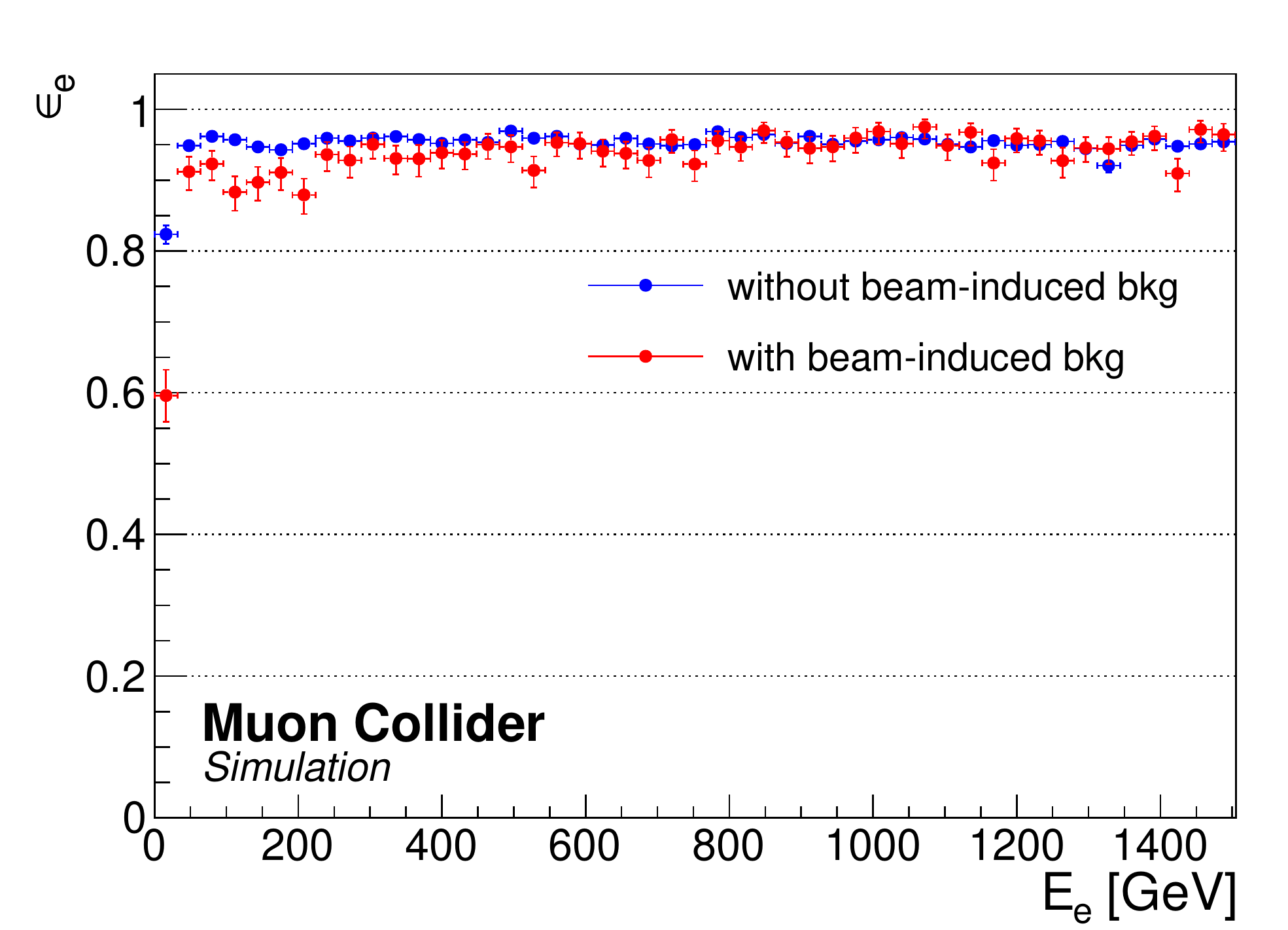}
\includegraphics[width=0.48\textwidth]{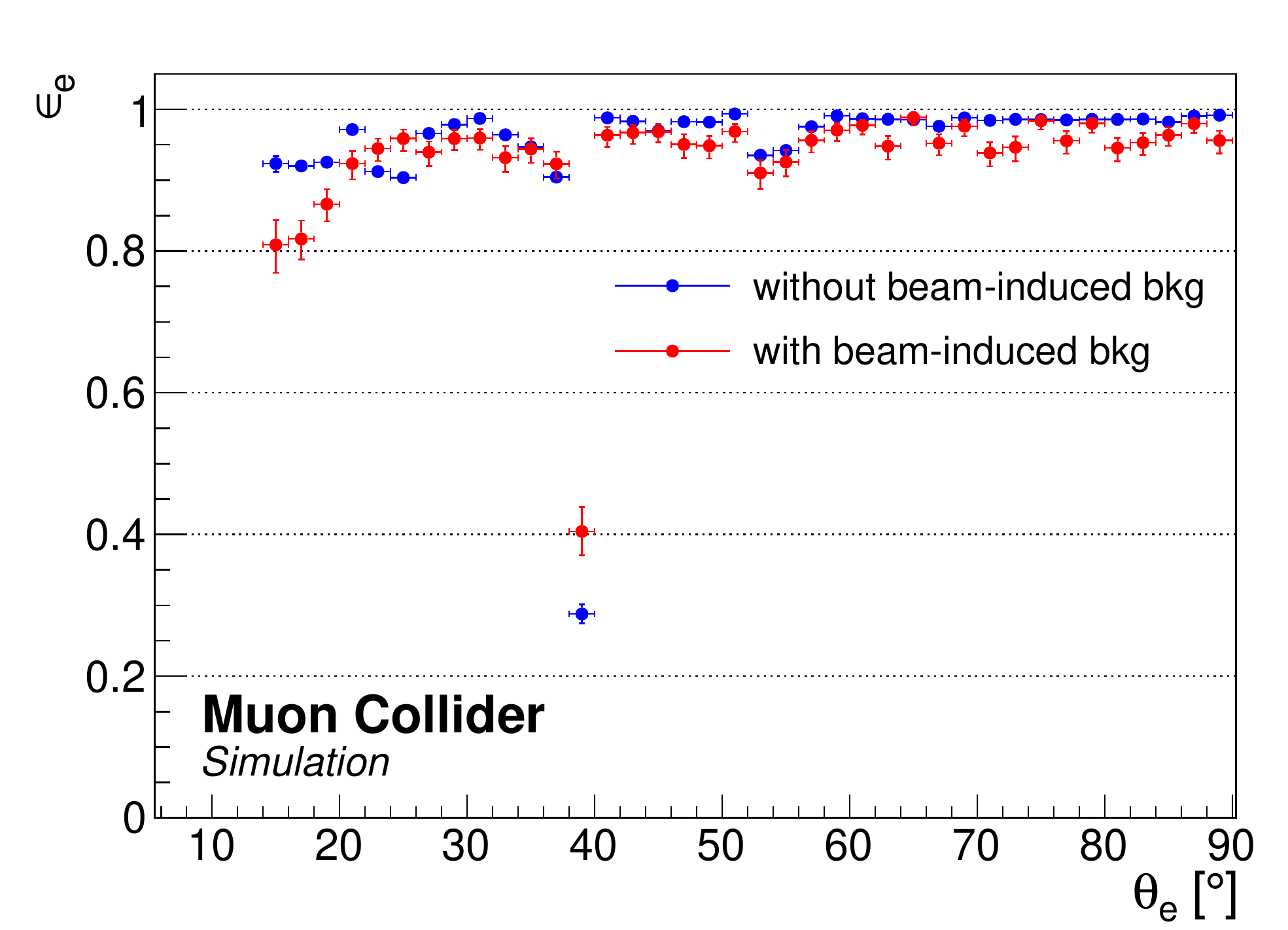}
\caption{Comparison of the electron reconstruction efficiency as a function of the electron 
    energy (left) and the electron polar angle (right) in the case of no beam-induced background and beam-induced background added to the event.
\label{fig:ele_efficiency}}
\end{figure}

The electron reconstruction and identification efficiencies as a function of the electron generated energy and polar angle are shown in Fig.~\ref{fig:ele_efficiency}.

The drop at $40\circ$ in the left panel is caused by a tracking inefficiency introduced by the
Double-Layer filter in the transition region between the VXD barrel and endcap.

%% file: muons.tex
Preliminarily, the muon reconstruction has been performed, following CLIC procedure, within the Pandora PFA framework~\cite{pandoramu}, that allows to investigate cluster topologies. A cluster is defined as a combination of hits (one hit per layer) inside a cone and on neighbouring layers. A detailed description of the muon reconstruction algorithm is reported in~\cite{clic.cdr}.

The efficiency, defined as the ratio between generated particles associated with a cluster and total generated particles, has been evaluated for single muons with transverse momentum uniformly distributed in the range \qty{100}{\MeV}-\qty{1}{\TeV} and polar angle $\qty{8}{\degree} < \theta < \qty{172}{\degree}$. No BIB was overlaid. The cluster efficiency is higher than 99\% for $\pT>\qty{10}{\GeV}$ and higher than 98\% for $\qty{8}{\degree} < \theta < \qty{172}{\degree}$.

\begin{figure}[!ht]
     \center
         \includegraphics[width=0.5\textwidth]{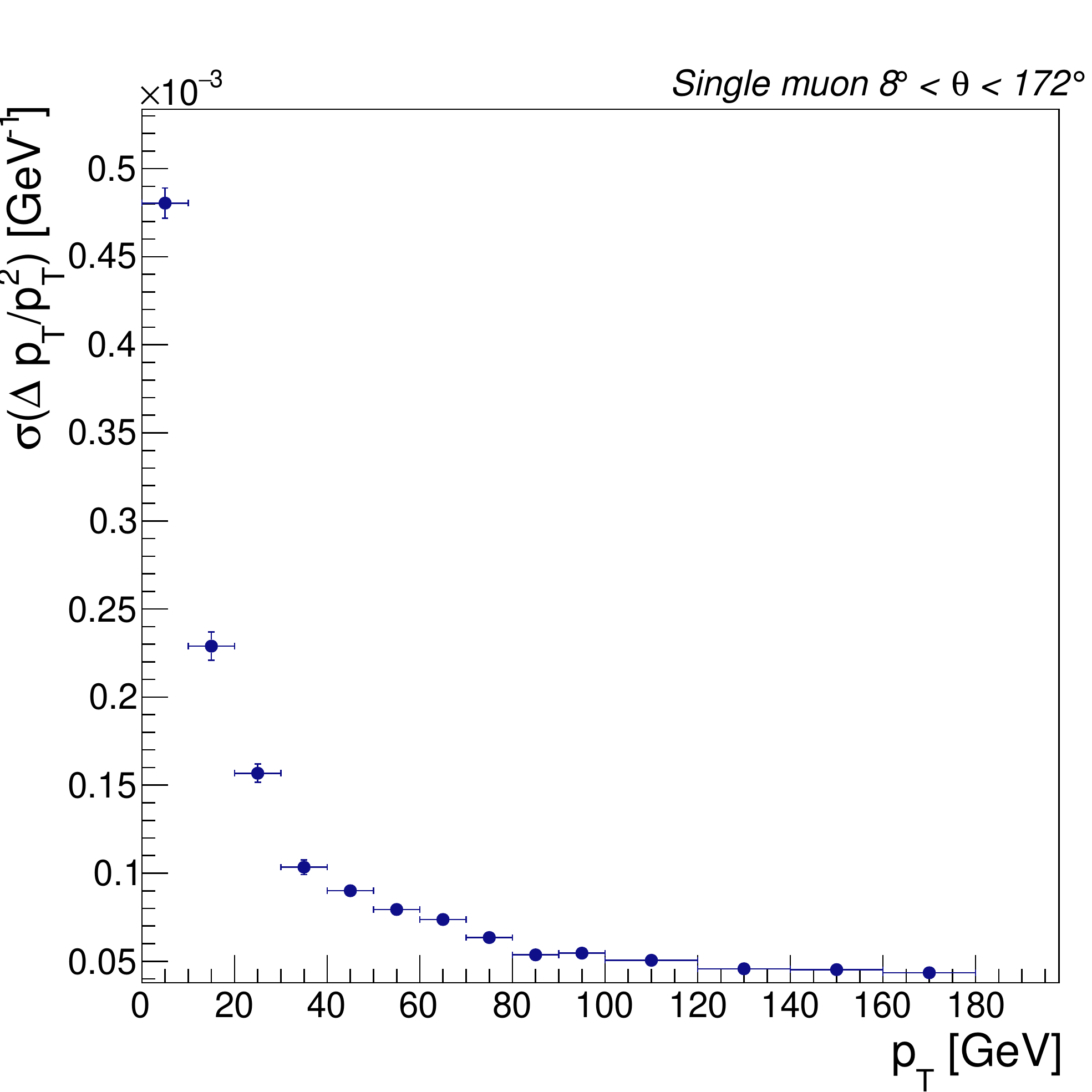}
         \vspace{-0.25cm}
    \caption{Muon track $p_{T}$ resolution as a function of $p_{T}$.}
        \label{fig:risoeZ}
\end{figure}

Muon track $p_{T}$ resolution is shown in Fig.~\ref{fig:risoeZ} where $\Delta p_{T}$ is the difference between the generated muon $p_T$ and the $p_T$ of the corresponding Pandora PFA reconstructed track. It results to be less than $10^{-4}$\,GeV$^{-1}$ for $p_{T}>30$\,GeV and around a factor of 7 better in the barrel region compare to the endcap \cite{aime_aps_poster}. 

\begin{figure}[!ht]
     \center
         \includegraphics[width=0.5\textwidth]{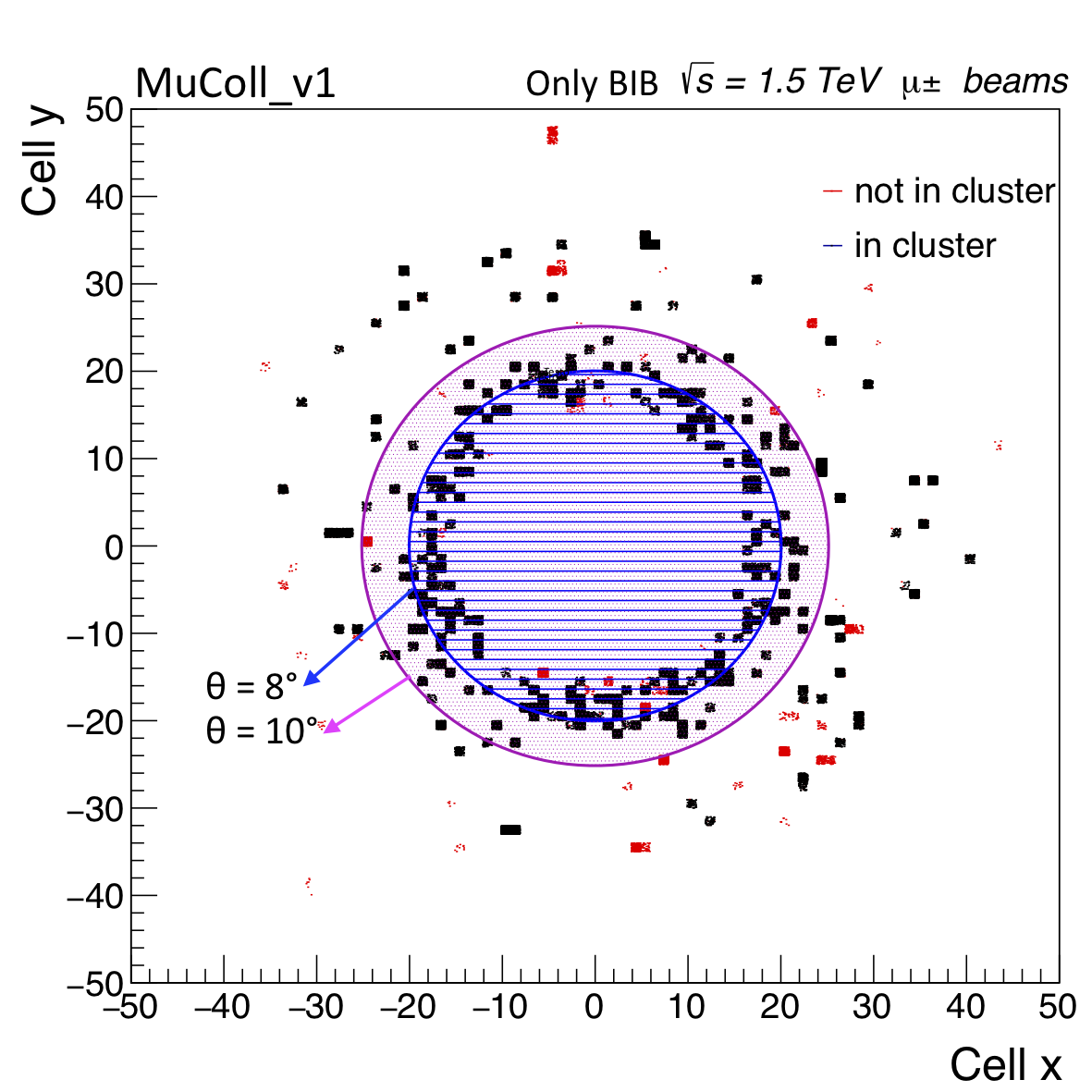}\includegraphics[width=0.5\textwidth]{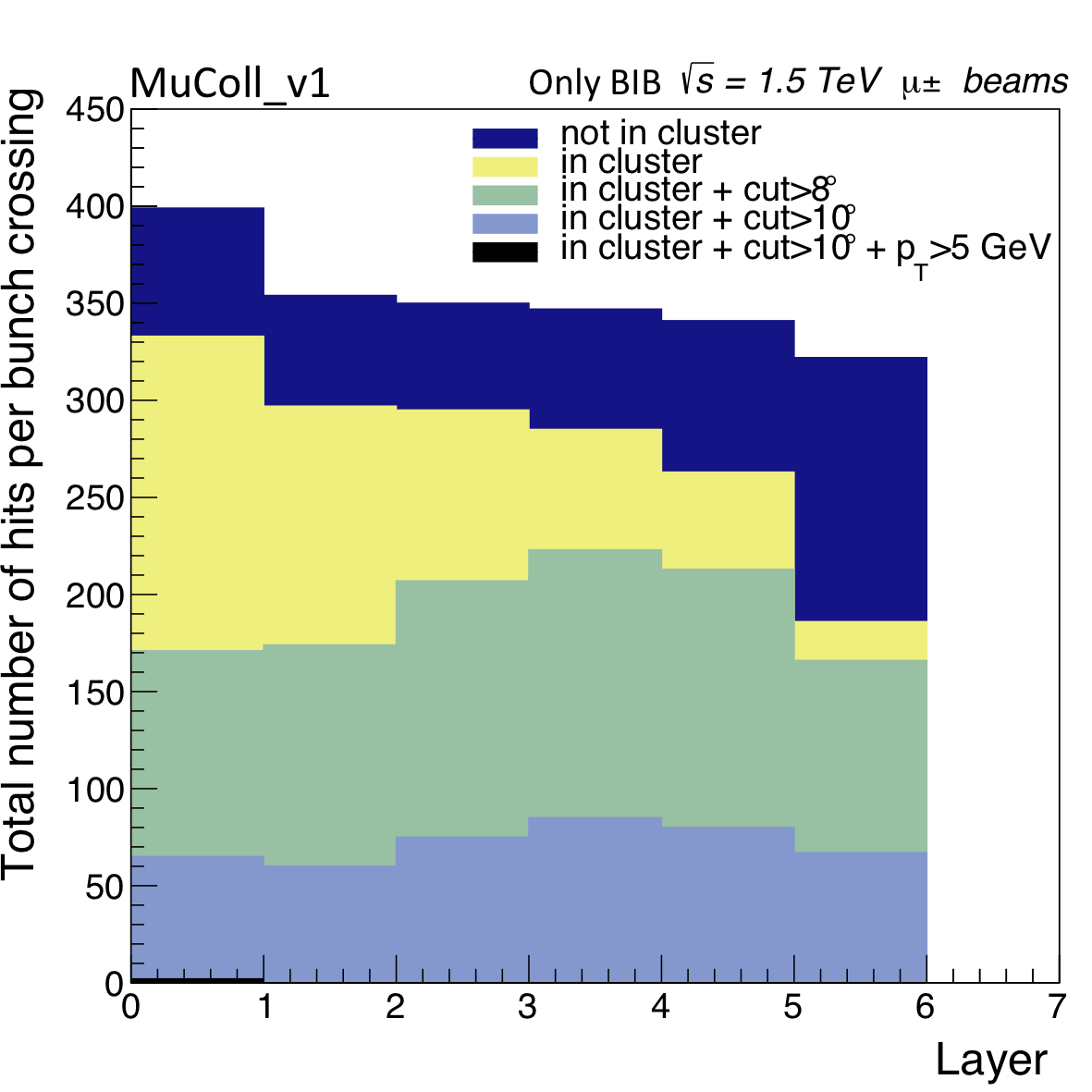}
         \vspace{-0.25cm}
 \caption{Left: BIB muon hit spatial distribution in the first layer of the muon endcap. In red the hits not associated to a cluster. The blue circle corresponds to region $\theta < 8^\circ$, while the purple to $\theta < 10^\circ$. Right: Number of hits per bunch crossing in each layer of the muon system. Different cuts are applied.}
        \label{fig:muonbib}
\end{figure}

Running the Conformal Tracking algorithm on a full event and Pandora requires an enormous computational effort to  reconstruct a single event when BIB is overlaid. Moreover, Pandora reconstruction is fully driven by the tracker where the BIB hit density is very high as already discussed before. At the same time, the BIB occupancy in the muon system is very low compared to other detectors. BIB hits are concentrated in the endcaps around the beam axis (Fig.\ref{fig:muonbib} left) and a simple geometrical cut, combined for example with a cut on the track transverse momentum, allows to get rid of almost all the BIB hits (Fig.\ref{fig:muonbib} right). This suggests using standalone muon objects to seed the global muon track reconstruction. 

A processor has been developed to clusterize muon hits inside a cone with a certain angular aperture $\Delta R_\mu = \sqrt{(\Delta \phi)^2 + (\Delta \eta)^2}$. At least hits on five layers are required to make a stand-alone muon track. Reconstructed hits in the Vertex Detector, Inner Tracker and Outer Tracker are then filtered using the region-of-interest approach described in Section~\ref{sec:trk-roi}: only hits in a cone of aperture $\Delta R_t $ (see below) around the muon stand-alone direction are selected.  These hits are then passed to the CT algorithm. At the moment no information coming from the ECAL and HCAL calorimeters is used.

\begin{figure}
     \center
         \includegraphics[width=0.5\textwidth]{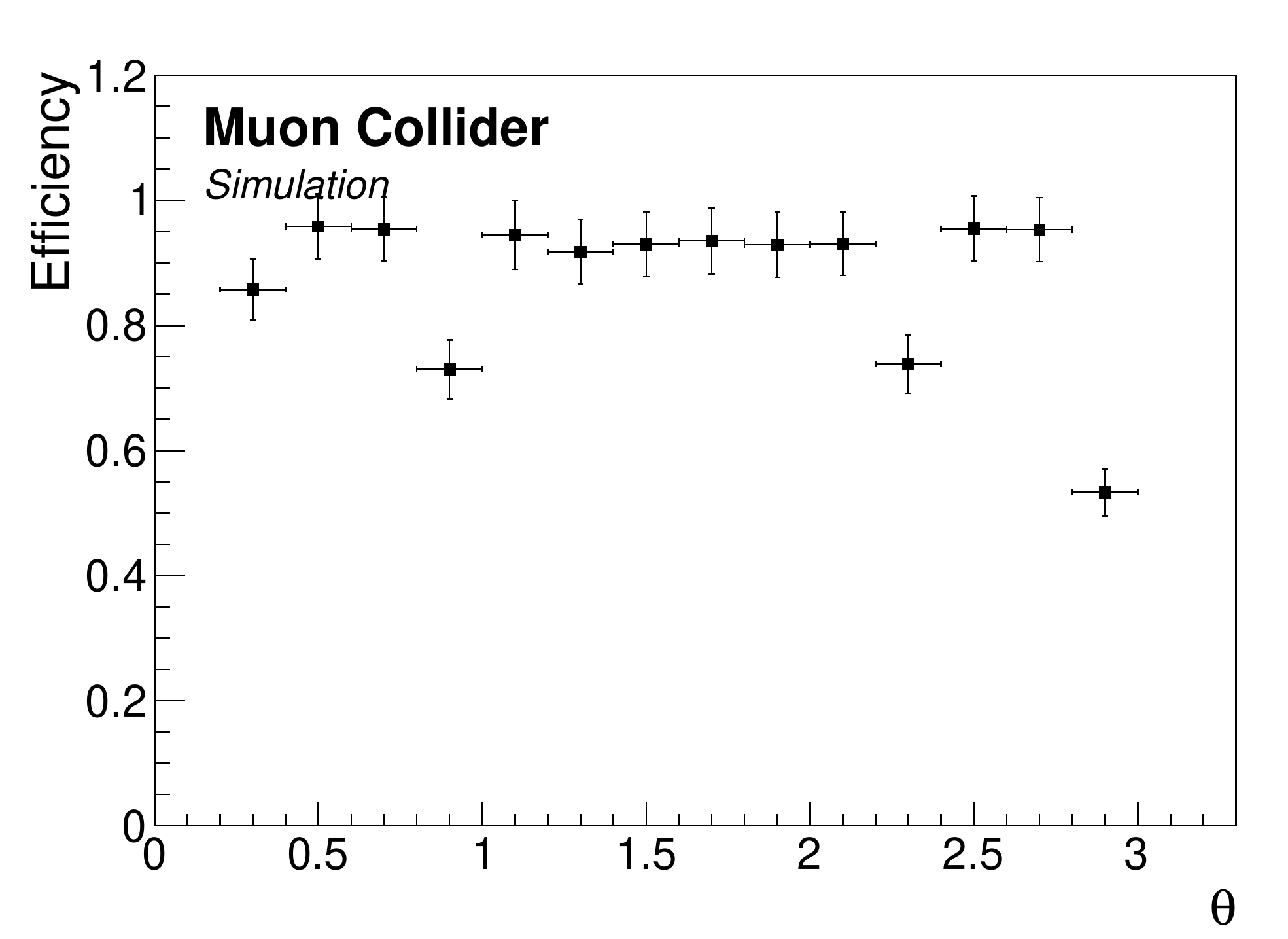}\includegraphics[width=0.5\textwidth]{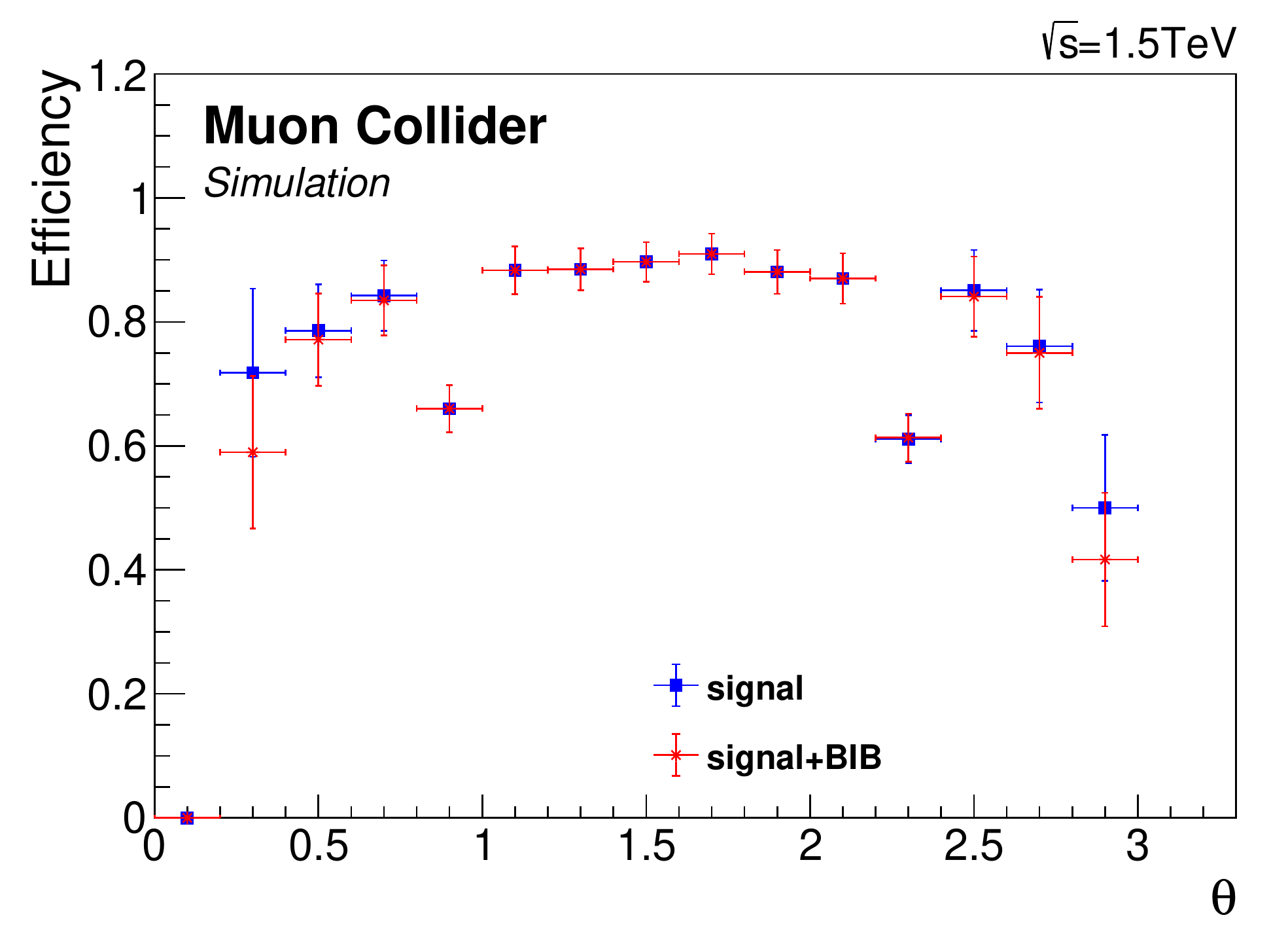}
         \vspace{-0.25cm}
 \caption{Muon reconstruction efficiency as a function of the polar angle in a sample of single muons with no BIB overlaid (left) and in a sample with multi-muons in the final state both with and without BIB (right).}
        \label{fig:muon_thetaefficiency}
\end{figure}

\begin{figure}
     \center
         \includegraphics[width=0.5\textwidth]{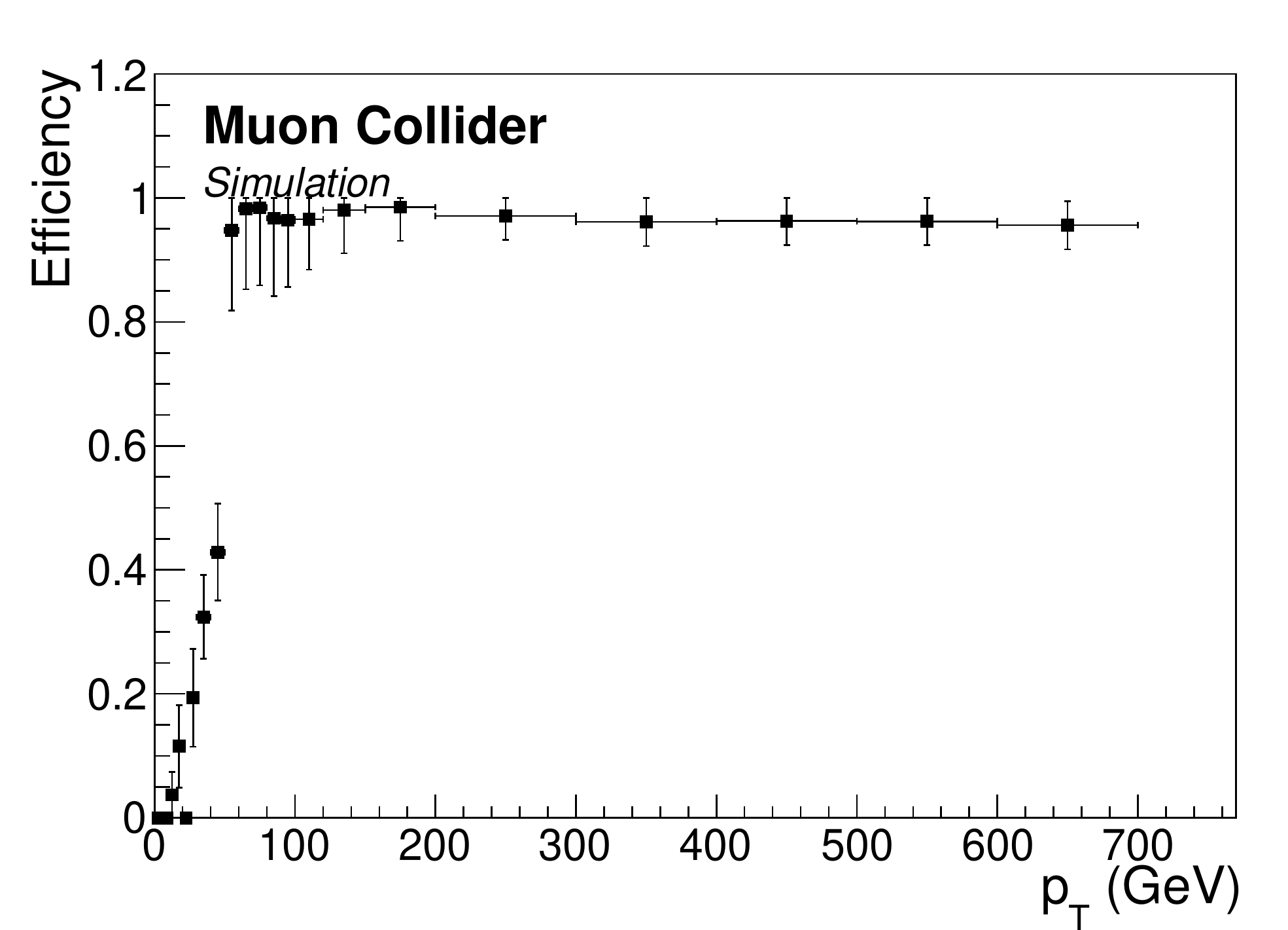}\includegraphics[width=0.5\textwidth]{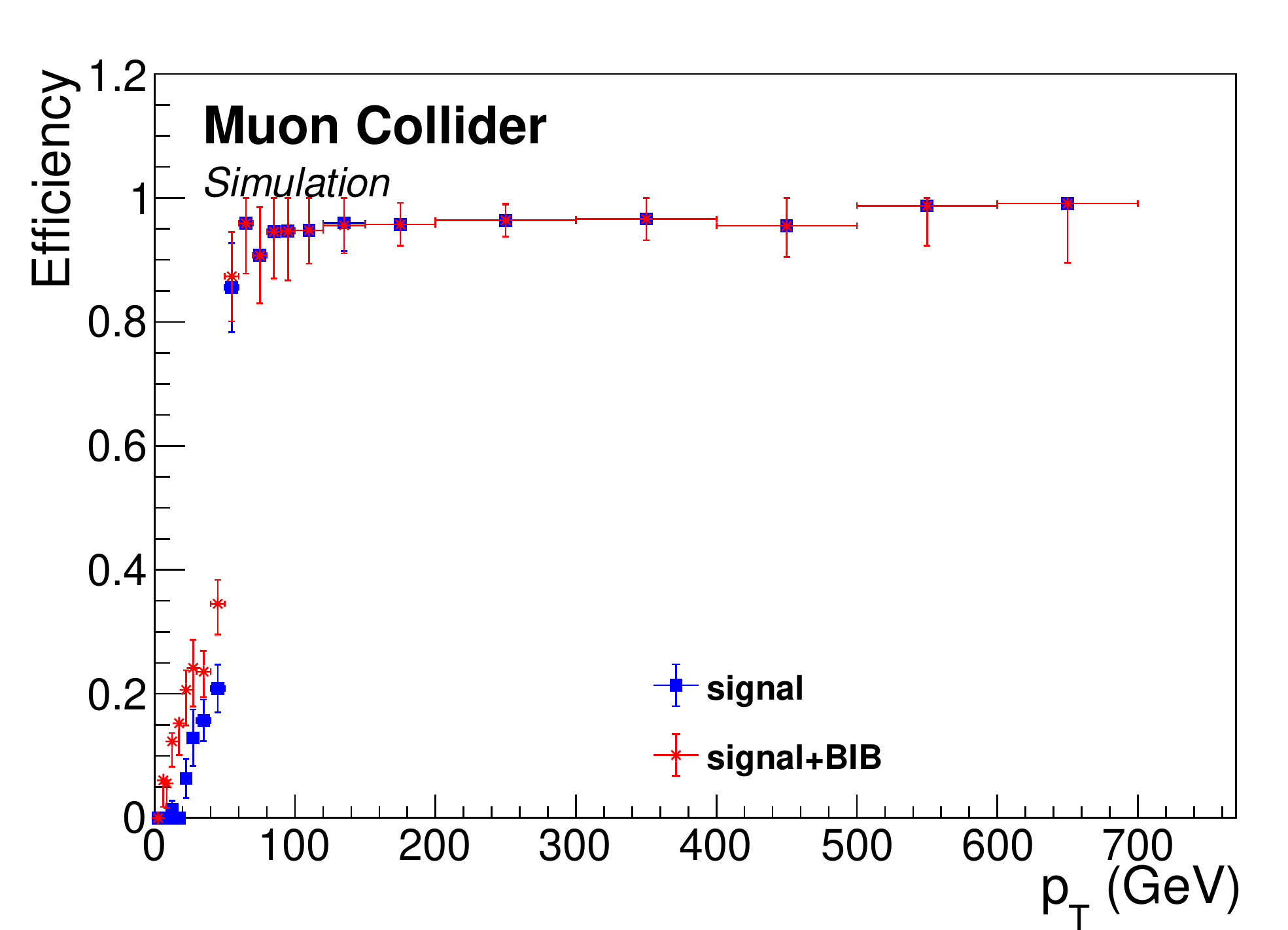}
         \vspace{-0.25cm}
 \caption{Muon reconstruction efficiency as a function of transverse momentum in a sample of single muons with no BIB overlaid (left) and in a sample with multi-muons in the final state both with and without BIB (right). \note{Why inefficient below ~80 GeV? checking if a matching of 0.01 in $\Delta R$ is too tight.}}
        \label{fig:muon_ptefficiency}
\end{figure}

Two different samples have been used to evaluate the performance of the processor:
\begin{enumerate}
    \item single muons with transverse momentum uniformly distributed in the range 100 MeV-700 GeV and polar angle $8^\circ <\theta < 172^\circ$ with no BIB overlaid
    \item HZ channel resulting in six muons in the final state at a center-of-mass energy of \SI{1.5}{TeV} both with and without BIB.
\end{enumerate}
The angular apertures chosen to maximize track purity and mitigate the background are $\Delta R_\mu =0.02$ and $\Delta R_t =0.05$.

\begin{figure}
     \center
         \includegraphics[width=0.5\textwidth]{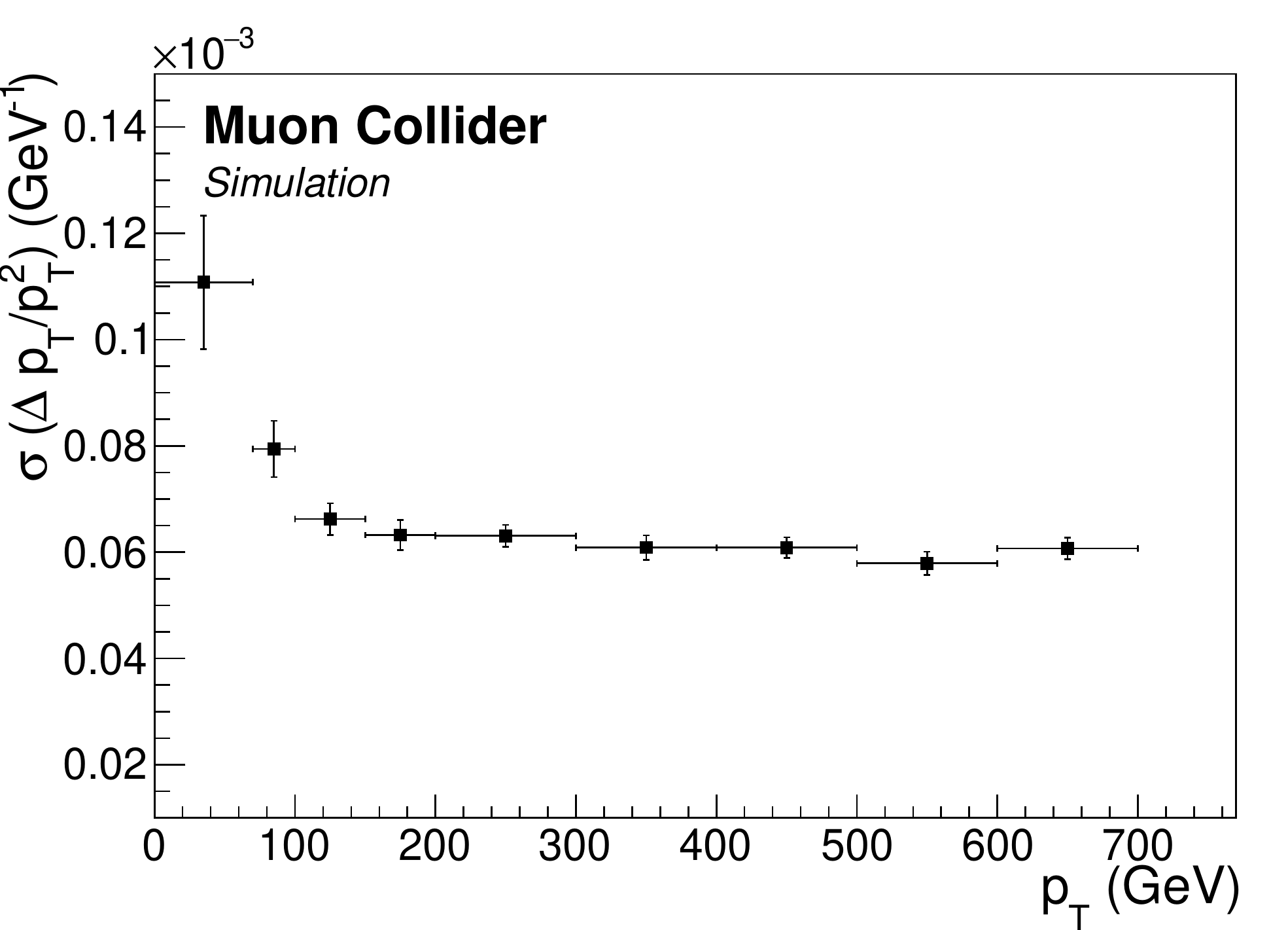}\includegraphics[width=0.5\textwidth]{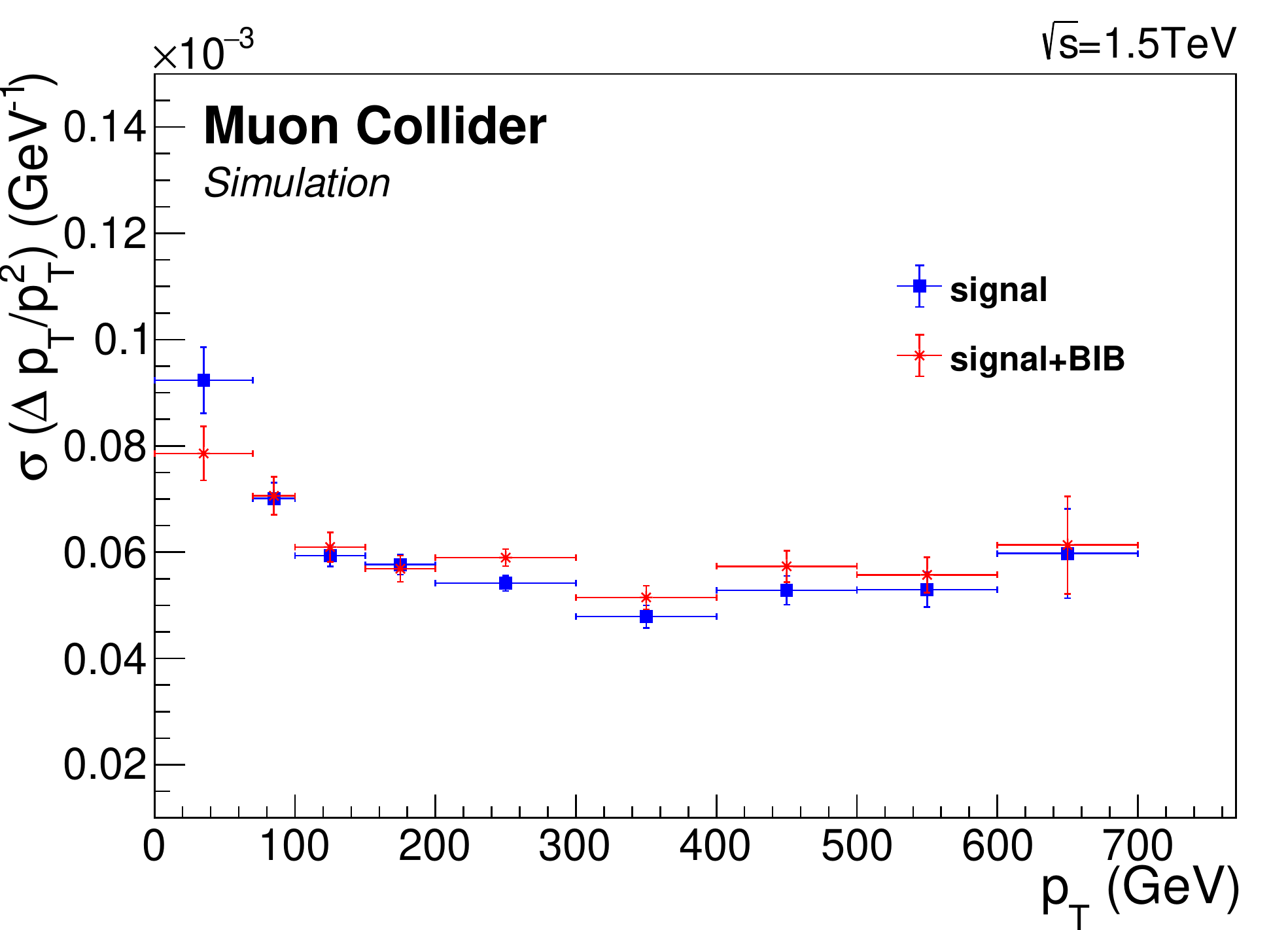}
         \vspace{-0.25cm}
 \caption{Muon track transverse momentum resolution as a function of \pT in a sample of single muons with no BIB overlaid (left) and in a sample with multi-muons in the final state both with and without BIB (right). }
        \label{fig:muon_ptresolution}
\end{figure}

Efficiencies are defined as the fraction of generated muons that are matched to a track within  a cone of aperture 0.01. 
Figure~\ref{fig:muon_thetaefficiency} shows the efficiency as a function of the polar angle $\theta$ both for the first (left) and second sample (right). An efficiency loss is evident in both cases at $\theta<15^\circ$ and $\theta>165^\circ$ due to a limit of the CT algorithm, while the two dips around $\theta\sim 52^\circ$ and $\theta\sim 132^\circ$ are due to tracks with hits both in endcap and barrel that require a specific approach, currently under study.

To study the efficiency as a function of transverse momentum the acceptance region was restricted to $15^\circ <  \theta < 165^\circ$ to take into account only the region with a high efficiency for CT. Figure~\ref{fig:muon_ptefficiency} shows an efficiency greater than 85\% for transverse momentum higher than \SI{80}{GeV} for both the samples. The drop of the curve at low $p_T$ is due to known effects currently under study, that mostly concern the inefficiency of reconstruction of muons in the region between barrel and endcap and tracks with a high curvature for which the choice of $\Delta R_t$ has to be optimized.  The muon track transverse momentum resolution (Fig. \ref{fig:muon_ptresolution}) is comparable with that obtained with CT and Pandora without BIB.

In general the presence of the BIB does not strongly affect the performance: the efficiency is lower only in the endcaps where all the BIB hits are concentrated, and the $p_{T}$ resolution is just a few percent worse. 

%% file: lumi.tex
A precise determination of the luminosity is of crucial importance for any physics absolute cross section measurement since it directly translates to its error.
LHC experiments have dedicated detectors, so-called luminometers, installed often the backward/forward region of the detector that are used in combination to the van Der Meer scan method to precisely measure the instantaneous luminosity. 
The $e^+e^-$ experiments like Belle2\cite{belle2} and BESIII\cite{BesIII}, measure the integrated luminosity by counting the number of events of a process whose cross section is theoretically known with high precision.
The most used one is the Bhabha scattering ($e^+e^- \rightarrow e^+e^-$) where, for example, the theoretical uncertainty on $\sigma$ at  $\sqrt{s}=1-10$ GeV is 0.1\% at large angle \cite{bhabha_cross_section}.
Due to the reduced acceptance in the backward/forward region because of the nozzles shielding structure at muon collider, the large angle muon Bhabha ($\mu^+$ $\mu^-$ $\rightarrow$ $\mu^+$ $\mu^-$) is proposed as a method for the luminosity measurement instead.
A first evaluation of the expected precision has been performed by using a sample of $\mu$-Bhabha generated with Pythia at the center-of-mass energy of \qty{1.5}{\TeV}. The events are then simulated through the detector and reconstructed by using the Pandora Particle Flow algorithm \cite{THOMSON200925}. 

Muons with emission angle $\theta$ respect to the beam-pipe $30^\circ <\theta< 150^\circ$ are selected. The beam-induced background effects in this angular region are negligible, as shown in Section~\ref{sec:muons}. Two opposite sign charge muons, both with transverse momentum $\pt > \qty{130}{\GeV}$ and angle between muons direction pair $\Delta\theta>3.08$ rad are required. Finally, the di-muon invariant mass has to be $1440<M_{\mu\mu}<1560$ GeV. These selection requirements reject almost all the physics background contributions and allow to keep the entire signal sample at $\sqrt{s}=1.5$ TeV. A different center of mass energy, muon reconstruction and sample selection need to be re-optimized.

By assuming an instantaneous luminosity $\mathcal{L} = 1.25 \times 10^{34} $ cm$^{-2}$s$^{-1}$ and considering a year of data taking ($10^7$ s) the statistical uncertainty obtained at $\sqrt{s}=1.5$ TeV is 0.2\% \cite{LHCP-Proc}.
This study demonstrates that the method can be applied at the muon collider from the experimental point of view. The expected total uncertainty on the luminosity measurement using this method strongly depends on the accuracy on the theoretical cross section of $\mu$-Bhabha scattering at large angles, and at high center of mass energy( several TeV). Further theoretical developments are needed in this area.

%% file: conclusions.tex
A \mumu collider offers a viable path towards a rich and scalable physics program at the energy frontier. 
One of the biggest challenges in designing a detector for the muon collider environment is the presence of beam-induced background. 
In this manuscript, we presented the performance of a detector design, including appropriate shielding near the interaction point, that can successfully cope with the expected beam-induced background and reconstruct with high accuracy the main physics observables needed for carrying out the expected physics program. Expected performance for reconstructing charged particles, jets, electrons, muons and photons has been presented, in addition to preliminary results on tagging heavy-flavor jets and measuring the delivered integrated luminosity.